\documentclass[11pt]{article}
\pdfoutput=1
\usepackage{jcapmod}
\usepackage{shorthand}
\usepackage[scaled]{helvet}
\usepackage[toc,page]{appendix}
\usepackage{mathtools}
\usepackage{booktabs}
\usepackage[english]{babel}
\usepackage{amsmath,amssymb,amsbsy,amstext, amsthm, simplewick, amsfonts}
\usepackage{hyperref}
\usepackage{graphicx}
\usepackage[small]{caption}
\usepackage{siunitx}
\usepackage{upgreek}
\usepackage{framed}
\usepackage{wrapfig}
\usepackage{multirow}
\usepackage{bbm}
\usepackage[svgnames,dvipsnames,x11names]{xcolor}
\usepackage[utf8x]{inputenc}
\usepackage{selinput}
\usepackage{bm}
\usepackage{float}
\usepackage{geometry}
\usepackage{yfonts}
\usepackage{caption}
\usepackage{subcaption}
\usepackage{sidecap}
\usepackage{longtable}
\usepackage{anyfontsize}
\usepackage{dsfont}
\usepackage{tikz}
\usepackage{relsize}
\usepackage{shuffle}
\usetikzlibrary{matrix}
\usepackage{bigstrut}
\usepackage{fnpct}
\usetikzlibrary{arrows.meta}
\usetikzlibrary{decorations.pathreplacing,calligraphy}
\usepackage{tensor}
\usepackage{mleftright}
\usepackage[most]{tcolorbox}
\usepackage{fancybox}

\usetikzlibrary{decorations.markings, decorations.pathmorphing, decorations.pathreplacing, decorations.shapes}

\usepackage{xcolor}
\usepackage{xparse}
\definecolor{mplBlue}{rgb}{0,0,1}   
\definecolor{mplRed}{rgb}{1,0,0}    
\definecolor{mplGreen}{rgb}{0,0.5,0}   

\usepackage{fontawesome5}

\makeatletter
\newcommand{\github}[1]{%
   \href{#1}{\faGithub}%
}
\makeatother

\NewDocumentCommand{\colornucleus}{omme{_^}}{%
  \begingroup\colorlet{currcolor}{.}%
  \IfValueTF{#1}
   {\textcolor[#1]{#2}}
   {\textcolor{#2}}
    {%
     #3
     \IfValueT{#4}{_{\textcolor{currcolor}{#4}}}
     \IfValueT{#5}{^{\textcolor{currcolor}{#5}}}
    }%
  \endgroup
}

\usepackage{array}
\newcolumntype{L}[1]{>{\raggedright\let\newline\\\arraybackslash\hspace{0pt}}m{#1}}
\newcolumntype{C}[1]{>{\centering\let\newline\\\arraybackslash\hspace{0pt}}m{#1}}
\newcolumntype{R}[1]{>{\raggedleft\let\newline\\\arraybackslash\hspace{0pt}}m{#1}}

\usepackage[framemethod=default]{mdframed}
\newmdenv[skipabove=7pt,
skipbelow=7pt,
rightline=true,
leftline=true,
topline=true,
bottomline=true,
backgroundcolor=gray!10,
linecolor=black,
innerleftmargin=5pt,
innerrightmargin=5pt,
innertopmargin=5pt,
innerbottommargin=5pt,
leftmargin=0cm,
rightmargin=0cm,
linewidth=1pt]{eBox}

\definecolor{Red}{RGB}{214, 39, 40}
\definecolor{Blue}{RGB} {31, 119, 180}
\definecolor{Orange}{RGB}{255, 153, 51}
\definecolor{Purple}{RGB}{178, 102, 255}
\definecolor{Green}{RGB}{44, 160, 44}
\definecolor{regal}{RGB}{90,0,120}
\definecolor{darkblue}{rgb}{0.15,0.35,0.55}
\definecolor{reddish}{rgb}{0.65, 0.2, 0.2}
\definecolor{darkgreen}{RGB}{50,150,0}
\definecolor{greyish}{rgb}{.90,.90,.90}
\definecolor{greyish2}{rgb}{.96,.96,.96}
\definecolor{greyish3}{rgb}{.37,.37,.37}
\definecolor{darkblue2}{rgb}{0.3,0.4,0.9}
\definecolor{Blue3}{RGB}{31, 119, 180}

\definecolor{blue3}{RGB}{31, 119, 180}
\definecolor{red3}{RGB}{	214, 39, 40}
\definecolor{orange3}{RGB}{255, 127, 14}
\definecolor{green3}{RGB}{44, 160, 44}
\definecolor{repBlue}{RGB}{31, 119, 180}
\definecolor{repRed}{RGB}{	214, 39, 40}
\definecolor{repGreen}{RGB}{44, 160, 44}

\newcommand\diff{\mathrm{d}}
\newcommand{\smin}{\rule[0.3ex]{0.2em}{0.1ex}}
\newcommand{\splus}{\scalebox{0.5}[0.5]{+}}
\newcommand{\ah}{\scalebox{0.5}[0.5]{\text{AH}}}
\newcommand{\ahe}{\scalebox{0.4}[0.4]{\text{AH}}}
\newcommand{\sz}{\scalebox{0.7}[0.7]{0}}

\newcommand{\ya}{y_{\raisebox{-0.5ex}{$\scriptscriptstyle \hspace{-0.2em}\cA$}}}

\newcommand{\ybcap}{y_{\raisebox{-0.2ex}{$\scriptscriptstyle \hspace{-0.15em}B$}}}

\newcommand{\xacap}{x_{\raisebox{-0.2ex}{$\scriptscriptstyle \hspace{-0.15em}A$}}}
\newcommand{\xacape}{x_{\hspace{-0.05em}\raisebox{-0.25ex}{\scalebox{0.45}{$A$}}}}
\newcommand{\xbcap}{x_{\raisebox{-0.2ex}{$\scriptscriptstyle \hspace{-0.15em}B$}}}
\newcommand{\xbcape}{x_{\hspace{-0.05em}\raisebox{-0.25ex}{\scalebox{0.45}{$B$}}}}
\newcommand{\phiacap}{\phi_{\raisebox{-0.2ex}{$\scriptscriptstyle \hspace{-0.15em}A$}}}
\newcommand{\phiacape}{\phi_{\hspace{-0.05em}\raisebox{-0.25ex}{\scalebox{0.45}{$A$}}}}
\newcommand{\phibcap}{\phi_{\raisebox{-0.2ex}{$\scriptscriptstyle \hspace{-0.15em}B$}}}
\newcommand{\phibcape}{\phi_{\hspace{-0.05em}\raisebox{-0.25ex}{\scalebox{0.45}{$B$}}}}
\newcommand{\yam}{y_{\raisebox{-0.5ex}{$\scriptscriptstyle \hspace{-0.2em}\cA^{\smin}$}}}
\newcommand{\yb}{y_{\raisebox{-0.5ex}{$\scriptscriptstyle \hspace{-0.1em}\cB$}}}
\newcommand{\ybm}{y_{\raisebox{-0.5ex}{$\scriptscriptstyle \hspace{-0.1em}\cB^{\smin}$}}}
\newcommand{\ybp}{y_{\raisebox{-0.5ex}{$\scriptscriptstyle \hspace{-0.1em}\cB^{\splus}$}}\hspace{-0.15em}}
\newcommand{\yc}{y_{\raisebox{-0.5ex}{$\scriptscriptstyle \hspace{-0.1em}\cC$}}}
\newcommand{\yu}{y_{\raisebox{-0.5ex}{$\scriptscriptstyle \hspace{-0.01em}\mathrm{U}$}}\hspace{-0.08em}}
\newcommand{\xa}{x_{\raisebox{-0.25ex}{$\scriptscriptstyle \hspace{-0.1em}\cA$}}}
\newcommand{\xc}{x_{\raisebox{-0.25ex}{$\scriptscriptstyle \hspace{-0.1em}\cC$}}}
\newcommand{\xcbar}{\bar{x}_{\raisebox{-0.25ex}{$\scriptscriptstyle \hspace{-0.1em}\cC$}}}
\newcommand{\xce}{x_{\hspace{-0.05em}\raisebox{-0.25ex}{\scalebox{0.45}{$\cC$}}}}
\newcommand{\xb}{x_{\raisebox{-0.25ex}{$\scriptscriptstyle \hspace{-0.1em}\cB$}}}
\newcommand{\xbe}{x_{\hspace{-0.05em}\raisebox{-0.25ex}{\scalebox{0.45}{$\cB$}}}}
\newcommand{\xbbar}{\bar{x}_{\raisebox{-0.25ex}{$\scriptscriptstyle \hspace{-0.1em}\cB$}}}
\newcommand{\xbebar}{\bar{x}_{\hspace{-0.05em}\raisebox{-0.25ex}{\scalebox{0.45}{$\cB$}}}}

\newcommand{\phia}{\phi_{\raisebox{-0.5ex}{$\scriptscriptstyle \hspace{-0.15em}\cA$}}}
\newcommand{\phic}{\phi_{\raisebox{-0.5ex}{$\scriptscriptstyle \hspace{-0.1em}\cC$}}}
\newcommand{\phiu}{\phi_{\raisebox{-0.5ex}{$\scriptscriptstyle \hspace{-0.05em}\mathrm{U}$}}}
\newcommand{\Ia}{I_{\raisebox{-0.5ex}{$\scriptscriptstyle \hspace{-0.2em}\cA$}}}
\newcommand{\Ib}{I_{\raisebox{-0.5ex}{$\scriptscriptstyle \hspace{-0.1em}\cB$}}}
\newcommand{\Ja}{J_{\raisebox{-0.5ex}{$\scriptscriptstyle \hspace{-0.2em}\cA$}}}
\newcommand{\Jb}{J_{\raisebox{-0.5ex}{$\scriptscriptstyle \hspace{-0.1em}\cB$}}}
\newcommand{\largeD}{large-\hspace{-0.1em}$D$ }

\definecolor{vio}{RGB}{19, 130, 164}
\definecolor{vioo}{RGB}{89, 2, 155}
\newcommand{\Comment}[1]{{}}
\usepackage{empheq}

\usepackage[linktocpage=true]{hyperref}
\hypersetup{
colorlinks=true,
citecolor=darkblue,
linkcolor=reddish,
urlcolor=darkblue,
pdfauthor={},
pdftitle={},
pdfsubject={}
}

\usepackage{colortbl}
\definecolor{lightgreen}{cmyk}{0.2, 0, 0.2, 0.2}
\definecolor{lightgray2}{cmyk}{0.1,0.1,0,0.1}
\definecolor{Red2}{RGB}{214, 39, 40}
\definecolor{Blue2}{RGB} {31, 119, 180}
\definecolor{Orange2}{RGB}{255, 127, 14}
\definecolor{Green2}{RGB}{44, 160, 44}

\makeatletter
\newlength{\apb@width}
\newcommand{\autoparbox}[2][c]{\settowidth{\apb@width}{#2}\parbox[#1]{\apb@width}{#2}}

\makeatother


\def\beq{\begin{equation}}
\def\eeq{\end{equation}}
\def\be{\begin{equation}}
\def\ee{\end{equation}}

\newcommand\ve{\varepsilon}
\newcommand{\order}[1]{\cO\left(#1\right)}
\DeclarePairedDelimiter\abs{\lvert}{\rvert}

\allowdisplaybreaks[1]
\newcommand\sqmatrix[2][c]{%
  \fixTABwidth{T}%
  \setbox0=\hbox{$\tabbedCenterstack{#2}$}%
  \setstackgap{L}{\dimexpr\maxTAB@width+\tabbed@gap}%
  \tabbedCenterstack[#1]{#2}%
}

\usetikzlibrary{shapes.misc}

\tikzset{cross/.style={cross out, draw=black, minimum size=2*(#1-\pgflinewidth), inner sep=0pt, outer sep=0pt},
cross/.default={1pt}}

\usetikzlibrary{calc}

\tcbset{
	mybox/.style={
		colframe=black,
		colback=white,
		boxrule=0.5pt,
		arc=3pt,
		width=\textwidth-25pt,
		boxsep=8pt,
		left=10pt, right=5pt, top=-10pt, bottom=5pt
	}
}

\begin{document}


\newgeometry{top=2cm, bottom=2cm, left=2cm, right=2cm}

\begin{titlepage}
\setcounter{page}{1} \baselineskip=15.5pt 
\thispagestyle{empty}

\begin{center}
{\fontsize{18}{18} \bf Black Hole Critical Collapse in Infinite Dimensions:}\vskip 4pt
{\fontsize{14}{14} \bf Continuous Self-Similar Solutions}
\end{center}

\vskip 30pt
\begin{center}
\noindent
{\fontsize{12}{18}\selectfont 
Craig R. Clark and Guilherme L.~Pimentel}
\end{center}

\begin{center}
\vskip 4pt
\textit{Scuola Normale Superiore and INFN, Piazza dei Cavalieri 7, 56126, Pisa, Italy}
\end{center}

\vspace{0.4cm}
\begin{center}{\bf Abstract}
		
\end{center}
\noindent We investigate the dynamics of black hole critical collapse in the limit of a large number of spacetime dimensions, $D$. In particular, we study the spherical gravitational collapse of a massless, scale-invariant scalar field with continuous self-similarity (CSS). The large number of dimensions provides a natural separation of scales, simplifying the equations of motion at each scale where different effects dominate. With this approximation scheme, we construct matched asymptotic solutions for this family, including the critical solution. We then compute the mass critical exponent of the black hole for linear perturbations that break CSS, finding that it asymptotes to a constant value in infinite dimensions. Additionally, we present a link between these solutions and closed Friedmann--Lema\^itre--Robertson--Walker (FLRW) cosmologies with a dimension-dependent equation of state and cosmological constant. The critical solution corresponds to an unstable Einstein-like universe, while subcritical and supercritical solutions correspond to bouncing and crunching cosmologies respectively. Our results provide a proof of concept for the \largeD expansion as a powerful analytic tool in gravitational collapse and suggest potential extensions to other self-similar systems.

\vskip 15 pt


\end{titlepage}
\restoregeometry

\newpage
\setcounter{tocdepth}{2}
\setcounter{page}{2}

\linespread{0.75}
\tableofcontents
\linespread{1.}

\newpage
\section{Introduction and Summary} \label{sec:Introduction}

Gravitational collapse is the fundamental mechanism responsible for the formation of large-scale bound structures throughout the universe. In many gravitational systems, there exists a critical threshold between dispersion and collapse that exhibits remarkable properties, including universality. This was first demonstrated by Choptuik in his numerical study of the spherical collapse of a massless scalar field \cite{Choptuik:1992jv}, where it was found that when a black hole forms near threshold, its mass follows a simple power-law scaling that is independent of the initial data. Another well-known example of this universality appears in the density profiles of cold dark matter halos, as observed by Navarro, Frenk, and White \cite{Navarro:1995iw,Navarro:1996gj}. Despite extensive numerical evidence for its prevalence across many gravitational systems, a complete theoretical understanding of this scaling behaviour remains elusive, as current analytic techniques fail to fully capture the dynamics across all length scales during collapse.

In this paper, we propose that taking the limit of a large number of spacetime dimensions \cite{Emparan:2020inr}, $D$, provides a new analytic technique that enables the study of collapse, even in the strong-field regime. This technique provides greater analytic control by exploiting the way in which the gravitational force changes with the number of dimensions, leading to a separation of the problem into regions in which different effects dominate. Then, by performing asymptotic expansions in $1/D$ within each of these regions, the problem is reduced to a set of simpler differential equations that provide a controlled approximation to the dynamics at every stage of collapse.

To illustrate this, we study a simple model: the spherical collapse of a massless, scale-invariant scalar field with continuous self-similarity (CSS)\cite{Roberts:1989sk,Christodoulou:1993BoundedVS,Brady:1994xfa,Oshiro:1994hd}. With this set of symmetries, the dynamics of gravitational collapse are fully determined by a single saddle point in a two-dimensional manifold of phase space. By leaving the initial amplitude of the scalar field as a free parameter, we can explore a one-parameter family of solutions within this manifold that includes dispersion, black hole formation and a critical solution that corresponds to the formation of a null singularity. 

When this manifold is embedded in the full phase space of a spherically symmetric scalar field that is minimally coupled to gravity, additional unstable directions emerge which break the continuous self-similarity \cite{Frolov:1997uu}. As a result, observing these solutions numerically requires a much higher degree of fine-tuning than the discretely self-similar (DSS) solution identified by Choptuik. Consequently, the critical solution within this family is \textit{not} the attractor leading to the observed universality in the scalar field system. 

Nonetheless, this CSS model is still interesting in its own right as it captures two of the essential features of the full critical collapse phase space within a finite-dimensional subspace of it. First, the CSS critical solution defines a manifold that separates dispersive and black hole solutions, while acting as an intermediate attractor for them. Second, perturbations of this critical solution \textit{within} the CSS manifold lead to a universal mass scaling exponent. Furthermore, the unstable perturbations that break CSS can give some insights into the DSS solutions seen by Choptuik \cite{Frolov:1997uu,Frolov:1999fv}.

This model also serves as an ideal testing ground for exploring critical collapse with \largeD techniques for several reasons:
\begin{itemize}
	\item Exact analytic solutions in $D=4$ are already known \cite{Roberts:1989sk,Christodoulou:1993BoundedVS,Brady:1994xfa,Oshiro:1994hd} and have been analysed extensively due to the absence of nonlinear terms that appear in higher dimensions. These solutions provide a solid foundation for extending the analysis to higher dimensions. 
	\item The system is solvable by direct numerical integration in all dimensions, enabling straightforward comparisons between the analytic results and numerical solutions.
	\item There are other systems, such as gravity coupled to a radiation-like perfect fluid, where the attractor solution exhibits continuous self-similarity \cite{Evans:1994pj,Neilsen:1998qc,Brady:1998ir,Koike:1999eg,Brady:2002iz}. These systems could be of particular phenomenological importance when considering the formation of primordial black holes, for example, during the radiation-dominated era of the universe \cite{Musco:2008hv,Carr:2020gox,LISACosmologyWorkingGroup:2023njw}.
\end{itemize}

Our analysis provides a proof of concept that the \largeD expansion is a powerful tool for understanding gravitational collapse. Although we focus on a specific model, we expect our methods to directly extend to other CSS systems, such as perfect fluids. With some modifications and additional techniques, they may also apply to other systems, including the DSS scalar field and possibly even dark matter halos in which a gas of dark matter particles virialises into a bound structure with very simple power-law behaviour. Finding analytic results for these systems would greatly enrich our understanding of critical phenomena and their applications in cosmology, high-energy astrophysics, and gravitational theory.

\vskip 20 pt
\noindent\textbf{A Large Number of Dimensions?}

As we will see, the number of dimensions provides a parameter that allows us to cleanly separate the various scales involved in gravitational collapse. This is essential because our ability to solve problems in physics typically \textit{relies} on a separation of scales. The underlying idea is simple: when a system has a clear hierarchy of scales, its behaviour at a given scale can be described using an effective theory that captures only the dominant contributions there. Subleading corrections can then be incorporated systematically through a controlled expansion in a small parameter that is set by the hierarchy between these scales.

Despite its simplicity, this idea is remarkably powerful because, in many systems, a small parameter \textit{naturally} emerges. For example, in multipole approximations, the small parameter is the ratio of the size of the source to the observation distance. In Effective Field Theories, the expansion parameter is the ratio of the energy of interest to a high-energy cutoff, and so on.

However, when there is no small parameter in a particular problem, a powerful strategy is to artificially `sneak' one in, allowing for the exploitation a new type of perturbation theory. This approach has been highly successful in large-\hspace{-0.1em}$N$ gauge theories, where the number of colour degrees of freedom, $N$, is taken to be large. Expanding in $1/N$ allows for controlled approximations in otherwise intractable problems. 

In the context of gravitational collapse, the difficulty faced by analytic methods can be traced back to precisely this absence of a small parameter that remains valid across all scales probed by the dynamics. In the early stages of collapse, gravity is weak, and an expansion in $\frac{GM}{r}$ is effective. However, as the system evolves to smaller scales, strong-field effects dominate and this parameter no longer remains small.

This is where the number of spacetime dimensions comes in. We propose that it can serve as an artificial expansion parameter, similar to $N$ in the large-\hspace{-0.1em}$N$ expansion of gauge theories. Expanding in the inverse number of dimensions leads to a separation of scales and clearly defined regions over which the gravitational force has different strengths. This effect is already evident from the Newtonian potential
\begin{equation}
	V(r) \sim -\left(\frac{r_{\sz}}{r}\right)^{D-3}.
\end{equation}
As the number of dimensions increases, the gravitational field weakens at large distances and becomes stronger at short distances. This results in the force ``turning on'' sharply at a characteristic radius set by the mass in a way that does not happen in $D=4$. The clearly distinguished regions, and the fact that $1/D$ remains small at \textit{all} scales, are the key features that makes this method useful. In fact, the \largeD expansion does not linearise the differential equation for a given system. Instead it appears to isolate the key non-linearity that is necessary to understand each region.

The \largeD expansion and its potential application to black hole physics was first recognised by Asnin, Gorbonos, Hadar, Kol, Levi, and Miyamoto \cite{Asnin:2007rw}, where the authors found that the near-horizon region was well-defined in this limit. This was later clarified by Emparan, Grumiller, Suzuki, and Tanabe \cite{Emparan:2013xia,Emparan:2013moa}, which enabled its exploitation in a variety of different contexts, including in black hole perturbation theory \cite{Emparan:2014aba}, fluid-gravity duality \cite{Bhattacharyya:2018iwt,Patra:2019hlq} and in the membrane paradigm \cite{Emparan:2016sjk,Bhattacharyya:2017hpj} --- see \cite{Emparan:2020inr} for a recent review. These successes suggest that similar methods could provide insight into the dynamics of gravitational collapse. An initial attempt in this direction was carried out by Rozali and Way \cite{Rozali:2018yrv}. Taking a massless scalar field minimally coupled to gravity in both asymptotically flat and anti-de Sitter spacetimes, they used the \largeD expansion to find solutions that corresponded to oscillating soliton stars.

A key motivation for applying \largeD techniques to \textit{critical} gravitational collapse in particular lies in the behaviour of the critical exponent as the number of dimensions changes. Numerical evidence suggests that as $D\rightarrow \infty$, this critical exponent could asymptote to the rational value, $\frac{1}{2}$ \cite{Bland2007}. Emparan and Herzog \cite{Emparan:2020inr} have discussed how \largeD methods exhibit ``mean-field-like'' characteristics, reinforcing this possibility. If this is the case, \largeD could provide a natural framework for developing analytic approximations to critical exponents in gravitational collapse.

For the use of \largeD techniques in other fields of physics, see, for example, \cite{Strominger:1981jg,Fitzpatrick:2013sya,Joshi_2015,Vollhardt:2022kwp}.

\vskip 20 pt
\noindent {\bf Summary of Results} 

We now briefly summarise the key results and the overall picture that emerges from our analysis.

We begin with our interpretation of the spacetimes described by the solutions to this model. The combination of spherical symmetry and self-similarity, which requires the entire spacetime to be filled by the collapsing fluid, allows us to think of the scalar field as a continuum of concentric shells, each at a fixed comoving radius. Enforcing \textit{continuous} self-similarity imposes a well-defined relationship between these shells, simplifying the problem to solving the dynamics of a single one of them.

Examining the evolution of one such shell, reveals a striking connection to Friedmann--Lema\^itre--Robertson--Walker (FLRW) cosmology. In particular, we find that the local dynamics of these shells are equivalent to that of a comoving observer in a closed ($k=+1$) FLRW universe with equation of state parameter and cosmological constant given by
\begin{equation}
	w= \frac{D-3}{D-1},~~ \Lambda =\frac{(D-1)(D-2)}{8}.
\end{equation}
Under this analogy:
\begin{itemize}
	\item Black hole formation corresponds to a crunching cosmology.
	\item Critical Collapse corresponds to an unstable Einstein-like universe.
	\item Dispersion corresponds to a bouncing cosmology.
\end{itemize}

Remarkably, this analogy is exact at the level of the governing equation of both systems. The conformal radius function in our model, $\cR$, plays the exact role of the cosmological scale factor and satisfies the first Friedmann equation 
\begin{equation}
	(\cR')^{2}= \frac{(2q)^{D-2}}{4\cR^{2(D-3)}}+\frac{1}{4}\cR^{2}-\frac{1}{2},
	\label{eqn:Friedmann1}
\end{equation}
where $q$ is the tunable amplitude of the incoming scalar field and sets the amplitude of the energy density term in this equation.

\begin{figure}[H]
	\centering
	\includegraphics[width=0.85\linewidth]{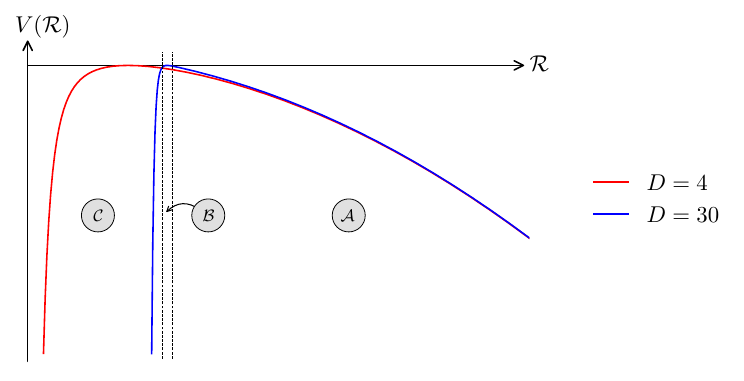}
	\caption{The effective potential $V(\cR)$, as defined by the right-hand side of the Friedmann Equation \eqref{eqn:Friedmann1}, at a value of $q$ corresponding to criticality in two different numbers of dimensions. Each distinct region has been labelled for the $D=30$ case.}
	\label{fig:friedmannpot}
\end{figure}

Although this analogy only holds locally, it provides a valuable framework for discussing the characteristic scales of the problem. Upon extracting the effective potential from the right-hand side of \eqref{eqn:Friedmann1}, we find that in critical collapse, the presence of additional geometric terms in the differential equation (the cosmological constant and spatial curvature), modifies the simple picture of the regions set by the Newtonian potential. As shown in Figure \ref{fig:friedmannpot}, there are now three distinct regions, determined by the relative magnitude of the terms in the effective potential. While these regions also exist in $D=4$, treating the number of dimensions as a free parameter allows us to measure their size, providing a clear definition of where one region transitions into another. This clarity is essential in order to analytically match the solutions to each region together.

Since these regions are central to our analysis, we summarise their key features below:
\begin{itemize}
	\item \textbf{Weak Gravity Region,} $\pmb{\cA}$: 
	\begin{itemize}
		\item At large distances, the density term is strongly suppressed in a large number of dimensions, leaving the geometric terms as the dominant effect.
		\item The characteristic time spent in this region is $\order{1}$.
		\item The spacetime in this region corresponds (up to exponentially small corrections in $1/D$) to exactly flat Minkowski space, where the scalar field evolves freely.
	\end{itemize}
	\item \textbf{Transition Region,} $\pmb{\cB}$: 
	\begin{itemize}
		\item The density term and the geometric terms approximately counterbalance each other. It is the most complicated region since it does not correspond to a single dominant effect. The large number of dimensions localises this to a narrow interval of the scale factor.
		\item The characteristic time spent in this region scales as $\order{\frac{1}{\sqrt{D}}}$. This is precisely the same scaling seen for the the time variable in the soliton star solution at \largeD \cite{Rozali:2018yrv}.
		\item Despite the short \textit{characteristic} time in a large number of dimensions, the \textit{actual} time spent in this region is more complicated as it is also approximately scales logarithmically in the distance of the initial conditions from criticality. An interplay between the initial conditions, the number of dimensions and the regions is an issue that will appear frequently in this paper and is a key challenge that we address.
	\end{itemize}
	\item \textbf{Strong Gravity Region,} $\pmb{\cC}$:
	\begin{itemize}
		\item At short distances, the density term completely dominates the geometric terms.
		\item This region is only accessed by the system if the initial amplitude of the scalar field is large enough for a black hole to form.
		\item The characteristic time spent in this region scales as either $\order{\frac{1}{\sqrt{D}}}$ or $\order{\frac{1}{D}}$, depending on how far the initial amplitude of the scalar field is from criticality.
	\end{itemize}
\end{itemize}
To construct solutions, we apply a \largeD method of matched asymptotics \cite{BenderOrszag1999} where the Friedmann equation, \eqref{eqn:Friedmann1}, is expanded differently in each region according to its characteristic timescale. This approach allows us to derive analytically matched asymptotic solutions for both critical collapse and far from critical black hole formation. We also give solutions to the dispersive and near-critical black hole outcomes as numerically matched perturbations of the critical solution. 

We then determine the Lyapunov exponents for deviations from the critical solution. These perturbations include both a unique self-similar mode and a continuous spectrum of non-self-similar modes. For the non-self-similar perturbations, we find that the spectrum of allowed Lyapunov exponents exhibits two critical dimensions: $D=6$ and $D=10$. The $D=6$ critical dimension marks the point beyond which the perturbed CSS critical solution no longer sits in the basin of attraction of the DSS critical solution, thereby preventing further insight into the DSS solution in higher dimensions. The $D=10$ critical dimension only affects subdominant perturbations and consequently lacks a clear physical interpretation. 

Using the Lyapunov exponents, we compute a scaling exponent for the mass of the black hole above criticality using renormalisation group methods. For the perturbations that preserve self-similarity, it is given by
\begin{equation}
	\gamma = \frac{D-3}{\sqrt{2(D-2)}},
\end{equation} 
in agreement with Soda and Hirata \cite{Soda:1996nq}, while for breaking self-similarity \footnote{We solve the perturbation equations asymptotically at the past and future time boundaries to determine the spectrum of allowed Lyapunov exponents. In this way, we obtain a solution that is exact in the number of dimensions. However, it has not yet been proven that there are no further restrictions from the `matching' of these asymptotics in a full solution to the equations. Nonetheless, comparison with the known results in $D=4$ provides evidence that the analysis is complete.} it is given by 
\begin{equation}
	\gamma =\frac{2(D-3)}{D-2}\approx 2-\frac{2}{D}\,.
\end{equation}
Interestingly, the critical exponent for breaking the self-similarity asymptotes to a finite value in infinite dimensions, in alignment with the numerical evidence for the discretely self-similar solution.

We conclude by emphasising a direct parallel between the system discussed thus far and the topology change that resolves the naked singularity at the pinch of the Gregory-Laflamme (GL) instability of black strings \cite{Gregory:1993vy} in the \largeD limit. This topology change can be effectively modelled by a double-cone geometry near the pinching point. In this framework, the fused-cone configuration corresponds to the non-uniform black string phase, whereas the separated-cone configuration represents the black hole phase. Interpolating between these two phases is a critical solution that arises precisely when the tips of the cones meet, corresponding to the pinched geometry. Due to the underlying conical structure, this critical configuration exhibits continuous self-similarity.

The presence of a CSS critical solution sitting between two distinct phases, mirrors the behaviour observed in the gravitational critical collapse. This connection was first elucidated by Kol \cite{Kol:2005vy}, who demonstrated that, upon dimensional reduction and a double Wick rotation, the actions governing both systems can be cast into the same form, but with different boundary conditions. The GL side of this correspondence is well-understood in the \largeD regime, allowing us to make the following observations when combined with our \largeD results for the CSS collapse of the massless scalar field:
\begin{itemize}
	\item The coordinate scalings with the number of dimensions are similar. For example, in both systems, criticality is associated with a characteristic scale of $\order{\frac{1}{\sqrt{D}}}$\cite{Emparan:2019obu}.
	\item The equations in the \largeD expansion are clearly related. In particular, taking the first non-trivial equation in the \largeD expansion governing Region $\cB$ \eqref{eqn:feq}, and performing the redefinitions $x\to i z$ and $f\to 4(\cP-\cP_{0})$, yields precisely equation (6.3) of \cite{Emparan:2015hwa}, which describes non-uniform black strings in the \largeD limit.
	\item Perturbations of the critical merger solution indicate that $D=10$ is a critical dimension that influences the stability \cite{Kol:2002xz}. This closely resembles our results for perturbations that break CSS, and may offer an explanation for the appearance of $D=10$ as a critical dimension in our system too. \footnote{Note that although there is this similarity, one should not expect the stability of perturbations of the two systems to be the same in general due to the Wick rotations.}
\end{itemize}

\vskip 4 pt
\noindent {\bf Outline} \\
In Section \ref{sec:Section 2} we will review critical collapse and the role of self-similarity in these spacetimes. This allows us to find the general dimensional equations for our model, and develop an intuitive picture of the collapsing shells. The connection that this picture has to cosmological solutions is also explored. We will then review the continuous self-similar solutions that are already understood in $D=4$ to give us a point of reference when developing the large dimension expansion.

In Section \ref{sec:3} we build, as an asymptotic expansion in the inverse number of spacetime dimensions, the family of spacetimes of this model. We specifically highlight how the separation into regions occurs, how the correct scalings of the time variable can be found for each region and the crucial role played by the dimensional dependence of the initial amplitude of the scalar field.

In Section \ref{sec:4} we study the spectrum of Lyapunov exponents of small perturbations about the critical solution in a general number of dimensions. These perturbations include both those that preserve the self-similarity, as well as those that break it. From these Lyapunov exponents, we then compute the critical exponent for the mass of the black hole above criticality. Finally in Section \ref{sec:5} we conclude and outline some possible future applications of the methods that we develop in this paper.

The appendices provide mostly technical details in the form of explicit calculations, but there are also some additional results. In Appendix \ref{app:MoR}, we discuss the alternative `Method of Regions' approach to finding solutions to this problem. Appendix \ref{app:SecondOrd} addresses a subtlety encountered when working with the second-order differential equation in Region $\cC$. Appendix \ref{app:HigherOrd} presents the next-order correction to Region $\cC$, including its matching to Region $\cA$. In Appendix \ref{app:Matching}, we carry out the explicit computations for the matching of Region $\cB$ to Region $\cA$ for both the metric function and the scalar field. Finally, in Appendix \ref{app:Notation}, we provide a comprehensive table of notation for the variables used in this work. 

We also provide a ``roadmap'' for the paper on page \pageref{page:Roadmap} to aid readers in navigating to specific parts of the paper that they are interested in. A summary of the solutions to the metric can be found on pages \pageref{page:MetricSolStart}--\pageref{page:MetricSolEnd}.

\vskip 4 pt
\noindent {\bf Notation and Conventions} 

We adopt natural units with $c \equiv 1$ and set the gravitational coupling constant $\kappa \equiv 8\pi G c^{-4} = 1$. The metric signature is mostly positive, with Greek letters ($\mu = 0, 1, 2, ..., D-1$) used for spacetime indices and Latin letters are used for other indices without structure. Quantities associated with criticality are denoted with a subscript $*$, e.g., $p_{*}$, while quantities specific to a particular region are labelled with the corresponding region identifier, e.g., $\xb$. In asymptotic expansions, region labels are generally omitted when their meaning is clear from the context.

\newpage
\begin{tikzpicture}[x=1cm, y=1cm]\label{page:Roadmap}
	
	\useasboundingbox (0,0) rectangle (15.7,22.6);
	
	
	
	\tikzstyle{block} = [rectangle, draw, text centered, minimum height=1.5cm, minimum width=3cm, inner sep=10pt,fill=white]
	\tikzstyle{block2} = [rectangle, draw, minimum height=1.5cm, minimum width=3cm,	inner sep=10pt, align=left,fill=white]
	\tikzstyle{diamond} = [diamond, draw, text centered, minimum height=1.5cm, minimum width=3cm, inner sep=10pt]
	\tikzstyle{arrow} = [thick,->,>=stealth]
	
	\draw [arrow] (-0.4+3.65+0.5+3.65/2-0.2,4.95) -- (-0.4+3,3.1);
	\draw [arrow] (-0.4+3.65+0.5+3.65/2+0.2,4.95) -- (-0.4+1+6+4.5,3.4);
	\draw [arrow] (2,9.2) -- (2,7.1);
	\draw [arrow] (2.6,9.2) -- (5,7.1);
	\draw [arrow] (3.2,9.2) -- (9,7.1);
	\draw [arrow] (4,9.2) -- (13,7.15);
	\draw [arrow] (6,9.2) -- (3,7.1);
	\draw [arrow] (7.5,9.2) -- (6,7.1);
	\draw [arrow] (9,9.2) -- (10,7.1);
	\draw [arrow] (12,9.2) -- (11,7.1);
	\draw [arrow] (14,9.2) -- (14,7.1);
	\draw [arrow] (5,13.93) -- (3,11.8);
	\draw [arrow] (10.5,13.93) -- (12.3,11.8);
	\draw [arrow] (-0.4+16.1/2,13.93) -- (-0.4+16.1/2,11.8);
	\draw [arrow] (-0.4+16.1/2,17.4) -- (-0.4+16.1/2,16.47);

	\node at (0.9,12.85) {
		\begin{tabular}{@{}p{3cm}@{}} 
			\multicolumn{1}{c}{\textbf{Analysis (\textsection \ref{sec:3.1})}}
		\end{tabular}
	};
	\node at (5.1,12.9) {$x\sim\order{1}$};
	\node at (8.7,12.9) {$x\sim\order{\sqrt{\ve}}$};
	\node at (12.35,12.9) {$x\sim\order{\ve}$};
	\node at (0.5,8.1) {
		\begin{tabular}{@{}p{3cm}@{}} 
			\multicolumn{1}{c}{\textbf{Matching}}
		\end{tabular}
	};
	\node at (1.2,4) {
		\begin{tabular}{@{}p{3cm}@{}} 
			\multicolumn{1}{c}{\textbf{Perturbations} \textbf{(\textsection \ref{sec:4})}}
		\end{tabular}
	};
		
	\node[rectangle, draw, rounded corners, thick, minimum height=1.5cm, minimum width=14cm, inner sep=10pt, align=left,fill=white] at (-0.4+16.1/2,20) {
		\begin{tabular}{@{}p{13cm}@{}} 
			\multicolumn{1}{c}{\textbf{Setup (\textsection \ref{sec:2.1})}}\\[0.5em]
			$\bullet$ $S=
			\frac{1}{2}\int\diff^{D}x\sqrt{-g}\left(R-g^{\mu\nu}\nabla_{\mu}\phi\nabla_{\nu}\phi\right)$\\[0.5em]
			$\bullet$ Spherical Symmetry\\[0.5em]
			$\bullet$ Continuous Self-Similarity (pp.~\pageref{page:CSSStart}--\pageref{page:CSSEnd})\\[0.5em]
			$\bullet$ Scale-invariant Scalar Field (pp.~\pageref{page:SIScalar})\\[0.5em]
			$\bullet$ Metric ansatz in null coordinates $\diff s^{2}=-2e^{2\Sigma(x)}\diff u\diff v+(-u v)y(x) \diff \Omega_{D-2}^{2}$\\[0.5em]
			$\bullet$ Flat space boundary conditions at $v=0$ (pp.~\pageref{page:BoundCond},\pageref{page:BoundCondStart})
		\end{tabular}
	};
	\draw[decorate,decoration={brace, mirror, amplitude=10pt}, thick] (8,19.1) -- (8,20.85);
	\node at (9,19.1+1.75/2-0.05) {$\implies$};
	\node at (11.7,19.1+1.75/2-0.05) {
		\begin{tabular}{@{}p{3cm}@{}} 
			\multicolumn{1}{c}{Fields are all a function}\\
			\multicolumn{1}{c}{of one variable}\\
			\multicolumn{1}{c}{$x=\log\left(\frac{v}{-u}\right)$ (pp.~\pageref{page:OneVarStart}--\pageref{page:OneVarEnd})}
		\end{tabular}
	};

	\node[rectangle, draw, rounded corners, thick, minimum height=1.5cm, minimum width=15cm, inner sep=10pt, align=left,fill=white] at (-0.4+16.1/2,15.2) {
		\begin{tabular}{@{}p{14cm}@{}} 
			\multicolumn{1}{c}{\textbf{Single Defining Equation}}\\[0.5em]
			\multicolumn{1}{c}{$2 \ve yy''+(1-2\ve)(y')^{2} + 2(1-\ve) y - y^{2} = 0, ~~ \ve = \frac{1}{D-2}$}\\[0.5em]
			$\bullet$ Integrate once $\to$ First Friedmann equation $\to$ Cosmology and shell picture (\textsection \ref{sec:2.2})
		\end{tabular}
	};

	\node[rectangle, draw, rounded corners, thick, minimum height=1.5cm, minimum width=5cm, inner sep=10pt, align=left,fill=white] at (-0.4+2.5,10.5) {
		\begin{tabular}{@{}p{4.5cm}@{}} 
			\multicolumn{1}{c}{\textbf{Region} $\pmb{\cA}$ (pp.~\pageref{page:RegionAStart})}\\[0.5em]
			$\bullet$ Flat Space \\[0.5em]
			$\bullet$ $(y_{0}')^{2}-y_{0}^{2}+2y_{0}=0$\vspace{0.18em}
		\end{tabular}
	};
	\node[rectangle, draw, rounded corners, thick, minimum height=1.5cm, minimum width=5cm, inner sep=10pt, align=left,fill=white] at (-0.4+5+0.5+2.5,10.5) {
		\begin{tabular}{@{}p{4.5cm}@{}} 
			\multicolumn{1}{c}{\textbf{Region} $\pmb{\cB}$ (pp.~\pageref{page:RegionBStart}--\pageref{page:RegionBEnd})}\\[0.5em]
			$\bullet$ Slow Dynamics\\[0.5em]
			$\bullet$ $f''+\frac{1}{4}(f')^{2}-f=0$\vspace{0.18em}
		\end{tabular}
	};
	\node[rectangle, draw, rounded corners, thick, minimum height=1.5cm, minimum width=5cm, inner sep=10pt, align=left,fill=white] at (-0.4+10+1+2.5,10.5) {
		\begin{tabular}{@{}p{4.5cm}@{}} 
			\multicolumn{1}{c}{\textbf{Region} $\pmb{\cC}$ (pp.~\pageref{page:RegionCStart}--\pageref{page:RegionCEnd})}\\[0.5em]
			$\bullet$ Runaway Collapse\\[0.5em]
			$\bullet$ $(Y_{0}')^{2}-\frac{q-1}{q}Y_{0}^{2}-\frac{2}{q}Y_{0}=0$
		\end{tabular}
	};

	\node[rectangle, draw, rounded corners, thick, minimum height=1.5cm, minimum width=3.65cm, inner sep=10pt, align=left,fill=white] at (-0.4+3.65/2,6) {
		\begin{tabular}{@{}p{3cm}@{}} 
			\multicolumn{1}{c}{\textbf{Marginally}}\\
			\multicolumn{1}{c}{\textbf{Subcritical}}\\
			\multicolumn{1}{c}{(pp.~\pageref{page:MSubStart}--\pageref{page:MSubEnd})}
		\end{tabular}
	};
	\node[rectangle, draw, rounded corners, thick, minimum height=1.5cm, minimum width=3.65cm, inner sep=10pt, align=left,fill=white] at (-0.4+3.65+0.5+3.65/2,6) {
		\begin{tabular}{@{}p{3cm}@{}}
			\multicolumn{1}{c}{%
				\parbox{2cm}{\centering
					\vspace{0.78em} 
					\textbf{Critical}\\
					(pp.~\pageref{page:CritStart}--\pageref{page:CritEnd})\\
					\vspace{0.78em}
				}
			}
		\end{tabular}
	};
	\node[rectangle, draw, rounded corners, thick, minimum height=1.5cm, minimum width=3.65cm, inner sep=10pt, align=left,fill=white] at (-0.4+2*3.65+1+3.65/2,6) {
		\begin{tabular}{@{}p{3cm}@{}} 
			\multicolumn{1}{c}{\textbf{Marginally}}\\
			\multicolumn{1}{c}{\textbf{Supercritical}}\\
			\multicolumn{1}{c}{(pp.~\pageref{page:MSupStart}--\pageref{page:MSupEnd})}
		\end{tabular}
	};
	\node[rectangle, draw, rounded corners, thick, minimum height=1.5cm, minimum width=3.65cm, inner sep=10pt, align=left,fill=white] at (-0.4+3*3.65+1.5+3.65/2,6) {
		\begin{tabular}{@{}p{3cm}@{}} 
			\multicolumn{1}{c}{\textbf{Very}}\\
			\multicolumn{1}{c}{\textbf{Supercritical}}\\
			\multicolumn{1}{c}{(pp.~\pageref{page:VSupStart}--\pageref{page:VSupEnd})}
		\end{tabular}
	};
	
	\node[rectangle, draw, rounded corners, thick, minimum height=1.5cm, minimum width=6cm, inner sep=10pt, align=left,fill=white] at (-0.4+3,1.7) {
		\begin{tabular}{@{}p{5.5cm}@{}} 
			\multicolumn{1}{c}{\textbf{CSS-Preserving}}\\[0.5em]
			$\bullet$ Unique Lyapunov exponent\\[0.5em]
			$\bullet$ $\gamma_{q}=\frac{D-3}{\sqrt{2(D-2)}}$ (pp.~\pageref{page:PresExp})
		\end{tabular}
	};
	\node[rectangle, draw, rounded corners, thick, minimum height=1.5cm, minimum width=9cm, inner sep=10pt, align=left,fill=white] at (-0.4+1+6+4.5,1.7) {
		\begin{tabular}{@{}p{8.5cm}@{}} 
			\multicolumn{1}{c}{\textbf{CSS-Breaking}}\\[0.5em]
			$\bullet$ Spectrum of Lyapunov exponents (pp.~\pageref{page:SpectrumLyap})\\[0.5em]
			$\bullet$ Critical dimensions, $D=6$ and $D=10$  (pp.~\pageref{page:CritDim})\\[0.5em]
			$\bullet$ $\gamma_{p} = \frac{2(D-3)}{D-2}$ (pp.~\pageref{page:BreakExp})
		\end{tabular}
	};

\end{tikzpicture}

\newpage
\section{Black Hole Critical Collapse} \label{sec:Section 2}
In this section we give a mostly self-contained introduction to black hole critical collapse, collecting key results from the literature and providing some interpretation that will be relevant for our generalisation to $D$ dimensions.

Choptuik's numerical simulations in 1992 provided the first evidence of black hole critical collapse \cite{Choptuik:1992jv}, where he studied the spherical collapse of a gravitating, massless scalar field numerically. By varying parameters such as the amplitude, $p$, of different shapes of shells, he observed that the scalar field would either disperse back to flat space or collapse to form a black hole. Then, by a shooting method, he approximately determined the \textit{critical} value of the amplitude, $p_{*}$, that described the transition point between these two different outcomes. For solutions with amplitude close to the critical value, he discovered a number of interesting properties:

\begin{itemize}
	\item Enhanced Symmetry --- They exhibited an additional approximate symmetry that had not been imposed on the solutions, called self-similarity. This symmetry can manifest in two different ways depending on the type of matter being studied: Continuous Self-Similarity (CSS) or Discrete Self-Similarity (DSS) \cite{Sornette:1997pb}. In the case of the massless scalar field, DSS was observed numerically.
	\item Critical Exponent --- The solutions that eventually formed a black hole were observed to have a mass proportional to the distance that the data was from criticality
	\begin{equation}
		M \sim (p-p_{*})^{\gamma}\,,
		\label{eqn:critexpdef}
	\end{equation}
	with \textit{no} mass gap. By fine-tuning only a single parameter in the initial data to its critical value, a naked singularity could be formed --- a known feature of some self-similar solutions \cite{Ori:1989ps}. This provided a numerically fine-tuned counterexample to the Weak Cosmic Censorship Conjecture \cite{Penrose:1969pc}, strengthening the claim that it holds only when the initial data is \textit{fully} generic in four dimensions\footnote{It can be violated in higher dimensions without fine-tuning as can be seen, for example, with the Gregory-Laflamme instability \cite{Gregory:1993vy}.}.
	\item Universality --- Regardless of the shape of the initial data, the same critical exponent and dynamical evolution was always found, but with a different $p_{*}$. Note that this universality is only true when considering the same type of matter.
\end{itemize}
This discovery sparked significant interest in understanding these special solutions that sit precisely on the verge between collapse and dispersion. For DSS critical solutions, analytic progress is made inherently difficult due to the non-trivial, periodic nature of the attractor, but some advances have been made \cite{Gundlach:1996eg,Garfinkle:1998df,Martin-Garcia:2003xgm,Gundlach:2003pg,Guo:2023cya}, including in finding another critical parameter related to the angle of lines from the origin that separate the regions of positive and negative density \cite{Ecker:2024haw}. Furthermore, this spacetime has been proven to exist (with numerical aid) \cite{Reiterer:2012hnr}, and the region exterior to the past light-cone of the singularity was recently constructed \cite{Cicortas:2024hpk}. However, an analytic solution for the full spacetime is still missing. For CSS critical solutions, analytic progress is easier due to a reduction of the system to ODEs, but other than for the massless scalar field, most solutions remain numerical \cite{Carr:2002me}.

Although we focus on the massless scalar field for simplicity in this paper, it should be noted that the same properties have been exhibited by many other types of matter including the Einstein-Maxwell-dilaton system \cite{Rocha:2018lmv}, the Einstein-axion-dilaton system \cite{Hatefi:2020gis} and axisymmetric pure gravity \cite{Abrahams:1993wa,Baumgarte:2023saw} to name just a few. It was also found that depending on the value of the coupling in some theories, the critical solution would change from exhibiting continuous self-similarity to discrete self-similarity \cite{Lechner:2001ng}. For a review, see \cite{Gundlach:2007gc} and the references within.

A useful intuition for how these features arise can be gained from the perspective of dynamical systems \cite{Gundlach:2002sx}. In General Relativity, the phase space is an infinite dimensional functional space with a vector field dictating the direction of time evolution. Each point in the phase space corresponds to a possible initial data set of the system --- a spacelike hypersurface within the full spacetime. However, despite extensive numerical results, analytic understanding remains limited, and much of the phase space structure in this problem is still conjectural. Consequently, and due to a few other conceptual challenges, constructing precise pictures of this phase space is difficult. Nonetheless, the qualitative insights that it gives are highly instructive.

Consider the gravitational collapse of generic matter. The phase space consists of three primary basins of attraction that correspond to physically distinct states of the system: dispersion to flat space; the formation of a transient state, such as a star; and the formation of a black hole. 

The issue of criticality concerns itself with the specific manifolds in the phase space that separate these basins of attraction from one another. There are two different types of critical manifolds that can be observed in gravitational collapse, and which type is relevant in a given system can be approximately predicted by simple considerations of the characteristic length scales in the initial data. To illustrate this, let us take a massive scalar field \cite{Jimenez-Vazquez:2022fix}. If the Compton wavelength of the scalar field exceeds the radial extent of the collapsing shell, the system exhibits Type II critical behaviour, where the critical solution coincides with the one observed by Choptuik, allowing for the formation of an infinitesimal mass black hole. However, if the Compton wavelength of the scalar field is smaller than the radial extent of the collapsing shell, the system exhibits Type I critical behaviour instead. In this case, the critical solution corresponds to a soliton star, and perturbations lead to black hole formation at a finite mass. 

In this paper, we focus specifically on the critical manifolds that are associated with Type II critical behaviour, where there is no `relevant' characteristic length scale in the initial data \footnote{Note that although there is no length scale in the initial data, one is dynamically generated corresponding to its distance from criticality.}. In this case, the critical manifold delineates the boundary between the basins of attraction corresponding to dispersion and black hole formation, exactly as implied by the results found by Choptuik.

The critical solutions are consequently unstable solutions that are simultaneously on the verge of dispersing and on the verge of collapsing to a black hole. The features highlighted previously can hence be qualitatively understood as follows:
\begin{itemize}
	\item Enhanced Symmetry --- On the critical manifold the symmetry becomes exact. Solutions close to criticality only exhibit the symmetry approximately during the phase of their evolution that brings them close to the critical manifold.
	\item Critical Exponent --- By analysing linear perturbations of the critical solution, one can determine Lyapunov exponents that characterise the rate at which the perturbed solution moves into the basins of attraction. Using Renormalisation Group techniques \cite{Koike:1995jm}, these Lyapunov exponents can be directly related to the mass critical exponent. This will be explored in detail in Section \ref{sec:4}.
	\item Universality --- The critical manifold has codimension 1 and contains a single, unique attractor. All collapsing shells that are tuned close to criticality are initially drawn to this attractor before a single growing mode ultimately drives the solution away and further into one of the basins of attraction.
\end{itemize}

In particular, this means that there is an alternative method to investigate these properties. Rather than numerically fine-tuning parameters to sit close to the critical manifold, one can instead look to find the critical solution directly by imposing self-similarity. Then, by taking perturbations of the critical solution, the local region away from the critical manifold can be understood, along with the critical exponent. This both simplifies the problem and bypasses the issue that the parameters can never be numerically tuned \textit{precisely} to criticality. This is the view that shall be taken within this paper, finding the critical solution in Section \ref{sec:3} and then the perturbations in Section \ref{sec:4}.

The particular matter model that we choose to study consists of a single massless scalar field that gravitates \footnote{This model has a long history, going back to Christodoulou's PhD thesis \cite{Christodoulou:1971}, with many subsequent developments due to Christodoulou himself \cite{Christodoulou:1986zr,Christodoulou:1987vv,Christodoulou:1993BoundedVS,Christodoulou:1994hg,Christodoulou:1999}. The DSS solution is only known approximately and numerically, being proven to exist (with computer aid) in \cite{Reiterer:2012hnr}. The CSS solutions were first described by Roberts \cite{Roberts:1989sk} and later rediscovered by BONT \cite{Brady:1994xfa,Oshiro:1994hd}.}. Although the numerics demonstrate that the critical solution is DSS in this model, critical CSS solutions are also known to share the properties mentioned above with the exception of universality. This means that, in the full phase space, there exists an additional, hidden critical manifold.

However, there are a few important differences between this CSS manifold and the DSS critical manifold discussed so far. First, this manifold contains both black hole and dispersive solutions. Restricting to CSS solutions is not sufficient to specify a \textit{critical} solution. Second, perturbations from the CSS critical manifold that break the symmetry are not unique \cite{Frolov:1997uu}. There is a full spectrum of Lyapunov exponents implying that it has codimension higher than 1. The CSS manifold is hence `more repulsive' than the DSS critical manifold, explaining why it is not seen numerically. Finally the nature of the singularity of the critical solution is different. The critical CSS solution corresponds to the formation of a null singularity rather than a naked one \footnote{Note that the distinction made here is specifically for the scalar field system. For example, the critical CSS solution of the perfect fluid corresponds to the formation of a naked singularity \cite{Ori:1989ps}.}

Although imposing this symmetry on the system does not reproduce the exact type of critical solution discussed above, we will see that the CSS manifold generates a finite dimensional sub-phase space that completely mirrors the full one. The critical manifold separates the dispersive and black hole basins of attraction, and there is a single unique critical exponent for perturbations within this subspace.\\

We now describe qualitatively the setup that we consider. The scalar field collapses from null past under its own gravitational interaction and the solutions that we seek are characterised by:
\begin{itemize}
	\item Asymptotic flatness in the past;
	\item Continuous self-similarity of both the metric and scalar fields;
	\item A scale-invariant scalar field profile.\footnote{There is a whole family of self-similar scalar field profiles, one of which is scale-invariant. The other cases are also interesting and were studied by Christodoulou \cite{Christodoulou:1994hg}.}
\end{itemize} 
The previous conditions leave one free parameter unconstrained, corresponding to the initial incoming amplitude of the scalar field. This is a natural consequence of the fact that the CSS symmetry was not sufficient to select a critical solution as mentioned before. Adjusting this parameter is what allows us to tune the initial energy of scalar field that collapses such that we can see the different possible end-states of the system within the CSS manifold:
\begin{itemize}
	\item If the amplitude of the incoming wave is small, the scalar field will initially collapse before dispersing due to its pressure. The system will asymptotically return to flat space. This is {\color{mplGreen} \textit{Subcritical}} Collapse.
	\item If the amplitude of the incoming wave is too large, the gravitational attraction will overcome the pressure of the scalar field such that it will fully collapse to form a black hole in a finite time. This is {\color{mplRed} \textit{Supercritical}} Collapse.
	\item If, however, we fine-tune this amplitude to a specific value, the system is stuck in a state of continuously trying to both collapse and disperse without ever fully achieving either, forming a null singularity in the future. This is {\color{mplBlue} \textit{Critical}} Collapse.
 \end{itemize}

In the first subsection, we formulate the problem and derive the governing equations in an arbitrary number of dimensions. We carefully analyse the symmetry properties of the resulting spacetimes, highlighting the distinction between continuous and discrete self-similarity. This lays the foundation for developing an intuitive shell-based picture of the CSS spacetime in the second subsection, where we also uncover a surprising connection to cosmology. In the final subsection, we review existing results in four dimensions to build intuition for the general, higher-dimensional case.

\subsection{The Setup and Self-Similarity} \label{sec:2.1}

In this section, we explain the setup in more detail and write the corresponding field equations. 
Consider a massless scalar field in $D$ dimensions, minimally coupled to gravity. The action is
\begin{equation}
	S = \frac{1}{2}\int\diff^{D}x\sqrt{-g}\left(R-g^{\mu\nu}\nabla_{\mu}\phi\nabla_{\nu}\phi\right).
	\label{eqn:action}
\end{equation}

We assume a spherically symmetric metric in doubly null coordinates $x^{\mu}=\left(u,v,\theta_{1},...,\theta_{D-2}\right)$, 
\begin{equation}
	\diff s^{2} = -2e^{2\Sigma(u,v)}\diff{u}\diff{v}+\rho^{2}(u,v)\diff{\Omega}^{2}_{D-2}\, , \label{eqn:metricp}
\end{equation}
where $u=\frac{t-r}{\sqrt{2}}$ and $v=\frac{t+r}{\sqrt{2}}$ are the standard retarded and advanced times, so that in flat space the metric is
\begin{equation}
	\diff{s}^2 = -2\diff{u}\diff{v}+\left(\frac{u-v}{\sqrt{2}}\right)^{2}\diff{\Omega}^{2}_{D-2}\, .
\end{equation}
\label{page:BoundCond}To ensure regular initial data for the collapse, we impose flat space boundary conditions at the advanced time $v=0$.\footnote{A more detailed justification for this choice of boundary conditions will be provided later.}
\begin{equation}
	\Sigma(u,0)=0,~~\rho(u,0) = \frac{-u}{\sqrt{2}},~~\phi(u,0)= 0.
	\label{eqn:boundcond1}
\end{equation}
The spherical symmetry allows us to define the Misner-Sharp mass \cite{Misner:1964}, $m(u,v)$, as a proxy for the `total energy' confined within some shell of radius $\rho$. It is given by
\begin{equation}
	\begin{aligned}
		1-\frac{2m(u,v)}{\rho^{D-3}}&=g^{\mu\nu}\partial_{\mu}\rho\partial_{\nu}\rho\, ,\\
		m(u,v) &=\frac{\rho}{2}\left(1+2e^{-2\Sigma}\partial_{u}\rho\partial_{v}\rho\right),
	\end{aligned}
	\label{eqn:massfunc}
\end{equation}
and is a useful diagnostic tool for the spacetime since we can test for the existence of an apparent horizon by the condition that 
\begin{equation}
	2m(u,v)=\rho(u,v)^{D-3}.
	\label{eqn:AHcond}
\end{equation}
\\

Now we present the equations of motion for the system in these coordinates. They are simply components of the Einstein equations, and the Klein-Gordon equation. Instead of keeping the dimension $D$, we find it useful to introduce the variable
\begin{equation}
	\ve \equiv \frac{1}{D-2},
	\label{eqn:defeps}
\end{equation}
such that our desired limit, $D\rightarrow \infty$ now becomes $\ve\rightarrow 0^{+}$.
The independent equations of motion are
\begin{subequations}\label{eqn:PDEs}
	\begin{align}
		\partial_{u}\rho\partial_{v}\phi+\partial_{v}\rho\partial_{u}\phi+2\ve \rho\partial_{uv}\phi &=0\label{eqn:PDE1} \\
		2\partial_{u}\rho\partial_{u}\Sigma-\partial_{uu}\rho-\ve \rho\partial_{u}\phi\partial_{u}\phi &=0\, ,\label{eqn:PDE2}\\
		(1-\ve)\partial_{u}\rho\partial_{v}\rho+\ve \rho\partial_{u}\partial_{v}\rho+\frac{1}{2}(1-\ve)e^{2\Sigma} &= 0\, , \label{eqn:PDE4}
	\end{align}
\end{subequations}
where \eqref{eqn:PDE2} is also valid for derivatives with respect to $v$ instead of $u$~\footnote{For completeness, there is also an equation that follows from the previous equations and Bianchi identities. Explicitly it is given by $$
	\frac{1}{2}(1-\ve)(1-2\ve)e^{2\Sigma}+(1-3\ve+2\ve^2)\partial_{u}\rho\partial_{v}\rho +4\ve(1-\ve)\rho\partial_{u}\rho\partial_{v}\rho+2\ve^{2}\rho^{2}\partial_{uv}\Sigma = -\ve^{2}\rho^{2}\partial_{u}\phi\partial_{v}\phi\, .
	\label{eqn:V}
$$}. Analytic solutions to these equations are nearly impossible to find without additional assumptions. To simplify the system and find a critical solution, we now impose self-similarity.\\

\label{page:CSSStart}Self-similarity is a geometric symmetry closely associated with fractals and belongs to the broader class of the similarity transformation --- the set of transformations that preserve the shape of an object \cite{Carr:1998at}. In particular, self-similarity refers to the special subset of similarity transformations where an object is rescaled \textit{uniformly} in all directions. 

These transformations are referred to as homothetic transformations and are characterised by a single fixed point in the space that remains invariant under their action. This point is called the scaling origin, or the centre of the transformation, and serves as the reference point about which the entire structure is scaled. Fundamentally, a homothety corresponds simply to the magnification or contraction of an object with respect to this given scaling origin.
\begin{figure}[H]
	\centering
	\includegraphics{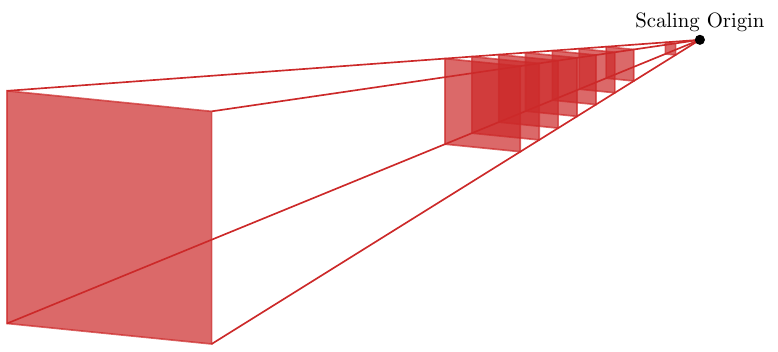}
	\caption{A sketch of a continuously self-similar spacetime. It ``looks the same", but rescaled, when the coordinates are rescaled by any amount.}
	\label{fig:magsqcss}
\end{figure}
We can define a spherically symmetric homothetic transformation as one for which under the equal scaling of the non-angular coordinates
\begin{equation}
	u\rightarrow e^{a} u,\qquad v\rightarrow e^{a} v,\qquad \theta_{i}\rightarrow \theta_{i},
	\label{eqn:rescaling}
\end{equation}
the metric rescales covariantly.

The distinction between a spacetime that is DSS or CSS can be made at this point. The CSS spacetimes can be thought of precisely as in Figure \ref{fig:magsqcss}. There is a centre of magnification and at every point in spacetime there is a direction along which the metric rescales continuously for all values of $a\in\R$.

\begin{figure}[H]
	\centering
	\includegraphics{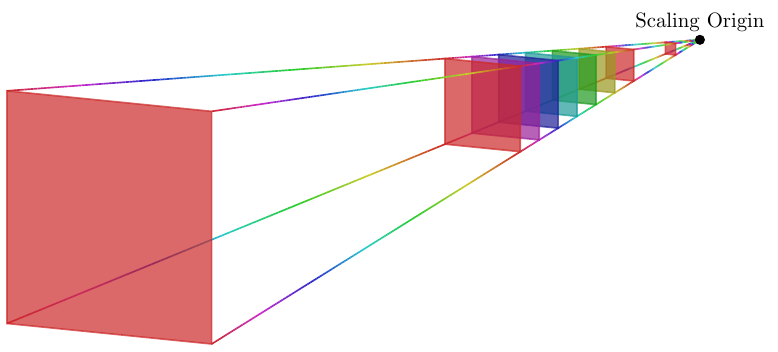}
	\caption{A sketch of a discretely self-similar spacetime. It only ``looks the same", but rescaled, when the coordinates are rescaled by a discrete set of values. There is a period associated to the magnification lines, indicated by the colour.}
	\label{fig:magsqdss}
\end{figure}

For DSS spacetimes, on the other hand, the metric only rescales for discrete steps in each of the coordinates, i.e for discrete values of $a=n\Delta$, $n\in \Z$. In terms of discrete sets, such as the Cantor Set, this type of self-similarity is intuitive as the system only looks the same again after a magnification by a particular set of discrete values. When it comes to continuous variables however, it is much harder to visualise DSS. It has been shown that these DSS spacetimes are equivalent to periodic boundary conditions in the right coordinates \cite{Gundlach:1995kd}, and so a simple way to visualise this could be to overlay a phase on the magnification lines of the CSS case, as we depict with colour in Figure \ref{fig:magsqdss}. For the square to be exactly the same again under the magnification, it must have the same phase as it originally did (the same colour), requiring a discrete rescaling of the coordinates and hence the shape.\\

The choice of a massless scalar field as our matter type means that to reproduce the results seen by Choptuik, we should impose the discrete version of the symmetry on our spacetime. However, for the reasons discussed previously we now impose CSS.

Using comoving coordinates, we can express this symmetry in an invariant form \cite{Cahill:1970ew}
\begin{equation}
	\mathcal{L}_{\xi}g_{\mu\nu}=2g_{\mu\nu},
	\label{eqn:hvfmet}
\end{equation}
where $\xi$ is called the homothetic vector field (HVF) and the choice of constant and sign on the right hand side of this equation is simply convention to fix the normalisation and direction of $\xi$ as away from the scaling origin to larger scales. With this definition we can now impose this symmetry on the equations of motion to restrict the solution space. \label{page:CSSEnd}

\label{page:OneVarStart}Let us assume that there exists a HVF, $\xi$, as given in \eqref{eqn:hvfmet}. Spherical symmetry, constrains the HVF such that it can only contain components in the $u$ and $v$ directions
\begin{equation}
	\xi = \xi^{u}(u,v)\pdr{}{u} +\xi^{v}(u,v)\pdr{}{v}.
	\label{eqn:HVF}
\end{equation}
Acting with the Lie Derivative on the metric along this vector and imposing \eqref{eqn:hvfmet}, we find that $\xi^{u} = \xi^{u}(u)$ and $\xi^{v} = \xi^{v}(v)$. This allows us to fix the gauge\footnote{In detail, it's given by 
$\bar{u}= e^{\int\frac{1}{\xi^{u}}\diff u},~\bar{v}= e^{\int\frac{1}{\xi^{v}}\diff v}.$}
such that the HVF takes the following simple form
\begin{equation}
	\xi = \bar{u}\pdr{}{\bar{u}} +\bar{v}\pdr{}{\bar{v}}.
	\label{eqn:HVFsimp}
\end{equation}
We also perform the redefinitions
\begin{equation}
	\begin{aligned}
		\Sigma &=\bar{\Sigma} -\frac{1}{2}\log\left(\frac{\alpha}{\bar{u}}\right)- \frac{1}{2}\log\left(\frac{\beta}{\bar{v}}\right),\\
		\rho &=\sqrt{-\bar{u}\,\bar{v}}~\cR,
	\end{aligned}
	\label{eqn:redefinedmet}
\end{equation}
to render the radius function dimensionless, so that it rescales trivially under \eqref{eqn:rescaling}, and to ensure that the form of the metric is unchanged after having fixed the gauge. We will refer to $\cR$ as the `conformal radius function' for reasons that will become clear shortly.

We construct a new time coordinate by demanding that it is invariant under the rescaling in \eqref{eqn:rescaling} and hence along the direction picked out by the HVF \eqref{eqn:HVFsimp}. Since self-similar problems are naturally described by logarithmic variables, we define our \textit{self-similar time} as 
\begin{equation}
	x\equiv\log\left(-\frac{\bar{v}}{\bar{u}}\right),
	\label{eqn:selfsimilartime}
\end{equation} 
where the minus sign inside the logarithm is to ensure that the variable will be real inside the region in which we look to solve the equations, $\bar{u}<0$ and $\bar{v}>0$. Upon changing coordinates to $(\bar{u},x)$ we find that the conditions from \eqref{eqn:hvfmet} fix $\bar{\Sigma}=\bar{\Sigma}(x)$ and $\cR=\cR(x)$. 
\begin{center}
	{\it The metric functions are only functions of the variable $x$}.
\end{center}

In order to preserve spacetime covariance under this scaling symmetry, we additionally need to impose a restriction on the scalar field. This restriction can be derived through the Einstein Field Equations to be
\begin{equation}
	\mathcal{L}_{\xi}(\partial_{\mu}\phi) =0.
	\label{eqn:hvfsca}
\end{equation}
In an analogous way to the computation for the metric functions before, we can find that this restricts the scalar field to be of the form
\begin{equation}
	\phi(\bar{u},x) = \bar{\phi}(x)-\bar{k}\log(-\bar{u}),
\end{equation}
where $\bar{k}$ is an integration constant. This additional coordinate dependence for the scalar field can be interpreted as allowing the scalar field to covary with scale \eqref{eqn:rescaling} and follows from the shift symmetry of the scalar field. Typically, this is used to introduce the \textit{scaling coordinate} \cite{Frolov:1997uu}
\begin{equation}
	s \equiv -\log(-\bar{u})\,.
	\label{eqn:scalingcoords}
\end{equation}
However, here we make the further assumption that, within CSS solutions, the scalar field profile is \textbf{scale-invariant}\label{page:SIScalar}. This sets $\bar{k}=0$, thus giving 
\begin{equation}
	\phi(\bar{u},x) = \overline{\smash{\phi}\vphantom{\phi}}(x)\,.
\end{equation}
For the rest of the paper we will drop the bar notation to avoid clutter.\label{page:OneVarEnd}
\\

Finally, imposing the symmetry on the four PDEs \eqref{eqn:PDEs} leaves us with the autonomous ordinary differential equations
\begin{subequations}\label{eqn:CSSeqns1}
	\begin{align}
		\ve \cR \phi''+\cR'\phi' &=0\label{eqn:CSSeqns1a} \\
		4\cR''-8\cR'\Sigma'-(1\pm4\Sigma'-4\ve (\phi')^{2})\cR &=0\, ,\label{eqn:CSSeqns1b}\\
		4\ve \cR\cR''+2(1-\ve)\left(2(\cR')^{2}+e^{2\Sigma}\right)-\cR^{2} &= 0\, , \label{eqn:CSSeqns1c}
	\end{align}
\end{subequations}
where the $\pm$ corresponds to having $v$ or $u$ derivatives in \eqref{eqn:PDE2}.

We have five degrees of freedom in total --- two for each of the functions controlled by second order differential equations $\cR$ and $\phi$, and one for $\Sigma$. Three of these are fixed by the boundary conditions, appropriately translated to these coordinates
\begin{equation}
		\lim_{x\rightarrow -\infty} \cR(x) = \frac{1}{\sqrt{2}}e^{-\frac{x}{2}}\,,~~ \lim_{x\rightarrow -\infty} \Sigma(x) = 0\,,~~ \lim_{x\rightarrow -\infty} \phi(x) = 0,
	\label{eqn:boundcond2}
\end{equation}
one is fixed by the constraint equation \eqref{eqn:CSSeqns1b} and the remaining degree of freedom is unfixed as mentioned previously. It will be the degree of freedom that determines if a black hole is formed or not. 

We can immediately simplify the equations, even in a general number of dimensions. A total derivative can be found from the equations by subtracting \eqref{eqn:CSSeqns1b} with one sign from \eqref{eqn:CSSeqns1b} with the other. After imposing the boundary condition, this gives
\begin{equation}
	\Sigma(x)=0,
\end{equation}
leading to a decoupled equation for the radius function \eqref{eqn:CSSeqns1c} and that the choice of sign in \eqref{eqn:CSSeqns1b} is irrelevant.\\

Since $\cR$ is a radius function and is positive definite, it is useful to define 
\begin{equation}
	y(x)=\cR(x)^{2},
	\label{eqn:ydef}
\end{equation}
giving us the equations
\begin{subequations}\label{eqn:CSSeqns2}
	\begin{align}
		2\ve y\phi''+y'\phi'&=0\,,\label{eqn:CSSeqns2a} \\
		4\ve^{2}y^{2}(\phi')^{2}-(1-\ve)(y')^{2}+(1-\ve)y(y-2) &=0\, ,\label{eqn:CSSeqns2b}\\
		2 \ve yy''+(1-2\ve)(y')^{2} + 2(1-\ve) y - y^{2} &= 0\, , \label{eqn:CSSeqns2c}
	\end{align}
\end{subequations}
where we have substituted \eqref{eqn:CSSeqns1c} into \eqref{eqn:CSSeqns1b} to manifestly write our constraint equation \eqref{eqn:CSSeqns2b} in terms of only first derivatives.

The decoupled equation for $y$, \eqref{eqn:CSSeqns2c}, is the defining equation of the problem. The radius function has decoupled from the scalar field meaning that its space of solutions will fully determine the system. The scalar field can then be determined by substituting this solution into the remaining equations.

\subsection{Interpreting The Spacetime --- Shells and Cosmology}\label{sec:2.2}

Before finding the $D=4$ solutions, we already have all of the tools necessary to physically interpret the spacetimes that arise from solutions to the equation \eqref{eqn:CSSeqns2}.

\begin{figure}[H]
	\centering
	\includegraphics{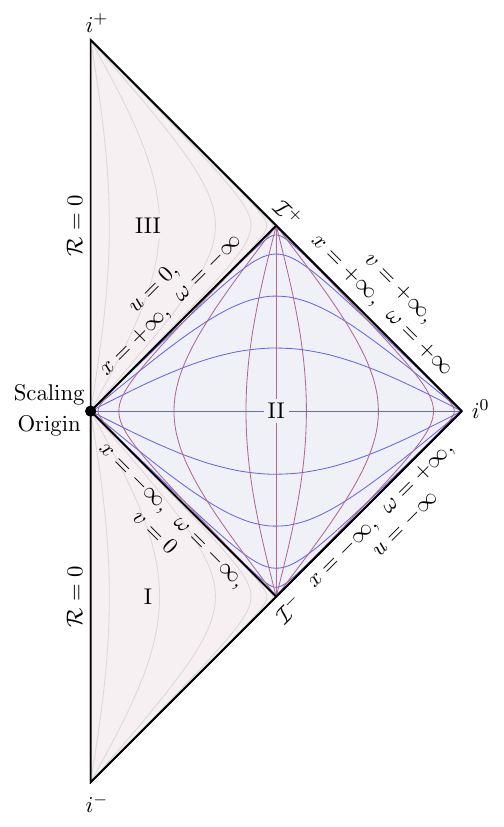}
	\caption{A Conformal Diagram demonstrating the fluid lines and HVF. Region I corresponds to flat Minkowski space due to our choice of boundary conditions. Region II is the dynamical region in which we solve the equations, with lines of constant $x$ indicated in blue and lines of constant $\omega$ in purple. The existence of Region III is dependent on the outcome of the dynamics in Region II. When it exists, it is also Minkowski space. Light grey lines in Region I and Region III indicate lines of constant $x$ beyond the self-similar horizon where the HVF becomes null.}
	\label{fig:PenDiag1}
\end{figure}

With the restriction to a scale-invariant scalar field, there is a more natural choice of scaling coordinate than \eqref{eqn:scalingcoords}, corresponding to the a coordinate over which the fluid velocity remains constant. To find it, we first note that the fluid velocity
\begin{equation}
	U^{\mu} \equiv \frac{\partial^{\mu}\phi}{\sqrt{-\partial^{\nu}\phi\partial_{\nu}\phi}}\,,~~ g_{\mu\nu}U^{\mu}U^{\nu}\equiv-1\,, 
	\label{eqn:fluidvel}
\end{equation}
is orthogonal to the HVF
\begin{equation}
	g_{\mu\nu}\xi^{\mu}\partial^{\nu}\phi =0\,.
\end{equation}
This observation is sufficient to determine the fluid velocity explicitly as
\begin{equation}
	U^{\mu}= -\left(\sqrt{\frac{-u}{2v}},\sqrt{\frac{v}{-2u}},0,0\right),
\end{equation}
with the comoving spacelike coordinate over which it remains constant
\begin{equation}
	\omega = \frac{1}{2} \log\left(-\frac{uv}{l^{2}}\right),
	\label{eqn:scalingcoordw}
\end{equation}
where $l$ is a constant dimensionful quantity. To visualise how these coordinates slice the spacetime, we sketch a conformal diagram in Figure \ref{fig:PenDiag1}. 
In these comoving coordinates, the metric is diagonal
\begin{equation}
	\diff s^{2} = \frac{1}{2}e^{2\omega}l^{2}\left(-\diff x^{2}+4\diff \omega^{2}+2 \cR(x)^{2}\diff \Omega^{2}\right).
\end{equation}
The coordinate $x$ provides a natural notion of time, increasing in value for any timelike observer from the past to the future. Meanwhile, $\omega$ serves as a spatial coordinate, increasing in value from the scaling origin out to spatial infinity, $i^{0}$. These coordinates also parameterise the two key vector fields discussed thus far: lines of constant $x$ corresponding to the HVF, and lines of constant $\omega$ tracing the fluid flow. Note that this information about the HVF is only accessible to a ``metaobserver" as it is purely spacelike in Region II.

The combination of spherical symmetry and self-similarity, which requires the entire spacetime to be filled by the collapsing fluid, allows us to interpret the scalar field as a continuum of concentric shells. Since $\omega$ is a comoving coordinate that remains constant along a given fluid line, each value of $\omega$ effectively parameterises the position of one of these shells. These concentric shells are all simply ``copies'' of one another at different scales with the relation between them determined by the self-similarity. This implies that the dynamics of just \textit{one} shell is sufficient to reproduce the full spacetime, simplifying the problem to a single dimension, $x$. In particular, the shell that is tracked by \eqref{eqn:CSSeqns2} is the one with $\omega=0$ --- a simple rescaling of $\cR$ allows us to track the others. 

Let us now consider a particular shell defined by the comoving coordinate $\omega=\omega_{\sz}$. Pulling back the metric onto their worldline gives
\begin{equation}
	\diff s^{2} = \frac{1}{2}e^{2\omega_{\scriptscriptstyle{0}}}l^{2}\left(-\diff x^{2}+2 \cR(x)^{2}\diff \Omega^{2}\right).
\end{equation}
If we compare this with the pull back of the FLRW metric onto the worldine of a comoving observer, $r=r_{\sz}$
\begin{equation}
	\diff s^{2} = -\diff{t}^{2}+a(t)^{2}r_{\sz}^{2}\diff \Omega^{2},
\end{equation}
then we can see that the pullbacks can be expressed in the same form after a redefinition of $x$.
\begin{center}
	\textit{A comoving observer in this self-similar spacetime experiences an effective FLRW-like expansion or contraction, where the comoving radius function $\cR(x)$ behaves like the cosmological scale factor.} 
\end{center}

Remarkably, although this interpretation only holds locally for comoving observers, it is actually exact at the level of the governing equation of both systems. With the substitution $z\equiv (y')^{2}$, and after using an integrating factor, the differential equation \eqref{eqn:CSSeqns2c} can be reduced to a first-order ``energy'' equation
\begin{equation}
	(y')^{2}= y^{2}-2y+y^{2}\left(\frac{2q}{y}\right)^{\frac{1}{\ve}},
	\label{eqn:1y2}
\end{equation}
where $q$ is an integration constant that will be given a physical interpretation in the next subsection. Without loss of generality, $q$ is chosen to be positive ($q>0$).

Returning to the radial function, $\cR(x)$ we find the differential equation
\begin{equation}
	(\cR')^{2}= \frac{(2q)^{D-2}}{4\cR^{2(D-3)}}+\frac{1}{4}\cR^{2}-\frac{1}{2},
	\label{eqn:Friedmann2}
\end{equation}
which has precisely the same form as the first Friedmann equation\footnote{A connection was also recently made between `Ordinary Wormholes' and the same types of FLRW universes arising from this equation \cite{Maloney:2025tnn}.}
\begin{equation}
	(\dot{a})^{2}=\frac{2}{(D-1)(D-2)}\rho(a) a^{2} +\frac{2\Lambda}{(D-1)(D-2)}a^{2}-k,
\end{equation}
for  a closed ($k=+1$) FLRW universe with equation of state parameter and cosmological constant given by
\begin{equation}
	w= \frac{D-3}{D-1},~~ \Lambda =\frac{(D-1)(D-2)}{8}.
\end{equation}
The density term in this equation can be seen to be of precisely the correct form to correspond to the density of our original spacetime
\begin{equation}
	\begin{aligned}
		\frac{M}{V} \sim \frac{q^{D-2}}{\cR^{D-2}}.
	\end{aligned}
\end{equation}
This suggests that the spatial curvature and cosmological constant terms (that we will collectively refer to as geometric terms) arise as a consequence of the self-similarity of the spacetime, the exact mechanisms of which are currently unknown. It is remarkable that these features can arise from such a simple matter model due to this choice of symmetry.

The evolution of the system is governed by two competing effects --- an outward force arising from the cosmological constant term, and an inwards force due to the energy density. The competition between these two will be an essential feature when finding the separation into regions caused by the \largeD expansion in the next section.

However, we can already see that that by varying $q$, the energy density of the universe changes. This gives rise to a family of different solutions, each corresponding to a different fate of the universe and each related to the three distinct outcomes for collapse that we discussed earlier in the following way:
\begin{itemize}
	\item Subcritical Collapse --- The outward force caused by the geometric terms always remains greater than the energy density. This causes the shells to slow down, before stopping and reversing their motion such that they begin to disperse outwards. A comoving observer in this scenario experiences a bouncing cosmology, where contraction is followed by a period of expansion. 
	\item Critical Collapse --- The energy density continues increases during the collapse to the point where it will eventually counterbalance the geometric terms exactly. Consequently, each shell approaches a fixed, finite physical radius. In this case a comoving observer experiences a collapsing cosmology that approaches a static state where the scale factor asymptotes to a constant. This static state is unstable, similar to the Einstein universe.
	\item Supercritical Collapse --- The density term eventually becomes larger than the geometric terms leading to a strong backreaction where every shell contracts to zero physical size at the same self-similar time. In this scenario a comoving observer will experience a crunching cosmology where the entire system collapses into a singularity.
\end{itemize}

\subsection{Review of the Four Dimensional Case}\label{sec:2.3}

\begin{figure}[H]
	\centering
	\includegraphics[width=0.85\linewidth]{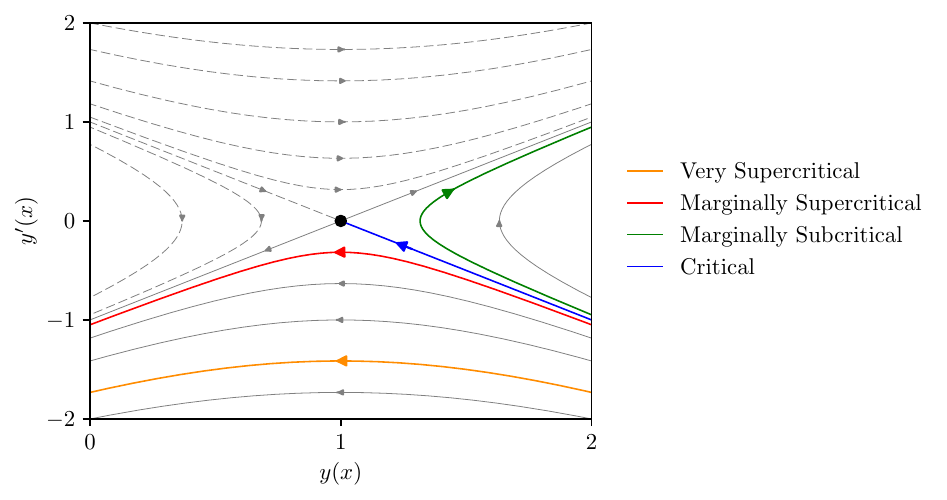}
	\caption{The phase portrait for \eqref{eqn:y4}. The physical flows are denoted with solid lines and some of these are colour coded as representatives of the flows that lead to the different end states detailed in the introduction to this section. In naming these flows, we have also introduced a notion of \textit{how} far from criticality the system is for later convenience.}
	\label{fig:phase4}
\end{figure}

Before taking the large dimension limit of this system, we first review the solutions in four dimensions that have been discussed previously \cite{Roberts:1989sk,Christodoulou:1993BoundedVS,Brady:1994xfa,Oshiro:1994hd} to gain some intuition. In four spacetime dimensions, $\ve = \frac{1}{2}$, the governing equation \eqref{eqn:CSSeqns2c} simplifies significantly to leave
\begin{equation}
	y''-y=-1\, .
	\label{eqn:y4}
\end{equation}
Although this is a very simple differential equation, let's go through the process of analysing its structure so that we have a point of comparison when looking at it in general dimensions. \\
Going to a first order formulation, we have
\begin{equation}
	\begin{aligned}
		y' =w,~~~ w' = y-1,
	\end{aligned}
\end{equation}
which has only one critical point given by $(y,w)=(1,0)$. The Jacobian is
\begin{equation}
	\begin{aligned}
		J&= \begin{pmatrix}
			\pdr{y'}{y} & \pdr{y'}{w}\\
			\pdr{w'}{y} & \pdr{w'}{w}
		\end{pmatrix}
		=\begin{pmatrix}
			0 & 1\\
			1 & 0
		\end{pmatrix},
	\end{aligned}
\end{equation}
which, when evaluated at the critical point, has two eigenvalues ($\lambda_{\pm} = \pm 1$) with opposite signs, signifying that it is a saddle point. By taking the initial boundary condition 
\begin{equation}
	\lim_{x\rightarrow -\infty}y\to \frac{1}{2}e^{-x},
	\label{eqn:bdcondy}
\end{equation}
we can see that all physical trajectories start at the far bottom right of the phase space, at large $y$ and large negative $y'$. The unfixed boundary condition gives us a free parameter that corresponds to the energy of the system, allowing us to start at different points relative to the critical manifold of the saddle. There are three very different possible trajectories depending no the value of this free parameter. In the phase portrait, Figure \ref{fig:phase4}, we identify these different trajectories with the names detailed in the introduction to this section --- Subcritical, Critical and Supercritical.

The critical solution represents the fine-tuned solution sitting at the boundary between solutions that disperse back to flat space and those that collapse to form a black hole. This is conceptually very similar to the situation studied by Choptuik \cite{Choptuik:1992jv} and is precisely the finite-dimensional phase space analogue mentioned previously. 

To discuss the explicit solutions, we now solve the differential equation to find
\begin{equation}
	y(x) = 1+ \frac{1}{2}e^{-x}+\frac{1}{2}\left(1-4q^{2}\right)e^{x},
	\label{eqn:ysol4}
\end{equation}
where the boundary condition \eqref{eqn:bdcondy} has been used to fix one of the integration constants. The remaining constant, $q$, has been chosen to match the first order form in \eqref{eqn:1y2}. It controls the late-time behaviour of the system and hence acts as the key free parameter that decides which end-state of the system will be reached. By simple comparison with the phase space picture, we can conclude that the critical value is $q_{*}=\frac{1}{2}$ and hence when
\begin{itemize}
	\item $q<q_{*}$ the system is \textcolor{mplGreen}{Subcritical},
	\item $q=q_{*}$ the system is \textcolor{mplBlue}{Critical},
	\item $q>q_{*}$ the system is \textcolor{mplRed}{Supercritical}.
\end{itemize}
It will be useful to introduce a measure of how close a solution is to criticality and so we define the small parameter
\begin{equation}
	\delta_{q} = q-q_{*}\,.
	\label{eqn:distcritq}
\end{equation}
\begin{figure}[H]
	\centering
	\includegraphics[width=0.9\textwidth]{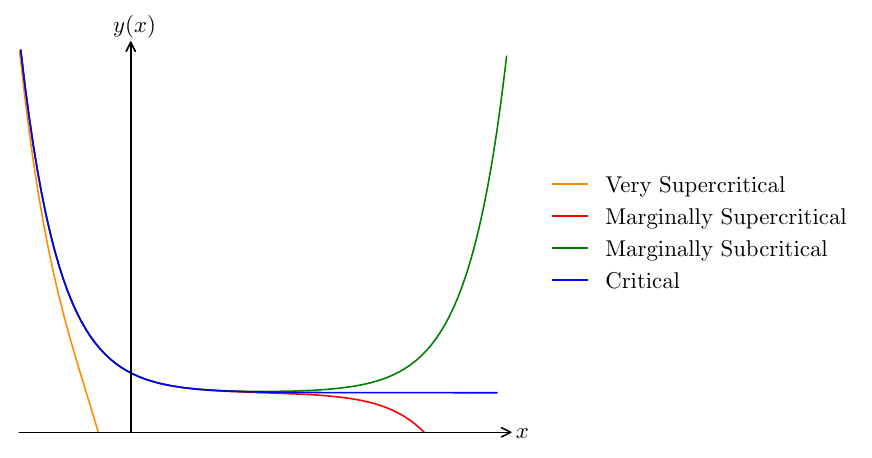}
	\caption{Solutions for the metric function in $D=4$, with various choices of $q$.}
	\label{Fig:Plotysol}
\end{figure}
In higher dimensions, the approximate shape of the solutions will not change dramatically, meaning that qualitatively the plot looks the same. The critical solution asymptotes to a constant value in the future --- the saddle point at $y=1$ and $y'=0$. If we are marginally above or below criticality, then the solution will hover near $y=1$ as it comes close to the saddle, before being repelled from it as the positive exponential in the solution dominates. If the initial data is too far from the critical solution, the solution quickly overshoots and doesn't `see' the asymptote as per the `Very Supercritical' solution in Figure \ref{Fig:Plotysol}.

\begin{figure}[H]
 	\centering
 	\includegraphics[width=0.9\textwidth]{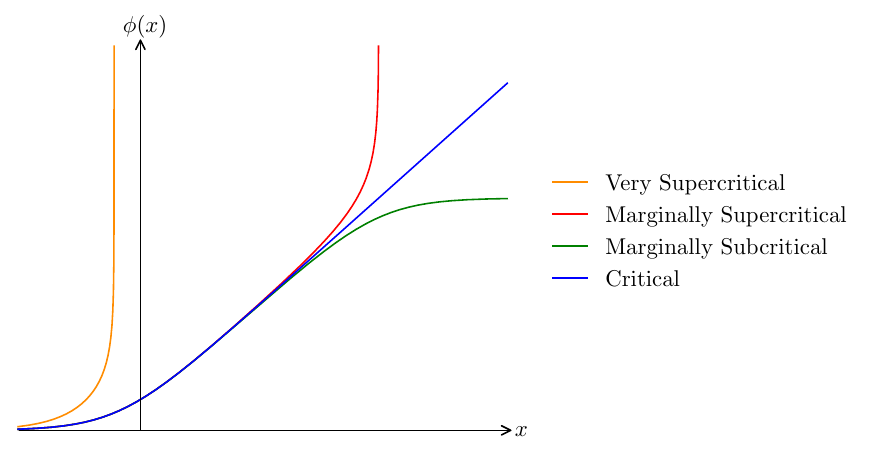}
 	\caption{Solutions for the scalar field in $D=4$, with various choices of $q$.}
 	\label{fig:plotphisol4}
 \end{figure}

We can then use \eqref{eqn:CSSeqns2b} and the boundary conditions \eqref{eqn:boundcond2} to solve for the scalar field
\begin{equation}
	\phi(x) = \frac{1}{\sqrt{2}}\log\left(\frac{1+e^{x}(1+2q)}{1+e^{x}(1-2q)}\right)\,.
\end{equation}
Plotting the result in Figure \ref{fig:plotphisol4} gives us behaviour that is analogous to the radius function.
The critical solution corresponds to a linear growth of the scalar field at late times, whilst the marginally subcritical and supercritical solutions follow the critical closely before asymptoting to a constant value or diverging respectively. When $q$ is too far away from the critical value, the solution does not `see' this linear growth once again. Note that the scalar field amplitude can reach arbitrarily large values whilst still ultimately dispersing provided it remains subcritical. The scalar field gradient is what determines the energy of the field and hence the strength of the back reaction. This is what the free parameter $q$ ultimately controls --- the initial amplitude of the scalar field gradient
\begin{equation}
	\rdr{\phi}{x} \sim 2\sqrt{2}q e^{x} + \order{e^{2x}} ~~\textit{as}~~ x\rightarrow-\infty.
	\label{eqn:scalaramp}
\end{equation}

In order to interpret the results physically, it is useful to write the solutions in null coordinates and compute some observables. The solutions are given by
\begin{equation}
	\begin{aligned}
		\Sigma(u,v) &=0,\\
		\phi(u,v) &=\frac{1}{\sqrt{2}}\log\left(\frac{(1+2q)v-u}{(1-2q)v-u}\right),\\
		\rho(u,v) &=\sqrt{\frac{u^{2}}{2}-uv+\frac{1}{2}(1-4q^{2})v^{2}}.
	\end{aligned}
\label{eqn:sold4}
\end{equation}
The spacetime structure of the solutions is clearly different depending on the value of the free parameter $q$. To see what they will correspond to, we will need the non-zero stress-energy tensor components
\begin{equation}
	\begin{aligned}
		T_{uu}&=\frac{2q^{2}v^{2}}{\rho^{4}}, ~~T_{vv}=\frac{2q^{2}u^{2}}{\rho^{4}}, ~~ T_{\theta_{1}\theta_{1}}=\frac{2q^{2}(-u)\,v}{\rho^{2}}, ~~
		T_{\theta_{2}\theta_{2}}&=\frac{2q^{2}(-u)\,v}{\rho^{2}}\sin^{2}(\theta_{1}),
	\end{aligned}
\end{equation}
the Ricci scalar
\begin{equation}
	R(u,v) = -\frac{4q^{2}(-u) v}{\rho^4},
	\label{eqn:ricci4}
\end{equation}
and the mass function as defined in \eqref{eqn:massfunc}
\begin{equation}
	\begin{aligned}
		m(u,v) =\frac{q^{2}(-u)v}{\rho}.
	\end{aligned}
\end{equation}
First, note that for $q=0$, all of these observables vanish and the full spacetime structure is simply that of flat space as can also be concluded directly from the solutions \eqref{eqn:sold4}. 

Secondly, extending the solution beyond either of the self-similarity horizons leads to a negative mass function in these regions. This indicates that our solution is only physically valid inside the domain in which we solved it. In fact, by extending the solution outside of it, there will be a time-like singularity at the origin where the Ricci Scalar diverges \eqref{eqn:ricci4}.\label{page:BoundCondStart}

This is the reason for our choice of boundary conditions. In order to construct a meaningful global solution, we must match our solution to a Minkowski spacetime along the $v=0$ null hypersurface since this choice will ensure regularity and asymptotic flatness of the spacetime in the past. We give a short justification that these boundary conditions work below, but a much more complete analysis for this exact problem was carried out by Christodoulou \cite{Christodoulou:1993BoundedVS}.

The normal (and tangent) vector to this null hupersurface is given by
\begin{equation}
	\hat{n}^{\mu}=(1,0,0,0)^{T},
\end{equation}
allowing us to see that the flux through it will vanish
\begin{equation}
	\begin{aligned}
		T_{\mu\nu}\hat{n}^{\mu}\rvert_{v=0} &= \left(0,T_{uv},0,0\right)\rvert_{v=0}\\
		&=(0,0,0,0),
	\end{aligned}
\end{equation} 
justifying the choice of matching to Minkowski space. We can also see this directly from the scalar field gradient normal to this surface
\begin{equation}
	\begin{aligned}
		\hat{n}^{\mu}\partial_{\mu}\phi\rvert_{v=0} &= \sqrt{2}q\left.\frac{v}{\rho^{2}} \right\rvert_{v=0}\\
		&=0.
	\end{aligned}
\end{equation}
The transition across the boundary is completely smooth and our imposed boundary conditions are regular. No jump means that there is no need for a source on this boundary according to the Israel Junction Conditions \cite{Israel:1966rt}.

The only non-zero stress energy tensor component at this surface is
\begin{equation}
	\begin{aligned}
		T_{vv}\rvert_{v=0} = \frac{8q^{2}}{u^{2}}\,,
	\end{aligned}
\end{equation}
which accounts for the fact that the scalar field derivative in the $v$ direction is non-zero along that boundary. Although this means that the scalar field's derivative is discontinuous here, making it only $\cC^{0}$, Christodoulou demonstrated that within the framework of Bounded Variation fields this is still physically well-defined. This framework relies on the use of weak derivatives, which are defined in an integral sense by requiring them to satisfy integration by parts conditions. In this way, weak derivatives can be well-defined even when the classical derivative is discontinuous.\label{page:BoundCondEnd}

With the issue of boundary conditions resolved, we can finally detail the spacetime structure for each of the different choices of $q$:\\

$\bullet$ \textbf{Subcritical --- \texorpdfstring{$\bm{0<q<\frac{1}{2}}$}}\\
\begin{minipage}{0.3\textwidth}
	\begin{figure}[H]
		\includegraphics[width=\textwidth]{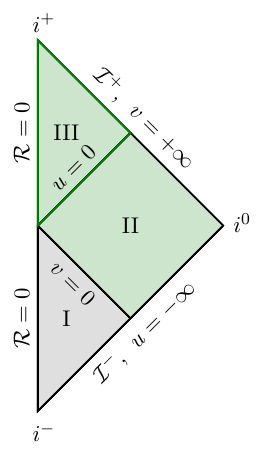}
	\end{figure}
\end{minipage} \hfill
\begin{minipage}{0.65\textwidth}
	The collapse of the scalar field does not produce a singularity. The scalar field disperses ensuring that there is no flux across the $u=0$ hypersurface and we can glue flat Minkowski space along this boundary for Region III.\\
	
	This follows from the same argument as for the $v=0$ boundary: If we go to $u=0$, the only non-vanishing component of the stress-energy tensor will be 
	\begin{equation}
		T_{uu}\rvert_{u=0} = \frac{8q^{2}}{(1-4q)^{2}v^{2}}
	\end{equation}
	and the mass function vanishes, $m(0,v)=0$.
\end{minipage}\\
\newpage
$\bullet$ \textbf{Critical --- \texorpdfstring{$\bm{q=\frac{1}{2}}$}}\\
\begin{minipage}{0.3\textwidth}
	\begin{figure}[H]
		\includegraphics[width=\textwidth]{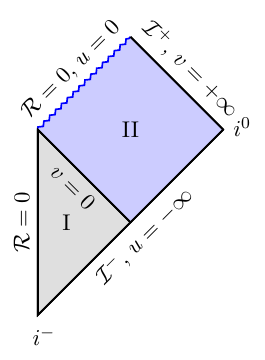}
	\end{figure}
\end{minipage} \hfill
\begin{minipage}{0.65\textwidth}
	At the critical value of $q$, the collapse of the scalar field leads to a null singularity at $u=0$.\\
	
	To see this, note that both the radius and mass functions vanish when $u=0$
	\begin{equation}
		\rho(u,v) = \sqrt{\frac{u^{2}}{2}-uv},
	\end{equation}
	\begin{equation}
		m(u,v) = \frac{1}{4}\sqrt{-uv},
	\end{equation}
	whilst the Ricci there diverges
	\begin{equation}
		R(u,v) = \frac{4 v}{u (u-2 v)^2}.
	\end{equation}
	This indicates a curvature singularity at $u=0$, but with no apparent horizon. Future null infinity, $\cI^{+}$, is geodesically incomplete and the singularity must be null. Although arbitrarily large values for the curvature can be observed, the singularity itself cannot be observed without actually reaching it.
\end{minipage}\\

$\bullet$ \textbf{Supercritical --- \texorpdfstring{$\bm{q>\frac{1}{2}}$}}\\
\begin{minipage}{0.3\textwidth}
	\begin{figure}[H]
		\includegraphics[width=\textwidth]{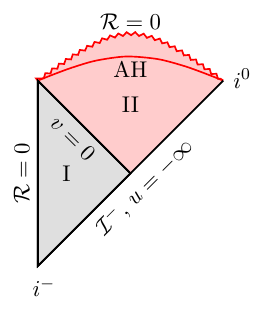}
	\end{figure}
\end{minipage} \hfill
\begin{minipage}{0.65\textwidth}
	For this range of $q$, the collapse of the scalar field leads to a spacelike singularity when the radius function vanishes as given by \eqref{eqn:ysol4} at
	\begin{equation}
		x_{0} = \log\left(\frac{1}{2\delta_{q}}\right).
	\end{equation}
	This singularity is preceded by an apparent horizon at a position determined by \eqref{eqn:AHcond},
	\begin{equation}
		x_{\ah}=\log\left(\frac{1}{4\delta_{q}(1+\delta_{q})}\right).
	\end{equation}
\end{minipage}\\

Note that this spacetime suffers from the problem that all self-similar spacetimes of this type do: \textit{the entire spacetime becomes trapped}. The mass function for such a spacetime diverges and the future is not asymptotically flat. However, this issue can be addressed by cutting off the scalar field influx at a finite time and gluing the resulting spacetime to an outgoing dust solution \cite{Wang:1996xh}. We will not worry about this here.\\

We are now ready to tackle the construction of the background spacetimes in a \largeD expansion.
\newpage

\section{Background Spacetime} \label{sec:3}

In this section, we derive the continuously self-similar solutions for the metric and scalar field, above, below and at criticality, using an expansion in the inverse number of spacetime dimensions.

Unlike the four-dimensional case, the governing equations do not admit a simple analytic solution in general dimensions. Nonetheless, the \largeD expansion will allow us to simplify this nonlinear system, by clearly defining regions in which different physical effects dominate. After obtaining solutions in each of these regions, we match them together to construct the full spacetime. We expect that many of the insights gained from this analysis will extend beyond this particular setup, thus providing a working example of how the \largeD expansion can be used to find solutions of gravitational collapse systems at criticality. Throughout the analysis, we validate key results using numerical methods.

We take the equations \eqref{eqn:CSSeqns2} derived in the previous section, expressed in terms of $\ve$ as a small parameter. As discussed there, the most important of these is
\begin{equation}
	2 \ve yy''+(1-2\ve)(y')^{2} + 2(1-\ve) y - y^{2}=0,
	\label{eqn:IVyeps}
\end{equation}
where $\ve$ is defined as in \eqref{eqn:defeps}.

We wish to find solutions in the limit $\ve\rightarrow 0^{+}$. In the strict $\ve=0$ limit, the differential equation becomes \textit{singular}, as its order reduces by one. In this case, there is not enough freedom to satisfy both boundary conditions at the same time for a generic solution. This typically implies that, for a period of time, the physical solutions must develop large second derivatives to compensate for the smallness of $\ve$, dictating the type of asymptotic analysis that we must perform \cite{BenderOrszag1999}.

\vskip 10pt
\noindent{\textit{Methodology: Matched Asymptotics}}

\noindent When a differential equation becomes singular in a particular limit of the control parameter, one typically uses the method of matched asymptotics (or boundary-layer analysis). Asymptotic solutions are constructed in separate regions of the domain, in the limit that a parameter in the equations is either very small or very large, before being matched together. In this way, instead of solving a difficult differential equation over the entire domain, the problem is reduced to solving simpler differential equations within subsets of this domain. These regions must have a small overlap with each other so that the solutions can be smoothly glued together to construct a global solution. The error of a matched asymptotic solution is generally $\order{\ve^{n+1}}$, where $n$ is the order at which the asymptotic expansion is truncated.

An outline of the method is as follows:
\begin{itemize}
	\item[1)] Identify the relevant timescales in the problem as a function of $\ve$. Each of these scales defines a distinct region of the system where different physical effects dominate.
	\item[2)] Rescale the independent variable according to the small parameter associated with each region e.g $X=\ve x$. 
	\item[3)] Make an asymptotic ansatz, e.g.
	\begin{equation}
		y(X) = \sum_n \ve^n y_n(X),~~\phi(X)= \sum_n \ve^n \phi_n(X), ~~ \text{as} ~~ \ve\rightarrow 0^{+}\,,
	\label{eqn:stdans}
	\end{equation}
	and substitute it into the differential equation. For convenience we will generally omit the region subscript on these ansatze, but it is clear from context to which region they belong.
	\item[4)] Solve the resulting simpler differential equations for $\{y_{n}(X), \phi_n(X)\}$ to the desired order in $\ve$.
	\item[5)] Repeat steps 1)--4) for all regions, and match the solutions at their boundaries or to the imposed boundary conditions. The final solution is constructed as a uniform solution $\yu$, using the matched asymptotic expansions:
		\begin{equation}
		\yu= \sum^{N}_{i=1}y_{i}-y_{\text{overlap}},
		\label{eqn:compositeform}
	\end{equation}
	where $y_{i}$ are the asymptotic approximations in each region and $y_{\text{overlap}}$ is the function that describes the total overlap between the regions.
\end{itemize}

This technique is well understood for linear differential equations, but the analysis can become more nuanced for nonlinear systems. In our problem, there is an additional complication due to an unfixed boundary condition, which, as we have seen, determines whether the system is above, below or at criticality. As a result, we must address the following challenges:
\begin{itemize}
	\item The relevant timescales cannot be identified solely through a method of dominant balance.
	\item The existence of certain regions depends on the boundary conditions. For example, the strong gravity region is absent for subcritical solutions.
	\item The size of the regions does not scale solely with the relevant timescale; it also depends strongly on the boundary condition. As a result, some regions may become unexpectedly large --- even infinitely large --- despite estimates based on the characteristic timescale suggesting that they should be small.
	\item The small parameter does not multiply the nonlinear terms. Consequently, expanding in the number of dimensions does not linearise the problem; instead, it isolates the essential nonlinearity of each region. The first non-trivial differential equation --- for example the equation for $y_{0}(X)$ --- remains nonlinear. Subsequent equations, however, are all inhomogeneous, linear differential equations.
\end{itemize}
This final point presents both a strength of our method and a challenge for it. On the positive side, it means that the nonlinearities are captured even at the leading-order by a large-dimension expansion. On the negative side, however, it means that we are still required to solve a nonlinear differential equation to find matched solutions, which is not always straightforward. As a consequence, there is one region governed by a deceptively simple nonlinear differential equation that we are unable to solve exactly. To make progress, we apply further approximations in the form of perturbations from criticality. 

The end result that we are only able to construct a uniform solution for the very supercritical case. The other outcomes are instead described by the step before this: a collection of formulae that describe the behaviour in each region to within the expected accuracy in terms of powers of $\ve$. 

We emphasise that this inability to find a composite solution is \textit{not} a failure of the separation of scales at \largeD\hspace{-0.35em}; rather, it reflects our inability to solve this particular non-linear differential equation. In fact, we are able to demonstrate that if a full analytic solution to this equation were available, a uniform solution could be constructed for every possible outcome.

With this in mind, let us outline the rest of this section. We first provide a qualitative discussion of the metric differential equation where we show how to understand the separation into regions and the characteristic timescales in each them. Then, we use this understanding to explicitly solve the metric differential equation in each of these regions, constructing each member of the one-parameter family as we go. After that, we use these solutions to solve the scalar field differential equation in an analogous way. 

\subsection{Analysis of the Metric Differential Equation} \label{sec:3.1}

Let us first qualitatively analyse the metric differential equation to identify the different regions and the relevant timescales in each of them. Once the different regions are fully understood, we will find explicit solutions for the metric and scalar field.

We begin by collecting some general information about the solutions in an arbitrary number of dimensions. As in $D=4$, we first study the phase space to demonstrate that some qualitative features of the system are crucially independent of the number of dimensions. To do this, we reduce the system to a first-order formulation by defining
\begin{equation}
	\begin{aligned}
		y' &= w\\
		 w' &=\frac{1}{2 \ve}y-\frac{1-\ve}{\ve}-\frac{(1-2\ve)}{\ve}\frac{w^{2}}{y}.
	\end{aligned}
\end{equation}
This system has a single critical point at $(y,w)=(2(1-\ve),0)$ with the Jacobian at this point given by 
\begin{equation}
	\begin{aligned}
		J&=\begin{pmatrix}
		0 & 1\\
		\frac{1}{2\ve} & 0
	\end{pmatrix}\end{aligned}.
\end{equation}
The Jacobian is invertible at the critical point, meaning that it is almost linear there provided $\ve\neq0$\footnote{This lack of invertibility in the strict $\ve\rightarrow0^{+}$ limit is related to the singular nature of the differential equation.}. The eigenvalues are real and have opposite signs
\begin{equation}
	\lambda_{\pm}=\pm\frac{1}{\sqrt{2\ve}},
	\label{eqn:eig}
\end{equation}
implying that the critical point remains a saddle point in any number of dimensions.

\begin{figure}[H]
	\centering
	\includegraphics[width=0.9\textwidth]{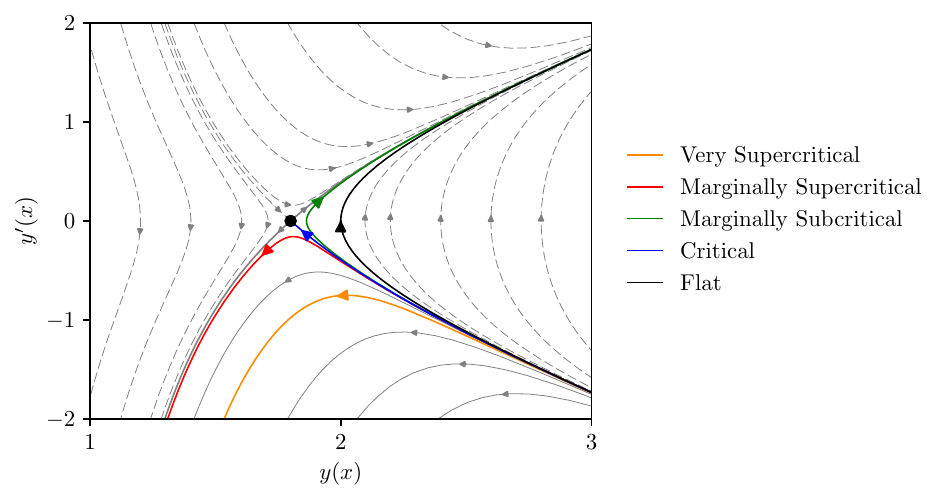}
	\caption{Phase space for $\ve=0.1$. As $\ve\rightarrow0^{+}$, the saddle point approaches the flat space trajectory before colliding with it in the strict limit $\ve=0$. As before, unphysical flows are shown with dashed lines, and representative flows are colour-coded as in $D=4$.}
	\label{fig:phaseD}
\end{figure}

Plotting the phase space in Figure \ref{fig:phaseD}, we observe broadly the same qualitative structure as before, with the critical manifold separating the trajectories that lead to black hole formation and dispersion.
Increasing the number of dimension has two noticeable effects on the structure of the phase space in the physical regions. First, the higher the number of dimensions, the \textit{sharper} the trajectories in phase space become --- that is, trajectories that deviate from the critical manifold move away from it more rapidly. 

Second, for supercritical trajectories with $D>4$, the formation of a singularity at $y=0$ corresponds to $y'=-\infty$ rather than a finite value. This difference is a direct consequence of the gravitational interaction becoming much stronger at short scales in higher dimensions. Note that the black hole is still formed in a finite self-similar time. 

To continue, we now turn to the first-order form of the equation derived in Section \ref{sec:2.2}. For convenience, we repeat the result here
\begin{equation}
	(y')^{2}= y^{2}-2y+y^{2}\left(\frac{2q}{y}\right)^{\frac{1}{\ve}},
	\label{eqn:1y}
\end{equation}
where $q$ is the initial amplitude of the incoming scalar field.

We can immediately determine the critical parameter $q_{*}$ in any dimension by evaluating the energy equation \eqref{eqn:1y} at the saddle point $y=2(1-\ve)$ and $y'=0$. Doing this gives
\begin{equation}
	\begin{aligned}
		q_{*}(\ve) &= \ve^{\ve}(1-\ve)^{1-\ve}\,,
	\end{aligned}
\end{equation}
where for $0<\ve\leq\frac{1}{2}$ we have $\frac{1}{2}< q_{*}\leq1$.

To understand the effect of changing the two free parameters $q$ and $\ve$ at the level of the differential equation, we have two options. The first is to examine the phase space plot, but it can be difficult to make precise statements using it. The second is to interpret \eqref{eqn:1y} as a zero-energy condition for a classical particle moving in an effective potential, $V(y)$, given by
\begin{equation}
	0 = \frac{1}{2}(y')^{2} + \left(2y-y^{2}-y^{2}\left(\frac{2q}{y}\right)^{\frac{1}{\ve}}\right)\equiv T(y')+V(y).
\end{equation}
This `particle' corresponds to a particular representative shell within the collapsing spacetime as discussed in Section \ref{sec:2.2}. For this reason, and due to the useful connection that we found between this differential equation and the first Friedmann equation\footnote{As a reminder of this connection, the first two terms in the effective potential are what we refer to as the ``geometric terms'' and correspond to the spatial curvature and cosmological constant, respectively. The third term arises from the energy density of the system and corresponds to the attractive gravitational force.}, we choose to emphasise this classical mechanics analogy.

\begin{figure}[H]
	\centering
	\includegraphics[width=0.85\textwidth]{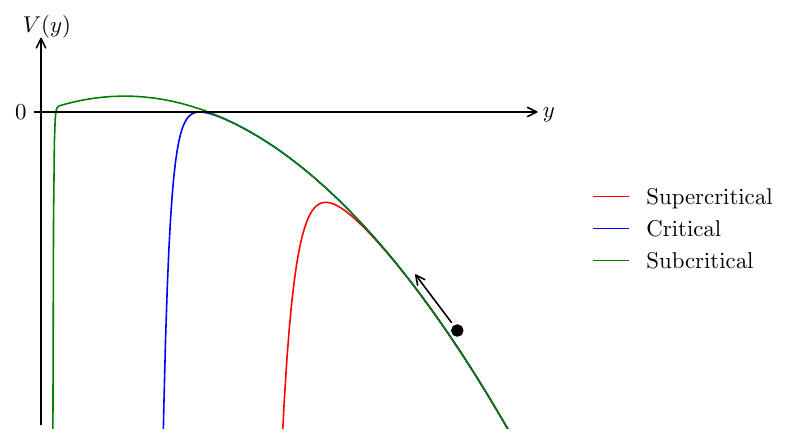}
	\caption{Effective potential for different values of the initial scalar field amplitude, $q$. A ball is shown moving along the potential to emphasise the analogy of a particle following a classical trajectory.}
	\label{fig:potentialplot}
\end{figure}

The geometric terms are fixed in both of these free parameters, but the density term is not. Figure \ref{fig:potentialplot} illustrates the effect that changing the initial amplitude of the scalar field, $q$, has on the potential --- it effectively changes the scale at which the the density term dominates.
From this, we can see that if the shell starts at $y=+\infty$, it can follow three qualitatively different trajectories, as in the phase space diagram, depending on the height of the potential barrier:
\begin{itemize}
	\item If the potential barrier is too high, the shell will `bounce' and disperse back to $y=+\infty$.
	\item If the peak of the potential is exactly at zero energy, the shell will asymptotically approach the saddle point at $y=2(1-\ve)$.
	\item If the peak is below this point, the particle has sufficient energy to cross the potential barrier and reach $y=0$, in a finite self-similar time. 
\end{itemize}

We now examine how the effective potential changes as the number of dimensions increases. We have already seen that the effect of changing the number of dimensions and $q$ are intertwined: the value of the critical parameter changes with the number of dimensions. This means that in order to isolate the effect of $\ve$ alone, we need to vary $q$ in a controlled way to ensure that the same physical solution is probed in every dimension. For this reason, we focus our attention on the situation where this dimension dependence is fully understood --- the critical solution. This is shown in Figure \ref{fig:friedmannpot}.

From the plot, and from the explicit form of the potential, we see that as the number of dimensions increases, the density term begins to dominate at larger scales and becomes significantly stronger at short scales. This leads to a sharper turn of the potential and hence more clearly defined regions where the different forces dominate. Although this isolated effect is harder to see for other values of $q$ where this dimensional dependence is not fully understood, the same qualitative behaviour persists. The task now for the remainder of this subsection is to understand these regions and how they translate into equivalent regions in the self-similar time $x$. Keep in mind that throughout this analysis, the non-trivial connection between the boundary conditions and the number of dimensions will repeatedly introduce complications, particularly in the time-scaling associated with each region.

The simplest of these regions is Region $\cA$, which corresponds to large scales where the geometric effects dominate. The attractive gravitational force effectively vanishes, and from \eqref{eqn:1y} we can immediately see that the characteristic timescale in this region is simply $\order{1}$.

The next region to consider is Region $\cB$, where the geometric and density terms approximately counterbalance each other, as expected for an intermediate boundary layer or transition region. For this reason, this region will be the most difficult to solve --- it has not reduced to a single dominant force. 

To gain insight into Region $\cB$, we note that it lies near the peak of the potential, or equivalently near the saddle point in the phase space. Therefore we can linearise about the saddle point to find
\begin{equation}
	y(x)\approx 2(1-\ve) + l_{1}(\ve)e^{-\frac{x}{\sqrt{2\ve}}}+l_{2}(\ve)e^{\frac{x}{\sqrt{2\ve}}}.
	\label{eqn:linearisation}
\end{equation}
The linearised solution unsurprisingly has the same form as the $D=4$ solution, given that in four dimensions the differential equation is linear. Although this solution is only valid near the saddle, it is clear that the positive exponential term is responsible for driving the system away from the critical point as $x\rightarrow \infty$. This allows us to identify $l_{2}\sim \delta_{q}$, where $\delta_{q}$ is the distance of the initial conditions from criticality introduced before \eqref{eqn:distcritq}. The dimensional dependence of $l_{1}$ and $l_{2}$ cannot be determined from this argument alone and will investigated later.

Importantly, the linearisation gives us the timescale associated with evolution close to the saddle point, $x \sim \sqrt{2\ve}$. Decreasing $\ve$ increases the rate at which the system moves along trajectories near the saddle, thereby reducing the time spent in this region.

From this, it is tempting to estimate that the time spent near the saddle will be $\order{\sqrt{2\ve}}$. However, this conclusion is premature as it assumes that all other factors in the problem are $\order{1}$. In particular, the timescale also depends on the potentially small free parameter $\delta_{q}$. Bringing the system closer to criticality, by tuning the value of $\delta_{q}$ towards zero, can significantly prolong the time spent near the saddle, with this time diverging at criticality. Despite gravity acting as a very short-range force in \largeD\hspace{-0.35em}, these fine-tuned solutions have very long tail effects \textit{in time} due to being near-criticality where the geometric and density terms counterbalance, leading to slow dynamics.

An important consequence of this is that when constructing asymptotic solutions in $\ve$ for the near-critical trajectories, we will encounter a second small parameter --- the distance from criticality $\delta_{q}$. Since it must implicitly depend on $\ve$, one must be careful to establish a hierarchy between $p$ and $\ve$. We will return to this point when constructing the solutions in the next subsection.

Finally, we have Region $\cC$ where the density term dominates. To analyse this region, we note that it is associated with the rapid runaway of the backreaction as a black hole forms when $y\rightarrow 0$. Keeping only the divergent terms in \eqref{eqn:1y}, we find
\begin{equation}
	\begin{aligned}
		y' &\approx-y\left(\frac{2q}{y}\right)^{\frac{1}{2\ve}}\\
		y &\approx 2q \left(-\frac{x+x_{0}(q,\ve)}{2\ve}\right)^{2\ve}\,,
	\end{aligned}
	\label{eqn:nearsing}
\end{equation}
where $x_{0}(q,\ve)$ is the \textit{exact} self-similar time at which the singularity forms. Note that this approximation only diverges for $\ve<\frac{1}{2}$, consistent with the change in phase space structure for $D>4$ as discussed earlier.

This seemingly simple approximate solution actually reveals three important insights about this region:
\begin{itemize}
	\item \textbf{The characteristic timescale changes with boundary conditions.} As discussed in Section \ref{sec:2.3}, the system's behaviour depends on \textit{how} supercritical it is, based on whether or not the solution `sees' the saddle. When $q \gg q_{*}$, the system is \textit{Very} Supercritical, and the collapse happens over a very short timescale $x\sim \ve$ --- exactly as expected for rapid collapse. However, when  $q \sim q_{*}$, the system is instead only \textit{Marginally} Supercritical. The dependence of $q_{*}$ on $\ve$ gives that the collapse happens over a longer timescale $x\sim \sqrt{\ve}$ --- the same as that of the near-saddle region. 
	\item \textbf{There is a convenient resummation in $\pmb{\ve}$.} For this particular region, it appears convenient to change variables to $y=Y^{\ve}$, simplifying the form of the solution.
	\item \textbf{The apparent horizon lies within this region.} Using equations \eqref{eqn:AHcond} and \eqref{eqn:1y} to solve for the radius of the apparent horizon, we find
	\begin{equation}
		y_{\ah} = \left(2^{1-2\ve}q\right)^{\frac{1}{1-\ve}}.
	\end{equation}
	By simple comparison with \eqref{eqn:nearsing}, we can see that this region \textit{includes} the apparent horizon.
\end{itemize}

\vskip 20 pt
This completes the analysis of the differential equation, allowing us to clearly identify the consequences of working at \largeD\hspace{-0.35em}: a natural separation of scales emerges, where the geometric and density terms in the differential equation dominate or balance in distinct regions, defined by an interval of the comoving scale factor $\cR$. These regions can be mapped onto equivalent regions in the self-similar time $x$, but care is needed due to the interplay between the boundary conditions and the number of dimensions. We summarise the key insights gained about each of these regions below, as they will form the foundation for the asymptotic analysis in the following subsections:
\begin{itemize}
	\item \textbf{Weak Gravity Region,} $\pmb{\cA}$: 
	\begin{itemize}
		\item At large distances, the density term is strongly suppressed in a large number of dimensions, leaving the geometric terms as the dominant effect.
		\item The characteristic time spent in this region is $\order{1}$.
	\end{itemize}
	\item \textbf{Transition Region,} $\pmb{\cB}$: 
	\begin{itemize}
		\item The geometric and density terms approximately counterbalance. The large number of dimensions localises this region to a narrow interval of the scale factor.
		\item The characteristic time spent in this region is $\order{\sqrt{\ve}}$.
		\item Despite the short characteristic timescale, the initial amplitude of the scalar field can be tuned to greatly extend the time spent in this region.
		\item A hierarchy must be established between the two small parameters, $\ve$ and the distance from criticality $\delta_{q}$, to have a consistent asymptotic expansion.
	\end{itemize}
	\item \textbf{Strong Gravity Region,} $\pmb{\cC}$:
	\begin{itemize}
		\item At short distances, the density term completely dominates.
		\item The characteristic time spent in this region scales as either $\order{\ve}$ or $\order{\sqrt{\ve}}$, depending on how far the initial amplitude of the scalar field is from criticality.
		\item The substitution $y=Y^{\ve}$ serves as an effective resummation in this region.
	\end{itemize}
\end{itemize}
In the next subsection, we will solve the simplified differential equations arising in each of these regions. However, we note that an alternative approach would be to use the integral representation coming from direct integration of the energy equation \eqref{eqn:1y}
\begin{equation}
	x = \pm\int \frac{\diff{y}}{y'},
	\label{eqn:intrep}
\end{equation}
together with the `Method of Regions'. We discuss this approach for completeness in Appendix \ref{app:MoR}.\footnote{We choose to focus on differential methods rather than integration for two key reasons. First, the integral representation is a luxury of this specific problem. Most differential equations in GR do not permit this simplification, as they remain PDEs or coupled in some non-trivial way. For this reason, we aim to minimise our reliance on this representation to keep the lessons learned here as general as possible. 

Second, the integral representation provides an implicit solution $x(y)$, which needs to be inverted to find an explicit solution for $y(x)$. This inversion is not generally straightforward and may become impractical or impossible when working perturbatively, particularly in the presence of nonlinearities or singular behaviour.}

\subsection{The Metric Function} \label{sec:3.2}
Before discussing the technical details of constructing the solutions to the differential equation, we summarise the different types of spacetimes that each solution corresponds to and present explicit formulas for $y(x)$.
\vskip 20pt
\noindent{\it Summary of results:}

\begin{figure}[H]
	\centering
	\begin{subfigure}[b]{0.45\textwidth}
		\includegraphics[width=\textwidth]{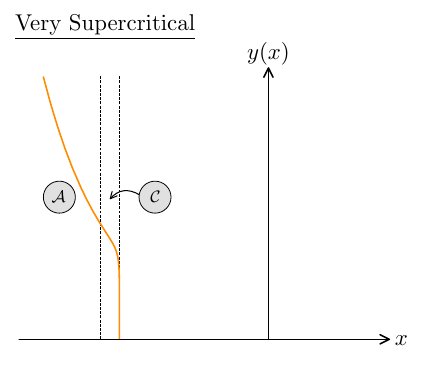}
	\end{subfigure}
	\begin{subfigure}[b]{0.45\textwidth}
		\includegraphics[width=\textwidth]{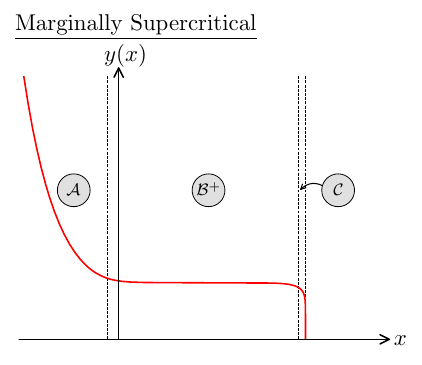}
	\end{subfigure}
	\\
	\begin{subfigure}[b]{0.45\textwidth}
		\includegraphics[width=\textwidth]{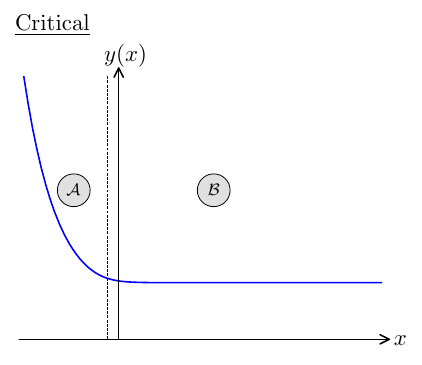}
	\end{subfigure}
	\begin{subfigure}[b]{0.45\textwidth}
		\includegraphics[width=\textwidth]{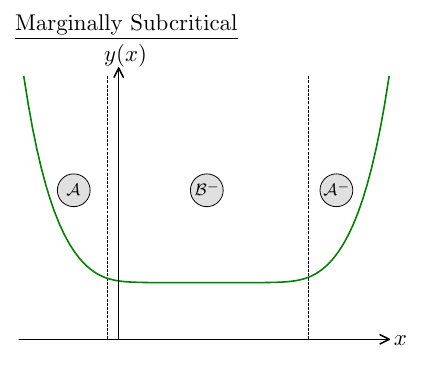}
	\end{subfigure}
	\caption{Numerical results for each qualitatively distinct solution, with the regions appropriately labelled.}
	\label{fig:NumericalD}
\end{figure}

There are four qualitatively distinct solutions that we explicitly construct, corresponding to the four different types of trajectories as labelled in the phase space diagram in Figure \ref{fig:phaseD}. Sketches of the metric functions in each of these cases are shown in Figure \ref{fig:NumericalD}. To distinguish between supercritical and subcritical versions of the same region, we have added a superscript $+$ or $-$ where appropriate. Despite this distinction, the underlying intuition about the regions remains the same. 
\label{page:MetricSolStart}

\newpage
\begin{itemize}
\item {\bf Very Supercritical}: The initial data is tuned such that the system is well above threshold, $q\gg q_{*}$. A black hole forms at \textit{early} self-similar times without the saddle point being probed. Consequently, this spacetime has a simple structure in terms of regions, consisting of only Regions $\cA$ and $\cC$. In this case, a uniform solution can be constructed. The metric, as $\ve \to 0^+$, is given by:
\begin{center}
	\begin{tcolorbox}[mybox]
		\begin{align*}
			&\ya(x)=1+\cosh(x)\notag \\
			&\yc(x) \sim 	(2^{1-2\ve}q)^{\frac{1}{1-\ve}}\left(\frac{1}{q-1}\left(\frac{1}{2}e^{\sqrt{\frac{q-1}{q}}\left(\frac{x+c}{\ve}\right)}+\frac{1}{2}e^{-\sqrt{\frac{q-1}{q}}\left(\frac{x+c}{\ve}\right)}-1\right)+\order{\ve^{2}}\right)^{\ve} 
			\notag\\
			&\yu(x) \sim 1+\cosh(x) + 2q\ve \log\left(\frac{q}{q-1} \sinh^{2}\left(\sqrt{\frac{q-1}{q}}\frac{x+c}{2\ve}\right)\right)\notag\\
			&\qquad~~~~~~-2q\ve \left(\log\left(\frac{q}{4(q-1)}\right)-\sqrt{\frac{q-1}{q}}\left(\frac{x+c}{\ve}\right)\right) +\order{\ve^{2}}\notag\\
			&c =\log\left(-1+2q+\sqrt{q(q-1)}\right)+\ve\sqrt{\frac{q}{q-1}}\log\left(\frac{q}{4(q-1)}\right).\notag
		\end{align*}
	\end{tcolorbox}
\end{center}
\item {\bf Marginally Supercritical}: The initial data is tuned such that the system is only slightly above criticality, introducing the second small parameter $\delta_{q}$, measuring this distance from criticality. This results in the formation of a black hole at \textit{late} self-similar times as the saddle point is now probed for a prolonged period of time. The dynamical solution passes through three Regions: $\cA$, $\cB^{+}$ and $\cC$ \footnote{Technically, Region $\cC$ does not exist. However, we include it here as a necessary region given our inability to solve the full differential equation in region $\cB^{+}$. We will discuss this in detail later.}. The double approximation to find this solution prevents analytic matching and the construction of a single, uniform solution. In terms of these regions, however, the metric as $\ve \to 0^+$ is given by:
\begin{center}
	\begin{tcolorbox}[mybox]
		\begin{align*}
			&\ya(x)=1+\cosh(x) \notag\\
			&\ybp(x) \sim 2(1-\ve) +\ve \sum_{n=1}^{\infty}\alpha_{n}\left(e^{-\frac{n}{\sqrt{2\ve}}(x+\sqrt{2\ve}b_{\scriptscriptstyle{0}})}-e^{\frac{n}{\sqrt{2\ve}}(x+\sqrt{2\ve}(b_{\scriptscriptstyle{0}}-d_{3}\log(\abs{\delta_{q}})))} \right) +\order{\ve^{2}} \notag\\
			&\yc(x) \sim 	(2^{1-2\ve}q)^{\frac{1}{1-\ve}}\left(\frac{1}{q-1}\left(\frac{1}{2}e^{\sqrt{\frac{q-1}{q}}\left(\frac{x+d_{4}}{\ve}\right)}+\frac{1}{2}e^{-\sqrt{\frac{q-1}{q}}\left(\frac{x+d_{4}}{\ve}\right)}-1\right)+\order{\ve^{2}}\right)^{\ve} \notag\\
			&\alpha_{n} =\begin{cases}
				-\frac{1}{4(n^{2}-1)}\left(\left(\frac{n}{2}\right)^{2}\alpha_{\frac{n}{2}}^{2}+\sum_{i=1}^{\frac{n}{2}-1}2i(n-i)\alpha_{i}\alpha_{n-i}\right) & \forall n\, \text{even}\\
				-\frac{1}{4(n^{2}-1)}\left(\sum_{i=1}^{\frac{n-1}{2}}2i(n-i)\alpha_{i}\alpha_{n-i}\right) & \forall n\, \text{odd}
			\end{cases}\notag\\
			&\alpha_{\scriptscriptstyle{1}}=1 \notag\\
			&b_{0} \approx-0.281221\,,\notag
		\end{align*}
	\end{tcolorbox}
\end{center}
where $d_{3}$ and $d_{4}$ are constants that must be matched numerically.
\item {\bf Critical}: The initial data is fine-tuned such that the system is at criticality. This causes the system to spend an infinite amount of self-similar time approaching the saddle point. This spacetime also has a simple structure in terms of regions, consisting of only Regions $\cA$ and $\cB$. We are able to analytically match these regions, but without a full solution to the differential equation controlling Region $\cB$ we are unable to construct a single, uniform solution. The metric, as $\ve \to 0^+$, is given by:
\begin{center}
	\begin{tcolorbox}[mybox]
		\begin{align*}
			&\ya(x)=1+\cosh(x) \notag\\
			&\yb(x) \sim 2 + \ve\left(-2+\sum_{n=1}^{\infty}\alpha_{n}e^{-\frac{n}{\sqrt{2\ve}}(x+\sqrt{2\ve}b_{\scriptscriptstyle{0}})}\right) +\order{\ve^{2}}\notag\\
			&\alpha_{n} =\begin{cases}
				-\frac{1}{4(n^{2}-1)}\left(\left(\frac{n}{2}\right)^{2}\alpha_{\frac{n}{2}}^{2}+\sum_{i=1}^{\frac{n}{2}-1}2i(n-i)\alpha_{i}\alpha_{n-i}\right) & \forall n\, \text{even}\\
				-\frac{1}{4(n^{2}-1)}\left(\sum_{i=1}^{\frac{n-1}{2}}2i(n-i)\alpha_{i}\alpha_{n-i}\right) & \forall n\, \text{odd}
			\end{cases}\notag\\
			&\alpha_{\scriptscriptstyle{1}}=1 \notag\\
			&b_{0}\approx-0.281221\,.\notag
		\end{align*}
	\end{tcolorbox}
\end{center}
\item {\bf Marginally Subcritical}: The initial data is tuned such that the system is only slightly below criticality, introducing a second small parameter $\delta_{q}$, measuring this distance from criticality. In this solution, the scalar field eventually disperses. The dynamical solution passes through three Regions: $\cA$, $\cB^{-}$ and $\cA^{-}$. The double approximation to find this solution prevents analytic matching and the construction of a single, uniform solution. In terms of these regions, however, the metric as $\ve \to 0^+$ is given by:
\begin{center}
	\begin{tcolorbox}[mybox]
		\begin{align*}
			&\ya(x)=1+\cosh(x) \notag\\
			&\ybm(x) \sim 2(1-\ve) +\ve \sum_{n=1}^{\infty}\alpha_{n}\left(e^{-\frac{n}{\sqrt{2\ve}}(x+\sqrt{2\ve}b_{\scriptscriptstyle{0}})}+e^{\frac{n}{\sqrt{2\ve}}(x+\sqrt{2\ve}(b_{\scriptscriptstyle{0}}-d_{1}\log(\abs{\delta_{q}})))} \right) +\order{\ve^{2}} \notag\\
			&\yam(x) = 1+\cosh\left(x+\sqrt{2\ve}\log\left(\frac{\ve}{\abs{\delta_{q}}}\right)d_{2}\right)\notag \\
			&\alpha_{n} =\begin{cases}
				-\frac{1}{4(n^{2}-1)}\left(\left(\frac{n}{2}\right)^{2}\alpha_{\frac{n}{2}}^{2}+\sum_{i=1}^{\frac{n}{2}-1}2i(n-i)\alpha_{i}\alpha_{n-i}\right) & \forall n\, \text{even}\\
				-\frac{1}{4(n^{2}-1)}\left(\sum_{i=1}^{\frac{n-1}{2}}2i(n-i)\alpha_{i}\alpha_{n-i}\right) & \forall n\, \text{odd}
			\end{cases}\notag\\
			&\alpha_{\scriptscriptstyle{1}}=1 \notag\\
			&b_{0} \approx-0.281221\,,\notag
		\end{align*}
	\end{tcolorbox}
\end{center}
where $d_{1}$ and $d_{2}$ are constants that must be matched numerically.
\end{itemize} \label{page:MetricSolEnd}

\subsubsection{The Very Supercritical Solution}

\label{page:RegionAStart}In order to construct the very supercritical spacetime, we first solve for Region $\cA$, which appears in all spacetimes and provides the simplest example of our methodology.

The solution in this region corresponds to the {\it `Outer Solution'} in the method of matched asymptotics and can be found by leaving the independent variable unscaled. Substituting the asymptotic ansatz \eqref{eqn:stdans}, into the differential equation \eqref{eqn:IVyeps}, we obtain the following hierarchy of equations
\begin{equation}
	\begin{aligned}
		&\ve^{0}: \, 2y_{0}-y_{0}^{2}+(y_{0}')^{2}=0\\
		&\ve^{1}: \, 2 y_{0}' y_{1}'-2 y_{0} y_{1}+2 y_{1}=2 y_{0}-2 y_{0} y_{0}''+2 y_{0}'^2\\
		&\ve^{2}: \,  2 y_{0}' y_{2}'-2 y_{0} y_{2}+2 y_{2}= 4 y_{0}' y_{1}'-2 y_{1} y_{0}''-2 y_{0} y_{1}''-(y_{1}')^2+y_{1}^2+2 y_{1}\,.
	\end{aligned}
\end{equation}
The zeroth-order equation has two possible solutions
\begin{equation}
	\begin{aligned}
		y_{0}(x) &= 1+\cosh \left(x+C_{0}\right)\\
		y_{0}(x) &= 2.
	\end{aligned}
\end{equation}
The first solution has the correct asymptotic behaviour to be identified with our flat space boundary conditions, while the second solution describes a state where the system sits at the saddle point indefinitely, which we discard. Imposing the boundary condition \eqref{eqn:bdcondy} fixes $C_{0}$ giving
\begin{equation}
	y_{0}(x) = 1+\cosh(x).
\end{equation}
Substituting this into the $\order{\ve^1}$ equation and solving for $y_1(x)$ gives
\begin{equation}
	y_{1}(x)=C_{1} \sinh(x).
\end{equation}
It can be shown that the boundary condition sets $C_{1}=0$, along with all higher order solutions, such that $y_{n}(x)=0~~ \forall n\geq1$. Therefore, the leading-order solution is correct to \textit{all} orders in perturbation theory, ensuring that the errors in Region $\cA$ are at most exponentially small
\begin{equation}
	\abs{\ya-y_{0}} = \order{\ve^{n}} \; \forall n \; \text{as} \; \ve \rightarrow 0^{+}.
\end{equation}
Thus the full solution to Region $\cA$ is
\begin{equation}
	\ya(x) = 1+\cosh(x).
	\label{eqn:flaty}
\end{equation}
Returning to the null $(u,v)$ coordinates, this solution can be seen to correspond precisely to flat space. Region $\cA$ therefore describes a regime where the gravitational interaction is suppressed at \largeD, and the scalar field evolves as though it were in flat spacetime, up to exponentially small corrections.\label{page:RegionAEnd}\\

For the very supercritical solution, where $q\gg q_{*}$, the only additional region to solve for is Region $\cC$. Since there is no second small parameter $\delta_{q}$ to consider, we choose to work with the first-order equation \eqref{eqn:1y}. For completeness, and to highlight a small subtlety, we discuss using the second order differential equation in Appendix \ref{app:SecondOrd}.\label{page:RegionCStart}

Using the information gathered about this region in the previous subsection, we now expand equation \eqref{eqn:1y} near the apparent horizon
\begin{equation}
	y(x)=(2^{1-2\ve}q)^{\frac{1}{1-\ve}}Y(x)^{\ve},
	\label{eqn:ycreparameter}
\end{equation}
giving
\begin{equation}
	(2^{1-2\ve}q)^{\frac{1}{1-\ve}}\ve^{2}(Y')^{2}=(2^{1-2\ve}q)^{\frac{1}{1-\ve}}Y^{2}-2Y^{2-\ve}+4Y.
\end{equation}
It is now clear that it is natural to introduce the fast time variable $z=x\ve$ as previously identified in equation \eqref{eqn:nearsing} so that all terms are balanced in the differential equation. This defines a \textit{`boundary layer'} --- a region where the formally small $\ve^{2}(Y')^{2}$ competes with the others as $Y'\sim \ve^{-1}$.

Substituting this fast time variable and the standard asymptotic ansatz, we find
\begin{equation}
	\begin{aligned}
		&\ve^{0}: \, 2q(Y_{0}')^{2}=2(q-1)Y_{0}^{2}+4Y_{0}\, ,\\
		&\ve^{1}: \, 2(qY_{0}'Y_{1}'-(1+(q-1)Y_{0})Y_{1})=-q \log\left(\frac{q}{2}\right)(Y_{0}')^{2}+\left(q\log\left(\frac{q}{2}\right)+\log(Y_{0})\right)Y_{0}^{2}.
	\end{aligned}
	\label{eqn:Yexp}
\end{equation}

The zeroth order equation can be solved to give
\begin{equation}
	Y_{0}(x) = \frac{1}{q-1}\left(\frac{1}{2}e^{\sqrt{\frac{q-1}{q}}\left(\frac{x+c}{\ve}\right)}+\frac{1}{2}e^{-\sqrt{\frac{q-1}{q}}\left(\frac{x+c}{\ve}\right)}-1\right),
	\label{eqn:Y0sol}
\end{equation}
where it is necessary to let $c=c_{0}+\ve c_{1}$ in order to match consistently with the solution in Region $\cA$. The next-order equation for $Y_{1}$ can also be solved explicitly and seen to improve the solution; we discuss it in Appendix \ref{app:HigherOrd}.

The asymptotic approximation for Region $\cC$ is given by
\begin{equation}
	\yc(x) \sim 	(2^{1-2\ve}q)^{\frac{1}{1-\ve}}\left(\frac{1}{q-1}\left(\frac{1}{2}e^{\sqrt{\frac{q-1}{q}}\left(\frac{x+c}{\ve}\right)}+\frac{1}{2}e^{-\sqrt{\frac{q-1}{q}}\left(\frac{x+c}{\ve}\right)}-1\right)+\order{\ve^{2}}\right)^{\ve} ~~\text{as}~~\ve\rightarrow 0^{+}.
	\label{eqn:yright}
\end{equation}\label{page:RegionCEnd}
\label{page:VSupStart}To match this smoothly to Region $\cA$, we first introduce the strong interaction time variable $\xc=\sqrt{\frac{q-1}{q}}\frac{x+c}{\ve}$ and expand $\ya$ in terms of it
\begin{equation}
	\begin{aligned}
		\ya(\xc) &= 1+\frac{1}{2}\left(e^{\ve\sqrt{\frac{q-1}{q}}\xce -c_{0}-\ve c_{1}}+e^{-\ve\sqrt{\frac{q-1}{q}}\xce -c_{0}-\ve c_{1}}\right)\\
		& \sim \frac{1}{2}e^{-c_{0}}(1+e^{c_{0}})^{2}+\frac{1}{2}e^{-c_{0}}(e^{2c_{0}}-1)\left(c_{1}-\sqrt{\frac{q}{q-1}}\xc\right)\ve+\order{\ve^{2}}.
	\end{aligned}
	\label{eqn:leftinright1}
\end{equation}
Similarly, expanding $\yc$ in terms of the outer variable $x$, or equivalently as $\xc\rightarrow -\infty$, we have
\begin{equation}
	\begin{aligned}
		\lim_{\xc\to-\infty}\yc(\xc) &\xrightarrow{\sim} (2^{1-2\ve}q)^{\frac{1}{1-\ve}}\left(\frac{1}{q-1}\left(\frac{1}{2}e^{\xce}+\frac{1}{2}e^{-\xce}-1\right)+\order{\ve^{2}}\right)^{\ve} \\
		&\sim 2q +2q\ve \left(\log\left(\frac{q}{4(q-1)}\right)-\xc\right)+\order{\ve^{2}}.
	\end{aligned}
	\label{eqn:rightinleft1}
\end{equation}
Comparing these expansions order-by-order in $\ve$ determines the matching coefficients to be
\begin{equation}
	\begin{aligned}
		c_{0} &= \log\left(-1+2q+\sqrt{q(q-1)}\right)\\
		c_{1} &= \sqrt{\frac{q}{q-1}}\log\left(\frac{q}{4(q-1)}\right).
	\end{aligned}
	\label{eqn:cmatching}
\end{equation} 
\begin{figure}[H]
	\centering
	\includegraphics[width=0.9\textwidth]{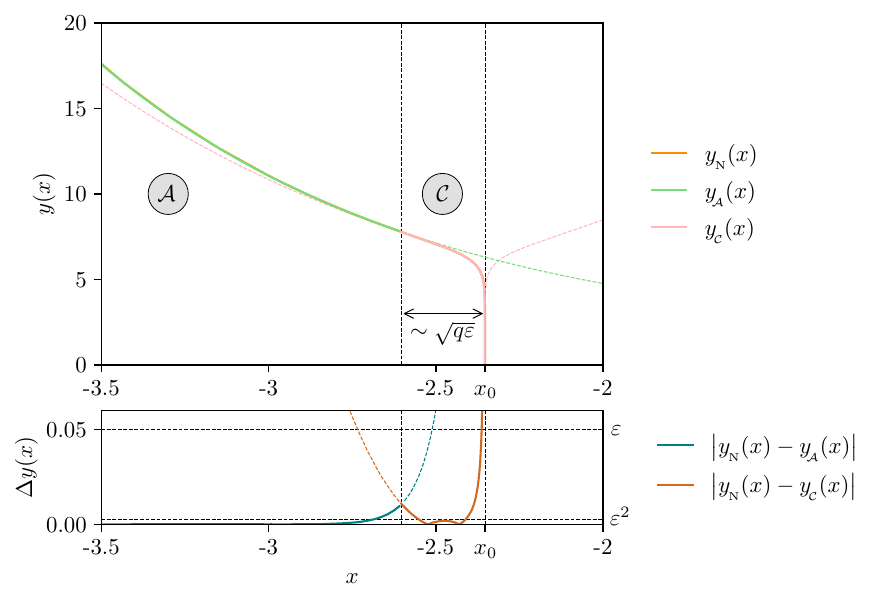}
	\caption{Comparison of the very supercritical solution approximations with the numerical result for $\ve=0.05$. The residuals are consistent with the error estimate, except near the singularity.}
	\label{fig:vsupcritapprox}
\end{figure}
Figure \ref{fig:vsupcritapprox} compares the analytical solutions in each region with the numerical solution. The residuals are consistent with the expected error of $\ve^{2}$. The size of Region $\cC$ is determined by the breakdown of perturbation theory, which occurs when
\begin{equation}
	\begin{aligned}
		Y_{0}(\xc) &\sim \ve Y_{1}(\xc) ~~\text{as}~~\ve\rightarrow 0^{+}, ~~ \xc \rightarrow -\infty.
	\end{aligned}
\end{equation}
Although the errors grow as the singular point $x_{0}$ is approached, this is \textit{not} a sign that a new region is missing. To clarify this statement, let us expand around the singular point
\begin{equation}
	\begin{aligned}
		\yc(x) &\sim (2^{1-2\ve}q)^{\frac{1}{1-\ve}}(2q)^{-\ve}\left(-\frac{x+c}{\ve}+\order{\left(\frac{x+c}{\ve}\right)^2}\right)^{2\ve}\\
		&\sim 2q\left(-\frac{x+c}{\ve}\right)^{2\ve} ~~\text{as} ~~ x \rightarrow -b.
	\end{aligned}
\end{equation}
This matches the approximate near-singularity solution previously obtained in \eqref{eqn:nearsing}. The functional form of $\yc$ is therefore correct to describe the singularity, but its position is not exactly determined; that is, $c\neq x_{0}$. We consequently conclude that
\begin{equation}
	x_{0}\sim c_{0}+\ve c_{1} + \order{\ve^{2}},
	\label{eqn:singapprox}
\end{equation}
and that computing the higher order corrections to $Y$ in $\ve$ will improve the approximation of the singularity's position. The growing error near the singularity arises because $y'(x)$ diverges there --- any small error in the identification of $x_{0}$ within the analytic solution translates into a large error in $y(x)$ itself.

\begin{figure}[H]
	\centering
	\includegraphics[width=0.85\textwidth]{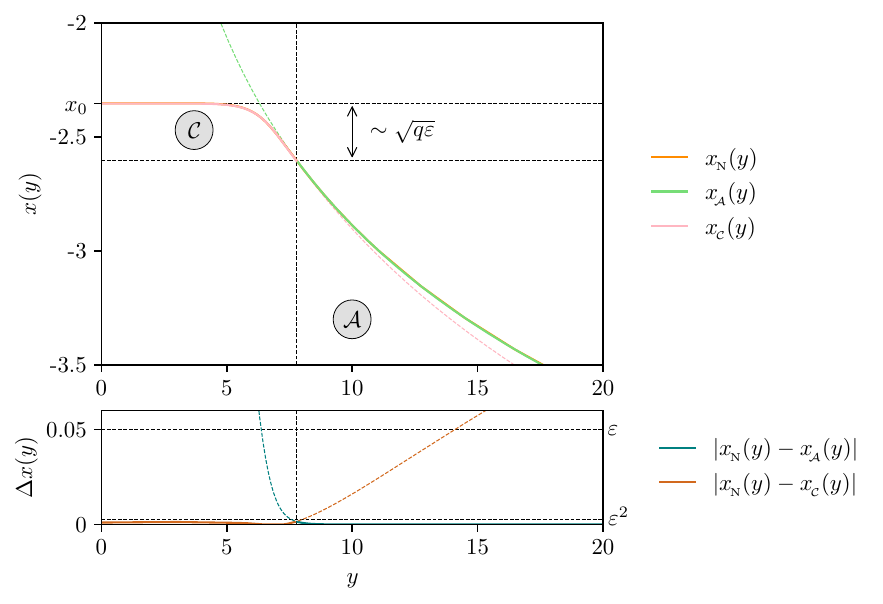}
	\caption{Comparison of the inverse function of the very supercritical solution approximations with the numerical result for $\ve=0.05$. The residuals are consistent with the error estimate over the entire domain.}
	\label{fig:vsupcritinv}
\end{figure}

This does \textit{not} imply a failure of perturbation theory at this order. If we examine the inverse function $x(y)$, as shown in Figure \ref{fig:vsupcritinv}, we see that the errors remain consistent with the expected order of the expansion across the entire domain. The conclusion is that the solution remains reliable even when $y'(x)$ diverges near the singularity --- the divergence simply reflects the physical nature of the singularity rather than a failure of the perturbative expansion.
The use of the resummation in $\ve$, introduced through the ansatz $y\sim Y^{\ve}$, leads to improved accuracy near the singularity compared to an expanded solution in the exponent. The reason for this is that expanding this solution is only valid when the quantity in parentheses is not too small. Assuming this, and expanding the solution gives
\begin{equation}
	y_{R,\text{exp}}(x) \sim 2q + 2q\ve \log\left(\frac{q}{q-1} \sinh^{2}\left(\sqrt{\frac{q-1}{q}}\frac{c+x}{2\ve}\right)\right)+\order{\ve^{2}}.
	\label{eqn:yrightexp}
\end{equation}
For this reason, we choose not to expand this exponent and keep the resummation. 

However, the expanded solution \eqref{eqn:yrightexp} has one notable advantage: it allows us to construct a composite solution with uniform errors across the entire domain. The overlap can be found to be given by
\begin{equation}
	y_{\text{overlap}}(x)=2q +2q\ve \left(\log\left(\frac{q}{4(q-1)}\right)-\sqrt{\frac{q-1}{q}}\left(\frac{x+c}{\ve}\right)\right)\,,
\end{equation}
allowing us to write the uniform solution as
\begin{equation}
	\begin{aligned}
		\yu(x) \sim 1+&\cosh(x) + 2q\ve \log\left(\frac{q}{q-1} \sinh^{2}\left(\sqrt{\frac{q-1}{q}}\frac{x+c}{2\ve}\right)\right)\\
		&-2q\ve \left(\log\left(\frac{q}{4(q-1)}\right)-\sqrt{\frac{q-1}{q}}\left(\frac{x+c}{\ve}\right)\right) +\order{\ve^{2}}.
	\end{aligned}
\end{equation}
\begin{figure}[H]
	\centering
	\includegraphics[width=0.85\textwidth]{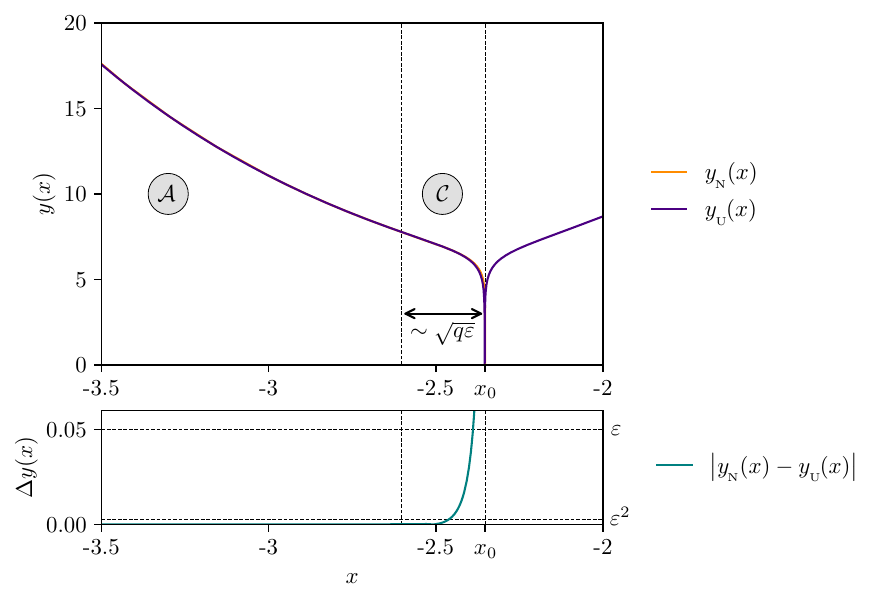}
	\caption{Comparison of the very supercritical uniform approximation with the numerical result for $\ve=0.05$. The residuals are consistent with the error estimate, except near the singularity.}
	\label{fig:vsupcritcomp}
\end{figure}

This completes the solution for the very supercritical regime. However, even without the previous analysis indicating the existence of the other regimes, we can see from this solution that it cannot be the end of the story, as the matching begins to break down as $q\to1$. This breakdown occurs when $c_{0} \sim \ve c_{1}$, or in terms of $q$ when
\begin{equation}
	q-1 \sim \ve \,\text{W}\left(\frac{1}{2\ve}\right)+\order{\ve^{2}} ~~\text{as}~~\ve\rightarrow 0^{+},
\end{equation}
where $\text{W}(x)$ is the Lambert function.\footnote{The Lambert function is defined implicitly as the principal branch solution of the functional equation $\text{W}(x) e^{\text{W}(x)}=x$.} 

This failure of the matching signals the \textit{formation of a new region}: Regions $\cA$ and $\cC$ can no longer be directly connected. This demonstrates the emergence of Region $\cB$ as the initial data $q$ is tuned --- the long-tailed region associated with evolution near the saddle point where $y'(x)\ll1$ and $y(x)\approx2(1-\ve)$. 

We now proceed to construct the solutions for the critical and near critical solutions, which incorporate this additional region.\label{page:VSupEnd}

\subsubsection{Critical and Near Critical Solutions}

\label{page:RegionBStart}To construct the solutions in Region $\cB$, we introduce the intermediate time variable $\xb\hspace{-0.1em}=\frac{x}{\sqrt{2\ve}}$. In this region it is more convenient to work with the second order differential equation.\footnote{This region is where the subtlety caused by the two small parameters needs to be addressed. Since $q$ is $\ve$-dependent, establishing a well-defined perturbation theory requires setting a hierarchy between $\delta_{q}$ and $\ve$. It turns out that setting 
	\begin{equation}
		q=q_{*}e^{\ve \Lambda}\,,
		\label{eqn:sclosecrit}
	\end{equation} 
with $\abs{\Lambda}< 1$ is a convenient way to do this within the first-order differential equation. Note that we are unable to be more specific about the value of $\Lambda$ since it still depends on $\ve$ in a non-trivial way. Intuitively, this amounts to tying the system to be close to criticality, where $\Lambda=0$ is the critical value, $\Lambda<0$ corresponds to the subcritical case and $\Lambda>0$ to the supercritical case.  Of course, once this is done, one can obtain the same final result using the first-order differential equation.} 

Substituting the intermediate time variable and the standard asymptotic series ansatz into the second-order differential equation gives
\begin{equation}
	\begin{aligned}
		&\ve^{0}: \, (y_{0}')^2=0\\
		&\ve^{1}: \,2 y_{0} y_{0}''+2 y_{0}' y_{1}'-2 (y_{0}')^2-y_{0}^2+2 y_{0}=0.
	\end{aligned}
\end{equation}
From these two equations, we conclude that $y_{0}=2$.

Proceeding to the next two orders in $\ve$ gives
\begin{equation}
	\begin{aligned}
		&\ve^{2}: \,  4y_{1}''+(y_{1}')^2-4 y_{1}-8=0\\
		&\ve^{3}: \,  2y_{2}'' +y_{1}'y_{2}'-2y_{2}=(y_{1}')^{2}+y_{1}^{2}-2y_{1}-y_{1}y_{1}''\,.
	\end{aligned}
\end{equation}
The equation for $y_{1}$ at $\order{\ve^{2}}$ remains nonlinear, indicating that the essential nonlinearity of this region has been isolated by expanding in the number of dimensions. 

We simplify the equation for $y_{1}$ by introducing the substitution  $y_{1}(\xb)= f(\xb)-2$ to obtain
\begin{equation}
	f''+\frac{1}{4}(f')^{2}-f=0\,.
	\label{eqn:feq}
\end{equation}
Unfortunately there is no known general solution to this differential equation in terms of elementary functions. Our inability to solve this equation analytically prevents us from constructing uniform solutions for the critical and near-critical regimes. Nevertheless, approximate analytic solutions with acceptable errors can still be obtained within this framework. We emphasise that this is not a failure of the \largeD expansion itself to give us a separation of scales --- if a solution to this equation were available, the construction of these uniform solutions would be straightforward, as we now demonstrate.

Integrating once gives us the equivalent first-order differential equation to \eqref{eqn:1y} 
\begin{equation}
	(f')^{2}=8e^{\Lambda}e^{-\frac{f}{2}}+4f-8,
\end{equation}
where $\Lambda$ is the integration constant corresponding to q as identified in \eqref{eqn:sclosecrit}. This admits the integral representation of a solution,
\begin{equation}
	\xb\hspace{-0.1em}= \pm\int\frac{\diff{f}}{f'},
	\label{eqn:fintrep}
\end{equation}
where the positive and negative signs correspond to dispersion and collapse, respectively.

To match into Region $\cA$, we take the negative branch and the limit $f\rightarrow \infty$. In this limit the integral simplifies to 
\begin{equation}
	\begin{aligned}
		\lim_{f\rightarrow+\infty} \left(\xb\hspace{-0.1em}+C_{2}\right) &\rightarrow -\int\frac{\diff{f}}{\sqrt{4f-8}}\\
		&\rightarrow \sqrt{f-2},
	\end{aligned}
\end{equation}
and hence 
\begin{equation}
	\lim_{\xb\to-\infty}\yb(\xb) \xrightarrow{\sim} 2+\ve\xb^{2} +\order{\ve^{2}}~~\text{as}~~ \ve \rightarrow 0^{+}.
\end{equation}
This precisely matches how $\ya$ scales towards this intermediate region
\begin{equation}
	\begin{aligned}
		\ya(\xb) &= 1+\cosh\left(\sqrt{2\ve}\xb\right)\\
		& \sim 2 + \ve\xb^{2} +\order{\ve^{2}}.
	\end{aligned}
\end{equation}
Thus, the full solution to this differential equation would smoothly connect to Region $\cA$. An analogous argument holds for matching into Region $\cA^{-}$ by selecting the other branch. 

Finally, we verify the matching into Region $\cC$, where  $f\rightarrow -\infty$ and $\Lambda>0$. In this limit the integral becomes
\begin{equation}
	\begin{aligned}
		\lim_{f\to -\infty} \left(\xb+C_{3}\right) &\rightarrow -\int\frac{\diff{f}}{\sqrt{8e^{\Lambda}e^{-\frac{f}{2}}}}\\
		&\rightarrow \sqrt{2}e^{\frac{1}{4}(f-2\Lambda)}\\
		\lim_{f\rightarrow-\infty} x+\sqrt{2\ve}C_{3} &\rightarrow 2\sqrt{\ve}e^{\frac{1}{4}(f-2\Lambda)}.
	\end{aligned}
	\label{eqn:justsupcritright1}
\end{equation}
If we take the near singularity solution, as given by \eqref{eqn:nearsing}, and solve for $x$, we find
\begin{equation}
	x+x_{0} \sim 2\ve\left(\frac{y}{2q}\right)^{\frac{1}{2\ve}} ~~\text{as}~~ y \rightarrow 0^{+}.
\end{equation}
We now rewrite this in terms of the intermediate variable $f=\frac{y-2(1-\ve)}{\ve}$ and set $q$ to be close to critical \eqref{eqn:sclosecrit} 
\begin{equation}
	\begin{aligned}
		x+x_{0} &\sim 2\sqrt{\ve} e^{\frac{\Lambda}{2}}  \left(\frac{(f-2) \ve +2}{2 (1-\ve )^{1-\ve } }\right)^{\frac{1}{2 \ve }}\\
		&\sim 2\sqrt{\ve}e^{\frac{1}{4}(f-2\Lambda)}+\order{\ve^{\frac{3}{2}}}~~\text{as}~~ \ve \rightarrow 0^{+}.
	\end{aligned}
\end{equation}
This matches our approximation of the integral \eqref{eqn:justsupcritright1} in this region \textit{exactly}. 

Notably, the solution in Region $\cB^{+}$ not only matches smoothly into Region $\cC$, but also appears to capture the entire behaviour of Region $\cC$ --- as could be expected by the fact that they have the same characteristic timescale. However, due to our reliance on a second approximation in order to solve the differential equation \eqref{eqn:feq} (essentially amounting to expanding about the saddle), we treat Region $\cC$ as a distinct region for the marginally supercritical solution. In practise, this ensures that Region $\cB$ is exclusively linked to the evolution near the saddle point.

Having established that \eqref{eqn:feq} is the correct equation for this region, we now seek approximate solutions. The first step is to systematically expand about the saddle point by treating the non-linearity as small. This means rewriting the equation as
\begin{equation}
    f''+\lambda(f')^{2}-f=0,
\end{equation}
where we treat $\lambda=\frac{1}{4}$ as a small parameter. Substituting a standard asymptotic series for $f$ as $\lambda\to0^{+}$ and solving at each order, we obtain a sum of positive and negative exponentials,
\begin{equation}
    \begin{aligned}
		f &\sim A_{0}e^{-\xbe}+B_{0}e^{\xbe}+\lambda\left(A_{1}e^{-\xbe}+B_{1}e^{\xbe}-\frac{A_{0}^{2}}{3} e^{-2\xbe}-\frac{B_{0}^{2}}{3} e^{2\xbe}-2A_{0}B_{0}\right) +\order{\lambda^{2}}.
    \end{aligned}
\end{equation} 

To estimate the size of Region $\cB$, we compare the largest terms in the $\order{\lambda}$ the $\order{\lambda^{0}}$ solutions. Since this is an intermediate solution, it must break down on both sides. For $\xb>0$, the breakdown occurs when
\begin{equation}
    \begin{aligned}
	B_{0}e^{\xbe} &\sim \frac{\lambda}{3}B_{0}^{2}e^{2\xbe}\\
	x &\sim \sqrt{2\ve}\log\left(\frac{3}{\lambda B_{0}}\right).
    \end{aligned}
\end{equation}
Similarly for $\xb<0$, the breakdown occurs when
\begin{equation}
    \begin{aligned}
	A_{0}e^{-\xbe} &\sim  \frac{\lambda}{3}A_{0}^{2}e^{-2\xbe}\\
	x &\sim -\sqrt{2\ve}\log\left(\frac{3}{\lambda A_{0}}\right).
    \end{aligned}
\end{equation}
The position of these breakdown points depend on the matching constants $A_{0}$ and $B_{0}$. Although we have not determined them precisely, we can estimate their order of magnitude. 

To match with Region $\cA$, we require $A_{0}\sim \order{1}$. Furthermore, since the positive exponential coefficients drive the system away from the saddle point, they must be related to the distance from criticality. If we assume that no $\ve$ dependence has been extracted from $\delta_{q}$, then $B_{0}\sim \frac{\delta_{q}}{\ve}\ll1$. This explains why the hierarchy set by \eqref{eqn:sclosecrit}, where $\delta_{q}\lesssim\ve$, is necessary to observe this region. The resulting range of validity for this intermediate solution is therefore
\begin{equation}
    -\sqrt{2\ve}\lesssim x\lesssim\sqrt{2\ve}\log\left(\frac{\ve}{\abs{\delta_{q}}}\right).
\end{equation}

This analysis of the linearised solution suggests that we can construct good approximate solutions to the differential equation in the limit that $\delta_{q}\ll1$.

We begin with the {\it critical solution}, which will serve as the foundation for the near-critical solutions. Unlike the very supercritical solution discussed previously, where the unfixed boundary condition is not known \textit{exactly}, the critical solution has a well-defined boundary condition since the evolution tends to the fixed point at $y=2(1-\ve)$ as $x\rightarrow +\infty$. This translates to the following boundary condition in $f$
\begin{equation}
	\begin{aligned}
		\lim_{\xb\rightarrow\infty}f\rightarrow0.
	\end{aligned}
	\label{eqn:boundcrit}
\end{equation}
As a consequence, the series of constants $B_{i}$ in the linearised solution \eqref{eqn:linearisation} must vanish. 

Using this, we propose a series ansatz consisting of only negative exponentials
\begin{equation}
	f(\xb) = \sum_{n=1}^{\infty} \alpha_{n}e^{-n(\xbe+b_{\scriptscriptstyle{0}})}\,,
	\label{eqn:critansatz}
\end{equation}
where the constant $b_{0}$ arises from the shift symmetry of the equation and has been separated from the numerical coefficients $\alpha_{n}$ such that we can set $\alpha_{1} = 1$ without loss of generality. 

Substituting this ansatz into the differential equation generates a recurrence relation for these coefficients
\begin{equation}
	\alpha_{n} =\begin{cases}
		-\frac{1}{4(n^{2}-1)}\left(\left(\frac{n}{2}\right)^{2}\alpha_{\frac{n}{2}}^{2}+\sum_{i=1}^{\frac{n}{2}-1}2i(n-i)\alpha_{i}\alpha_{n-i}\right) & \forall n\, \text{even}\\
		-\frac{1}{4(n^{2}-1)}\left(\sum_{i=1}^{\frac{n-1}{2}}2i(n-i)\alpha_{i}\alpha_{n-i}\right) & \forall n\, \text{odd}.
	\end{cases}
	\label{eqn:alphares}
\end{equation}
This defines an exact solution to this differential equation that satisfies the boundary condition of criticality. Moreover, the coefficients $\alpha_{n}$ decay rapidly as $n$ grows, so only a few terms are required for high numerical accuracy (depending on the size of $\ve$). 

The full solution in the original variables is
\begin{equation}
    \begin{aligned}
	\yb(x)&\sim 2 + \ve\left(-2+\sum_{n=1}^{\infty}\alpha_{n}e^{-\frac{n}{\sqrt{2\ve}}(x+\sqrt{2\ve}b_{\scriptscriptstyle{0}})}\right) +\order{\ve^{2}} \\
	\alpha_{1}&=1\,.
    \end{aligned}
    \label{eqn:critsol}
\end{equation}
Including higher-order exponential terms reveals that the solution rapidly diverges at $x=-b_{0}\sqrt{2\ve}$, setting the domain of validity of this approximate solution as
\begin{equation}
    -\sqrt{2\ve}\lesssim x\leq+\infty.
\end{equation}
This is an infinitely long tail in self-similar time created by fine-tuning, despite the short characteristic time anticipated by the large number of dimensions. \label{page:RegionBEnd}

\label{page:CritStart}To complete the construction of the critical solution, we must match this solution to Region $\cA$. Unfortunately, the domain of validity of this approximate solution does not overlap with Region $\cA$, meaning that we cannot use conventional methods that rely on an overlapping region. Instead we look to use the integral representation. We present the key idea and the final results below, leaving the detailed derivation for Appendix \ref{app:MatchingMet}.

By splitting the solution into regions, we have introduced matching coefficients that account for the ignorance that one region has about the boundary conditions imposed in another. However, the integral representation propagates boundary information continuously throughout the entire solution, without being restricted to the definitions of the regions. 

Explicitly, writing out \eqref{eqn:intrep} as a definite integral between a point in Region $\cA$, denoted by `$\xacap$', and a point in Region $\cB$, denoted by `$\xbcap$', gives
\begin{equation}
    \begin{aligned}
	\xbcap-\xacap &=-\int_{y(\xacape)}^{y(\xbcape)}\frac{1}{\sqrt{z^{2}-2z+z^{2}\ve\left(\frac{2(1-\ve)^{1-\ve}}{z}\right)^{\frac{1}{\ve}}}}\diff{z}\,.
    \end{aligned}
    \label{eqn:defint}
\end{equation}
The quantity computed by this integral is the relative shift in $x$ between the solutions in each region --- precisely the quantity that corresponds to the matching coefficient $b_{0}$. Consequently, we seek an asymptotic solution to Region $\cB$ through the integral representation to fix this undetermined matching coefficient.

Carrying out this procedure, we find that
\begin{equation}
    \ybcap \sim 2(1-\ve) +\ve e^{-\frac{1}{\sqrt{2\ve}}\left(\xbcape+\sqrt{2\ve}b_{0}\right)}~~ \text{as}\,,~~ x_{b}\rightarrow \infty\,,
\end{equation}
where the asymptotic, but divergent sum for $b_{0}$ is given by
\begin{equation}
    \begin{aligned}
	b_{0} &=\frac{1}{\sqrt{2}}\bigg(2 -\sum_{n=1}^{\infty}\binom{-\frac{1}{2}}{n}e^{-n}  E_{n+\frac{1}{2}}(n)+\sqrt{e}\sum_{n=0}^{\infty}\frac{1+(-1)^{n}}{2^{n}(1+n)n!}\gamma_{n}\\
	&~~~~~~~~~~~~~~+\sqrt{2}\sum_{n=0}^{\infty} (-1)^{n}\frac{1+(-1)^{n}}{1+n} \bigg)-2\log(2)\\
	\gamma_{n}&=-1+\sum_{i=1}^{n}(-1)^{i+1}\left(\frac{(2i+1)^{n-i-1}(n+1-i)\binom{n}{i-1}(2i+1)!!}{i}\right)e^{i}\,,
    \end{aligned}
\end{equation}
where $E_{n}(x)$ is the generalised exponential integral.
This can be optimally truncated to extract the matching coefficient as
\begin{equation}
    \begin{aligned}
	b_{0} &\approx -0.281221\,.
    \end{aligned}
\end{equation}

We consequently have the critical solution, matched up to $\order{\ve}$. This matching procedure can be straightforwardly extended to compute higher-order corrections in $\ve$, which will also take the form of exponentials. 

When working with the approximation \eqref{eqn:critsol} numerically, the order of truncation in the exponential series is important to maintain accuracy. As $\ve$ decreases, the number of terms retained in the sum must increase to ensure that the errors arising from the neglected higher-power exponentials remains smaller than the $\ve^{2}$ correction. 

\begin{figure}[H]
    \centering
    \includegraphics[width=0.85\textwidth]{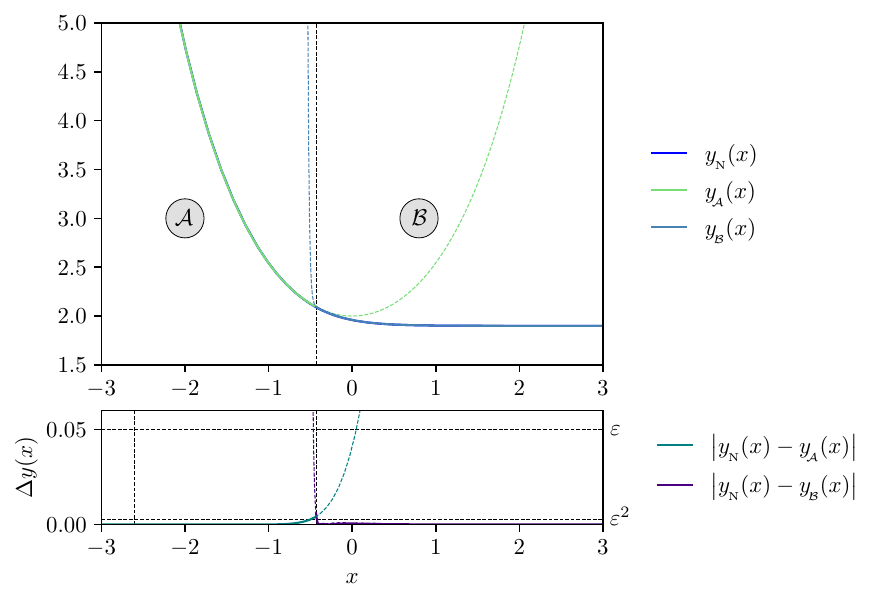}
    \caption{Comparison of the critical solution approximations with the numerical result for $\ve=0.05$. The residuals are consistent with the error estimate of $\ve^{2}$ over the entire domain.}
    \label{fig:critapprox}
\end{figure}

Figure \ref{fig:critapprox} demonstrates that this approximation agrees well with the numerical solution, confirming the validity of the approximation and matching.
\label{page:CritEnd}

From this solution, it is not particularly difficult to move slightly away from criticality. In principle, we could perform a double expansion in both $\ve$ and $\delta_{q}$, but there is a more efficient argument when sufficiently close to criticality ($\delta_{q}<\ve$) that leads to slightly better approximations.\label{page:MSubStart}

We begin with the marginally subcritical regime, as it represents the simplest modification of the critical solution. From the perspective of the first-order differential equation, the subcritical solution corresponds to the case where the derivative $y'(x)$ reaches zero and then changes sign. This implies that the solution should exhibit a symmetry about this turning point --- we can approximately construct the marginally subcritical solution by extending the critical solution up to the point where $y'(x)=0$, before reflecting it at that point.

This approximation can be made smooth by noting that it approximately corresponds to combining two critical solutions: one describing the approach to the minimum (involving negative exponentials) and the other describing its departure from it (involving positive exponentials). These positive exponentials must become large at the appropriate time to drive the solution from near the saddle point accordinat to how close to criticality the system is. This is achieved by introducing a matching constant that is modified by adding the log of the distance from criticality $\delta_{q}$ as predicted by the perturbative analysis of differential equation
\begin{equation}
    \ybm \sim 2(1-\ve) +\ve \sum_{n=1}^{\infty}\alpha_{n}\left(e^{-\frac{n}{\sqrt{2\ve}}(x+\sqrt{2\ve}b_{\scriptscriptstyle{0}})}+e^{\frac{n}{\sqrt{2\ve}}(x+\sqrt{2\ve}(b_{\scriptscriptstyle{0}}-d_{1}\log(\abs{\delta_{q}})))} \right)\,,
    \label{eqn:margsub}
\end{equation}
where $d_{1}=\order{1}$ is a constant that must be matched numerically. The shift of the constant by this value multiplies each of the positive exponentials by $\left(\delta_{q}\right)^{n}$, thereby decoupling them from the negative exponentials, provided that $\delta_{q}$ is sufficiently small, $\delta_{q}\ll\ve$. 

The final step is to smoothly glue this to Region $\cA^{-}$ with a function given by
\begin{equation}
    \yam(x) = 1+\cosh\left(x+\sqrt{2\ve}\log\left(\frac{\ve}{\abs{\delta_{q}}}\right)d_{2}\right),
\end{equation}
where $d_{2}=\order{1}$ is a constant that must be matched numerically again. 

\begin{figure}[h]
    \centering
    \includegraphics[width=0.85\textwidth]{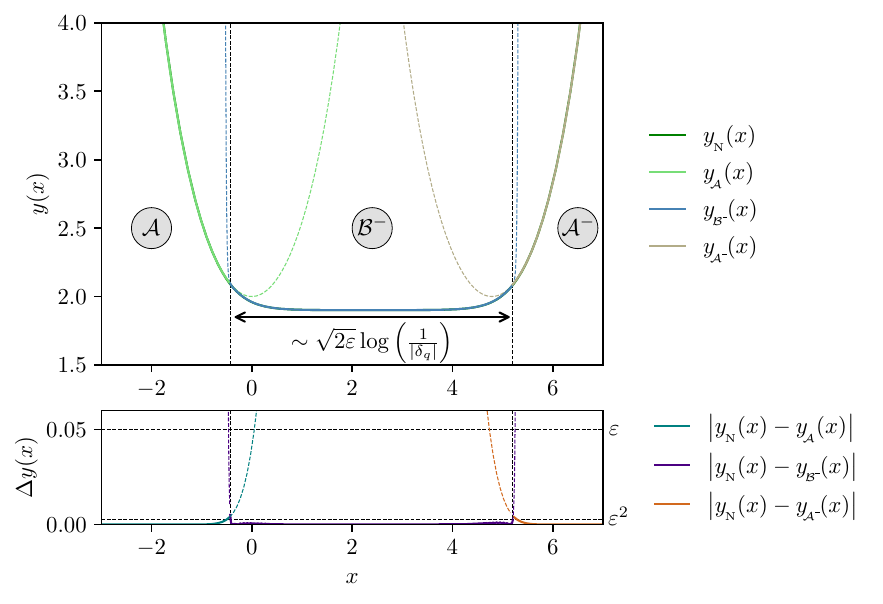}
    \caption{Comparison of the marginally subcritical solution approximations with the numerical result for $\ve=0.05$, with numerically matched coefficients. The residuals are consistent with the error estimate of $\ve^{2}$ over the entire domain.}
    \label{fig:subcritapprox}
\end{figure}

Figure \ref{fig:subcritapprox} shows that this solution is in excellent agreement with the numerical result.

\label{page:MSupStart}The final step in constructing our family of solutions to the metric function is to address the Marginally Supercritical Regime. Unlike with the subcritical case, we cannot use the symmetry argument that justified the ansatz \eqref{eqn:margsub}, since allowing $f$ to fall to negative values would trigger the runaway collapse found in \eqref{eqn:fintrep} due to the negative exponential term in the integral. 

\begin{figure}[H]\label{page:MSupEnd}\label{page:MSubEnd}
    \centering
    \includegraphics[width=0.85\textwidth]{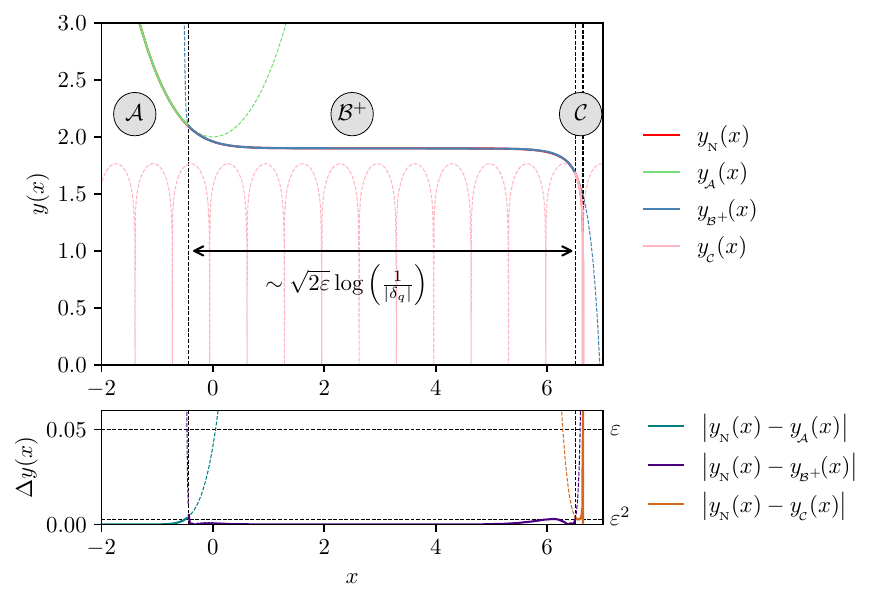}
    \caption{Comparison of the marginally supercritical solution approximations with the numerical result for $\ve=0.05$, with numerically matched coefficients. The residuals are consistent with the error estimate of $\ve^{2}$ over the entire domain.}
    \label{fig:supcritapprox}
\end{figure}

However, since we have said that we will treat this collapse region separately, we can adopt the same ansatz as in Region $\cB^{-}$ for the marginally subcritical case, but with a negative coefficient for the positive exponentials 
\begin{equation}
    \ybp \sim 2(1-\ve) +\ve \sum_{n=1}^{\infty}\alpha_{n}\left(e^{-\frac{n}{\sqrt{2\ve}}(x+\sqrt{2\ve}b_{\scriptscriptstyle{0}})}-e^{\frac{n}{\sqrt{2\ve}}(x+\sqrt{2\ve}(b_{\scriptscriptstyle{0}}-d_{3}\log(\abs{\delta_{q}})))} \right).
\end{equation}
We are now left with matching this solution to Region $\cC$, \eqref{eqn:yright}, which we do numerically. The full solution by regions is shown in Figure \ref{fig:supcritapprox}, which demonstrates that the approximation remains consistent with the expected error of $\order{\ve^{2}}$ across the entire domain.

This completes the construction of the family of solutions to the metric function. The analysis presented here is subtle due to complexity of the differential equation controlling Region $\cB$ and the dimension-dependent nature of the boundary conditions. Despite the fact that one of the resulting differential equations is only approximately solved, the advantage of working at small $\ve$ is clear: the solution naturally separates into distinct regions corresponding to particular balances of the forces resulting in simpler differential equations. As we will see in the following subsection, this remains true for the scalar field solution, as expected, given that it is tied to the metric through the gravitational constraint equations.

\subsection{The Scalar Field} \label{sec:3.3}
\begin{figure}[H]
    \centering
    \begin{subfigure}[b]{0.45\textwidth}
	\includegraphics[width=\textwidth]{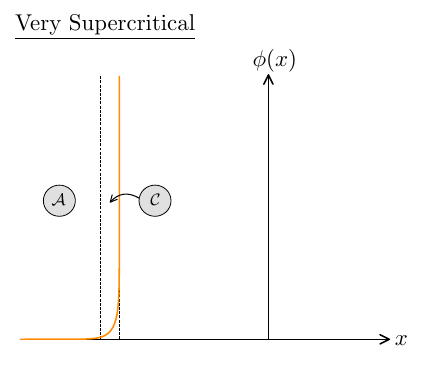}
    \end{subfigure}
    \begin{subfigure}[b]{0.45\textwidth}
	\includegraphics[width=\textwidth]{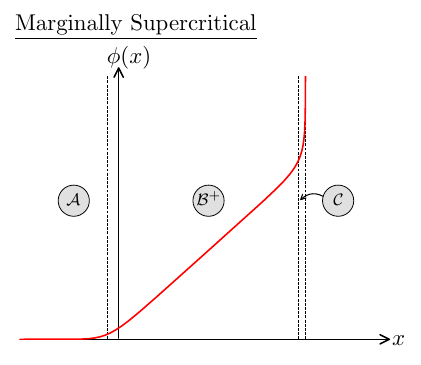}
    \end{subfigure}
\\
    \begin{subfigure}[b]{0.45\textwidth}
	\includegraphics[width=\textwidth]{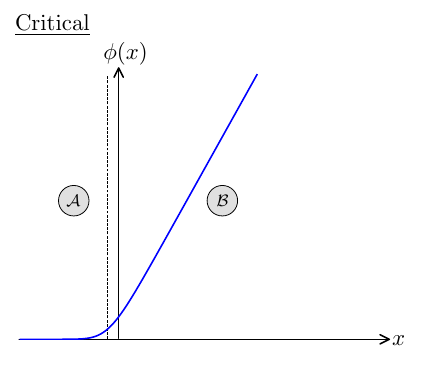}
    \end{subfigure}
    \begin{subfigure}[b]{0.45\textwidth}
	\includegraphics[width=\textwidth]{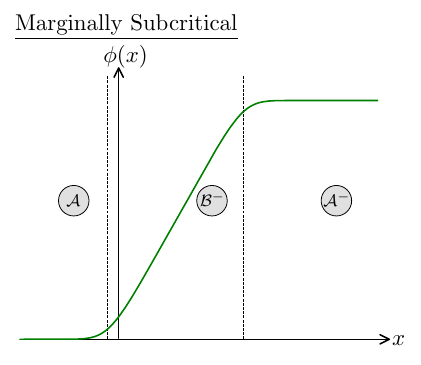}
    \end{subfigure}
    \caption{Numerical results for each qualitatively distinct solution of the scalar field, with the regions appropriately labelled.}
    \label{fig:NumericalDSc}
\end{figure}
Having completed the construction of the metric function in each of the different regimes, we now turn to the scalar field. If our objective is only to determine the gradient of the scalar field, the problem reduces to a straightforward calculation. It is simply given by a power of $y$, as can be seen after integrating the equation of motion for the scalar field. However, determining the full field profile $\phi$ requires a more detailed analysis, which we now present.
 
From Figure \ref{fig:NumericalDSc}, we observe that the structure of the regions used in the computation of the metric function are preserved in the scalar field solution. Thus, the methods employed to compute the metric will carry over directly to the scalar field with only minor modifications. However, there are two key subtleties that require careful attention:
\begin{itemize}
	\item The constraint equation must be applied carefully in Region $\cA$ to avoid it setting the solution there to be trivial.
	\item The metric function and scalar field are not on equal footing. To determine the scalar field at $\order{\ve^{n}}$, the metric function must be computed to $\order{\ve^{n+1}}$.
\end{itemize}
We now address these points explicitly. 

The scalar field is governed by its equation of motion and the constraint equation 
\begin{subequations}\label{eqn:scalareqns}
	\begin{align}
		2\ve y\phi''+y'\phi' &=0, \label{eqn:Iy2} \\
		4\ve^{2}y^{2}(\phi')^{2}-(1-\ve)(y')^{2}+(1-\ve)y(y-2) &=0.\label{eqn:IIy3}
	\end{align}
\end{subequations}

The flat space boundary condition, translated into the self-similar coordinate $x$, gives
\begin{equation}
    \lim_{x\rightarrow-\infty}\phi\rightarrow 0\,.
\end{equation}

The equation of motion possesses both a \textit{scaling symmetry} and a \textit{shift symmetry} in the scalar field $\phi$. In contrast, the constraint equation only has the \textit{shift symmetry} in $\phi$, implying that the constraint equation fixes the amplitude of the scalar field. As previously shown in \eqref{eqn:scalaramp}, this amplitude determines whether the system will undergo collapse or dispersion. Thus, the constraint equation ties the amplitude of the scalar field to the constant $q$ from the previous sections.

We now compute this amplitude in terms of $q$ explicitly. Integrating the equation of motion \eqref{eqn:Iy2} once yields
\begin{equation}
	\phi'(x) = \frac{\hat{A}(\ve)}{y^{\frac{1}{2\ve}}},
	\label{eqn:phiampexpl}
\end{equation}
where $A(\ve)$ is the dimension-dependent amplitude of the scalar field.

Substituting this into the constraint equation \eqref{eqn:IIy3} leads to a first-order differential equation for the metric function
\begin{equation}
	(y')^{2}=y^{2}-2y+\frac{4\hat{A}^{2}\ve^{2}}{1-\ve}y^{2-\frac{1}{\ve}}\,.
\end{equation}
By comparing this to the first-order differential equation for $y(x)$ derived in Section \ref{sec:2.2}, we directly match the amplitude $\hat{A}(\ve)$ to the constant $q$ as follows
\begin{equation}
	\hat{A}(\ve)=\frac{\sqrt{1-\ve}}{2\ve}(2q)^{\frac{1}{2\ve}}.
	\label{eqn:scalarampd}
\end{equation}

Secondly, we observe that the scalar field and metric function are not treated on an equal footing when it comes to the large dimension expansion. If we take the constraint equation \eqref{eqn:IIy3} and rewrite it such that all of terms of $y$ are quadratic we obtain
\begin{equation}
    2 y(x) y''(x)-y'(x)^2+y(x)^2 \left(4 \epsilon  \phi '(x)^2-1\right)=0.
    \label{eqn:constraintord}
\end{equation} 
Substituting the standard asymptotic ansatzae \eqref{eqn:stdans}, we find the following hierarchy of equations
\begin{equation}
\begin{aligned}
	&\ve^{0}:~ 2 y_{0} y_{0}''-y_{0}'^2-y_{0}^2=0\\
	&\ve^{1}:~ y_{1} y_{0}''-y_{0}' y_{1}'+y_{0} \left(y_{1}''-y_{1}\right)+2 y_{0}^2 \phi_{0}'(x)^2=0.
\end{aligned}
\end{equation}
At leading order in $\ve$, the scalar field completely decouples from the gravitational sector. This implies that in the strict $\ve=0$ limit, the spacetime becomes insensitive to the presence of the scalar field --- the range of the gravitational interaction is completely suppressed in infinite dimensions, reducing to a contact force that has infinite strength. This decoupling provides a physical interpretation for the singular nature of the differential equation in the infinite-dimensional limit: the system never actually comes into `contact'. The only solution to the system is flat space over the whole domain.

Therefore, to construct physically meaningful solutions where the scalar field is evaluated at $\order{1}$, it is necessary to compute the metric function to $\order{\ve}$. In fact, determining the scalar field term $\phi_{n}$ always requires the metric field term $y_{n+1}$ to be computed first.

We now compute the solution in Region $\cA$, which we previously identified as corresponding to flat space. Taking the solution for the metric function in this region \eqref{eqn:flaty}, and substituting it into the equation of motion for the scalar field, we find
\begin{equation}
	2\ve \phia''+\tanh\left(\frac{x}{2}\right)\phia'=0\,.
\end{equation}
Introducing the change of variables $z=\cosh\left(\frac{x}{2}\right)$, this equation reduces to a hypergeometric differential equation. The exact solution in $\ve$ solution to this region (up to exponentially small corrections from the metric function) is given by
\begin{equation}
	\phia(x) = \hat{B}(\ve)+\frac{\hat{A}(\ve)}{2^{\frac{1-2\ve}{2\ve}}} \sinh \left(\frac{x}{2}\right) \,_2 F_1 {\small\left.\left[ \begin{matrix} \ \frac{1}{2}, & \frac{1+\ve}{2\ve} \\ \multicolumn{2}{c}{\frac{3}{2}} \end{matrix}\ \right| -\sinh ^2\left(\frac{x}{2}\right) \right]}.
	\label{eqn:scalarina}
\end{equation}
The corresponding solution in Region $\cA^{-}$ is simply a translation of this result in $x$.

Direct substitution of this solution into the constraint equation gives
\begin{equation}
	\hat{A}^2 2^{1/\ve } \ve ^2 \left(\sinh (x) \text{csch}\left(\frac{x}{2}\right)\right)^{4-\frac{2}{\ve }}=0,
\end{equation}
which should set $\hat{A}(\ve)=0$. This reflects the fact that the constraint equation treats the flat space metric function as exact. However, since we know that Region $\cA$ is not \textit{truly} flat space and that asymptotic expansions are never strictly zero, this must be interpreted as an artefact of the expansion.  Consequently, we instead resolve this issue by using the constraint \eqref{eqn:scalarampd} to determine $\hat{A}(\ve)$. The plots below confirm that this procedure produces the correct solution within the expected error bounds.

Finally, we fix the constant associated with the shift symmetry of the scalar field. Taking the limit $x\rightarrow -\infty$ of the solution allows us to find
\begin{equation}
	\hat{B}(\ve)= \frac{\sqrt{\pi } \hat{A}(\ve)  \Gamma \left(\frac{1}{2 \ve }\right)}{2^{\frac{1}{2 \ve }}\Gamma \left(\frac{1+\ve}{2 \ve }\right)}\,.
\end{equation}
We can now compute the solutions in the other regions using the same methods as for the metric function. For brevity, we omit the details since the procedure is identical.

\subsubsection{The Very Super Critical Solution}

We now compute the solution in Region $\cC$ by taking the solution to the metric function in this region \eqref{eqn:yright} and substituting it into the constraint equation \eqref{eqn:IIy3}. After substituting in the asymptotic ansatz and simplifying various factors, we find
\begin{equation}
    2 \phi_{0} '(\xc)\sim\frac{2^{\frac{1}{2-2 \ve }} \sqrt{1-\ve } q^{\frac{\ve }{2 (\ve -1)}}}{\sqrt{\cosh (\xc)-1}} +\order{\ve}  \,,
\end{equation}
where $\xc=\sqrt{\frac{q-1}{q}}\frac{x+c}{\ve}$, and the $\order{\ve}$ terms require the higher-order corrections to the metric function. 

We can solve this equation exactly while keeping $q$ as a free parameter to find
\begin{equation}
    \phi_{0}(\xc)=-\frac{\sqrt{2} \sinh \left(\frac{\xce}{2}\right) \log \left(\tanh \left(-\frac{\xce}{4}\right)\right)}{\sqrt{\cosh (\xc)-1}}+C_{4} \,.
\end{equation}
We now look to match this solution to Region $\cA$. To do this, we expand this solution in Region $\cC$ in the outer variable $x$, or equivalently in the limit that $\xc\rightarrow -\infty$. Expanding in $\ve$, we find
\begin{equation}
    \lim_{\xc\to-\infty}\phic(\xc) \xrightarrow{\sim} C_{4}+2 e^{\xce/2}+\order{\ve} ~~\text{as}~~\ve\rightarrow0^{+}.
\end{equation}
Next, we take the solution in Region $\cA$ and expand it in terms of the inner variable $\xc$. This is more subtle due to the hypergeometric function, but the result is that 
\begin{equation}
    \phia(\xc)\sim 2 e^{\xce/2}+\order{\ve} ~~\text{as}~~\ve\rightarrow0^{+}.
\end{equation}
From these expansions, we can straightforwardly read off that $C_{4}=0$, giving the final solution in Region $\cC$ as
\begin{equation}
    \phic(\xc)\sim-\frac{\sqrt{2} \sinh \left(\frac{\xce}{2}\right) \log \left(\tanh \left(-\frac{\xce}{4}\right)\right)}{\sqrt{\cosh (\xc)-1}} +\order{\ve} ~~\text{as}~~\ve\rightarrow0^{+}\,.
\end{equation}

\begin{figure}[H]
    \centering
    \includegraphics[width=0.85\textwidth]{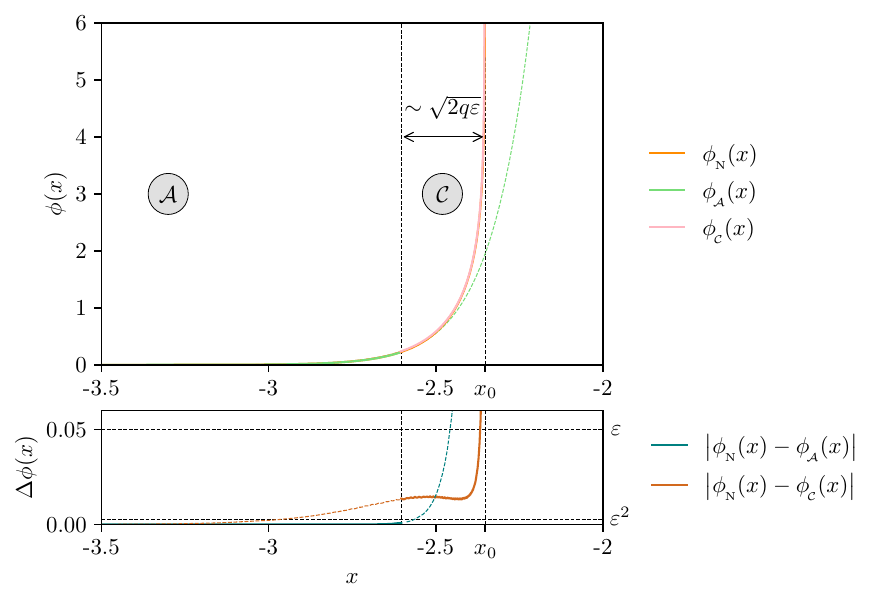}
    \caption{Comparison of the very supercritical scalar field solution approximations with the numerical result for $\ve=0.05$. The residuals are consistent with the error estimate, except near the singularity.}
    \label{fig:vsupcritapproxsc}
\end{figure}

As shown in Figure \ref{fig:vsupcritapproxsc}, this solution suffers diverging errors near the singularity for the same reason as with the metric function, but the solutions remain within the expected error bounds across the rest of the domain.

We can also compute a uniform solution, given by 
\begin{equation}
	\phiu(x) \sim \phia(x)+\phic(x)-e^{\sqrt{\frac{q-1}{q}}\frac{x+b}{2\ve}},
\end{equation}
which we plot in Figure \ref{fig:vsupcritcompsc} below.

\begin{figure}[H]
    \centering
    \includegraphics[width=0.85\textwidth]{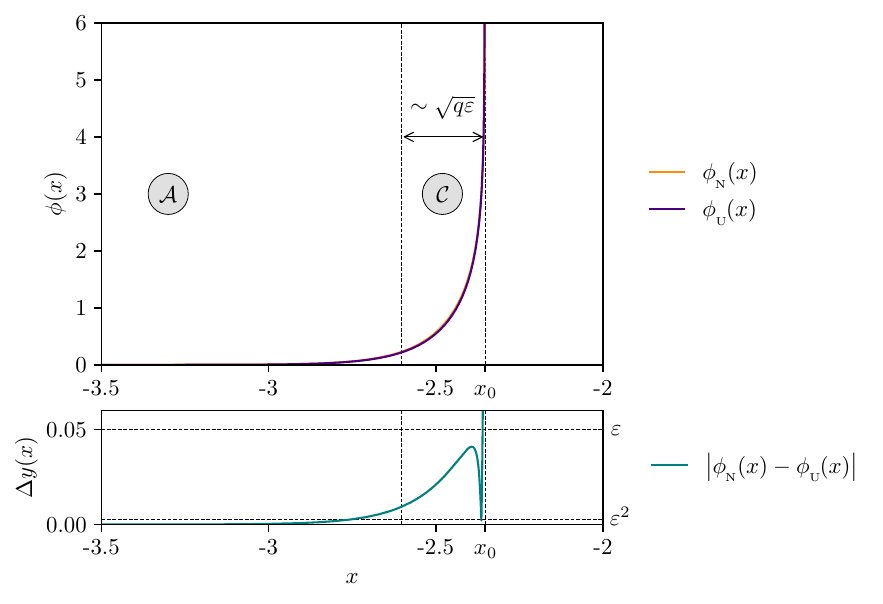}
    \caption{Comparison of the very supercritical uniform approximation for the scalar field with the numerical result for $\ve=0.05$. The residuals are consistent with the error estimate, except near the singularity.}
    \label{fig:vsupcritcompsc}
\end{figure}

\subsubsection{The Critical Solution}

To find the scalar field solution in Region $\cB$, we use the constraint equation in the form of \eqref{eqn:constraintord} as it turns out to be more convenient to solve for $\phi$ recursively using this equation.

We first rewrite the differential equation into a more manageable form. From the previous section, we have that the metric at criticality can be written in the following way
\begin{equation}
    \yb(\xbbar)\sim 2(1-\ve) + \ve f(\xbbar)+\order{\ve^{2}} ~~\text{as}~~\ve\rightarrow 0^{+}\,,
\end{equation}
where $\xbbar=\frac{x+\sqrt{2\ve}b_{0}}{\sqrt{2\ve}}$ and $f(\xbbar)$ is the sum of exponentials previously obtained for the metric function approximation
\begin{equation}
    f(\xbbar) = \sum_{n=1}^{\infty}\alpha_{n}e^{-n \xbebar}\,.
\end{equation}

\begin{figure}[H]
\centering
\includegraphics[width=0.85\textwidth]{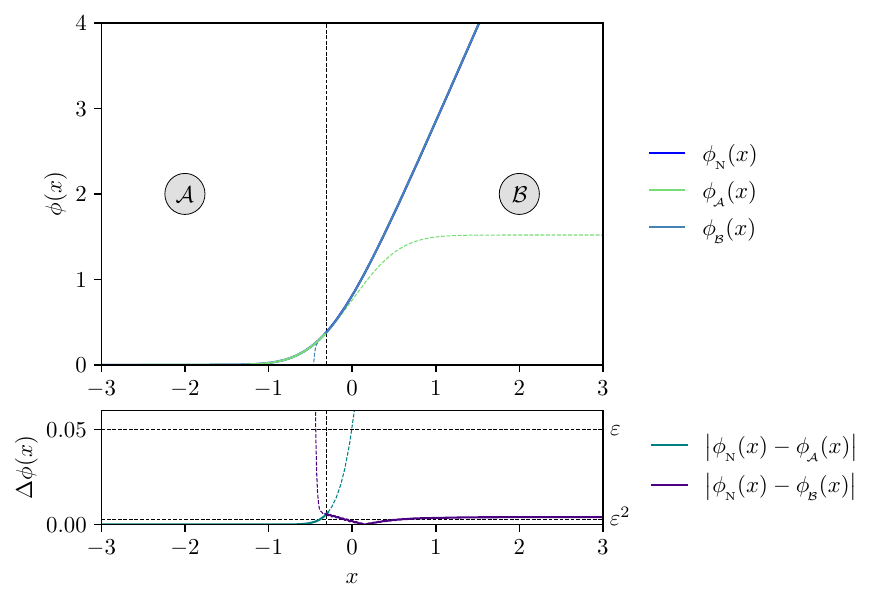}
\caption{Comparison of the critical solution approximations for the scalar field with the numerical result for $\ve=0.05$. The residuals are consistent with the error estimate over the entire domain.}
\label{fig:critapproxsc}
\end{figure}

Substituting the asymptotic ansatz for $\phi$ into \eqref{eqn:constraintord}, we find at leading order that the equation reduces to
\begin{equation}
    \begin{aligned}
	f''+4 (\phi_{0}')^2-2=0\,.
    \end{aligned}
\end{equation}
A suitable ansatz for solving this equation is
\begin{equation}
\begin{aligned}
	\phi_{0}(\xbbar) &= a_{0}+\frac{1}{\sqrt{2}}\left(\xbbar+\sum_{n=1}^{\infty}\beta_{n}e^{-n \xbbar}\right),
	\label{eqn:scalarcritans}
\end{aligned}
\end{equation}
where the linear factor is included since
\begin{equation}
    \lim_{\xbbar\rightarrow+\infty}\left(f''+4 (\phi_{0}')^2-2=0\right)\rightarrow 2(\phi_{0}')^{2}-1=0.
\end{equation} 
The coefficients $\beta_{n}$ are determined by recursion relations analogous to those found for the metric function coefficients $\alpha_{n}$
\begin{equation}
	\beta_{n} =\begin{cases}
		\frac{1}{n}\sum _{i=1}^{\frac{n}{2}-1} i (n-i) \beta_{i} \beta_{n-i}+\frac{n c_n}{4}+\frac{1}{8} n \beta_{\frac{n}{2}}^2 & \forall n\, \text{even}\\
		\frac{1}{n}\sum _{i=1}^{\frac{n-1}{2}} i (n-i) \beta_{i}\beta_{n-i}+\frac{n c_n}{4} & \forall n\, \text{odd}.
	\end{cases}
	\label{eqn:betacoef}
\end{equation}
Therefore, we once again obtain a formally resummed solution to Region $\cB$, where the only undetermined coefficient is the matching coefficient $a_{0}$. This can be fixed by matching with the left solution using the same procedure used for the metric function --- tracking the constant through an integral representation. We demonstrate the computation explicitly in Appendix \ref{app:MatchingSca}, but the result of this is simply that
\begin{equation}
	a_{0} \approx 0.801973\,.
\end{equation}
Figure \ref{fig:critapproxsc} shows that this solution matches well with the numerical result once again.

Finally, using the same arguments as before, we can construct the solutions for the marginally subcritical and supercritical solutions without significant difficulty. As the analysis is essentially identical to the metric function case, we omit it.
\newpage
\section{Perturbations}\label{sec:4}
In this section, we take linear perturbations of the CSS critical solution derived in Section \ref{sec:3} to compute the mass scaling exponent in a general number of dimensions for this system. Unlike the DSS case, these perturbations are not uniquely defined because the solution can be perturbed to form a black hole in a number of different ways. We separate these perturbation into two distinct categories, based on whether or not CSS is broken. This leads, in turn, to two distinct possible critical exponents, depending on which category the perturbation falls under:
\begin{itemize}
	\item \textbf{CSS-Preserving Perturbations:} A small perturbation of the initial data, $\delta_{q}$, that remains within the CSS manifold leads to the marginally supercritical solution discussed in the previous section. The resulting scaling law for the mass of the black hole can be expressed as
	\begin{equation}
		M \sim (q-q_{*})^{\gamma_{q}}\,.
	\end{equation}
	The critical CSS manifold is codimenion 1 in the restricted phases space of the CSS manifold, meaning that this critical exponent is universal for all initial data that is \textit{within} it. After performing this perturbation, we find that
	\begin{equation}
		\gamma_{q}= \frac{D-3}{\sqrt{2(D-2)}}.
	\end{equation}
	\item \textbf{CSS-Breaking Perturbations:} A small perturbation to the initial data, $\delta_{p}$, that breaks the continuous self-similarity leads to the formation of a non-self-similar black hole. If the imposition of CSS is viewed as having fixed some initial data to a particular value $p=p_{*}$, then the mass scaling law for a perturbation away from this value is given by
	\begin{equation}
		M \sim (p-p_{*})^{\gamma_{p}}\,.
		\label{eqn:massscalp}
	\end{equation}
	The critical CSS manifold is not codimenion 1 in the full phase space, implying that the perturbations that break the symmetry are not unique. There are infinitely many directions in which one can perturb, resulting in a continuous spectrum of allowed Lyapunov exponents. Consequently this critical exponent is not universal. After performing the perturbations, we find that \footnote{Where we have taken the largest real part from the spectrum of Lyapunov exponents.}
	\begin{equation}
		\gamma_{p}= \frac{2(D-3)}{D-2}.
	\end{equation}
\end{itemize}

\begin{figure}[H]
	\centering
	\includegraphics[width=0.7\textwidth]{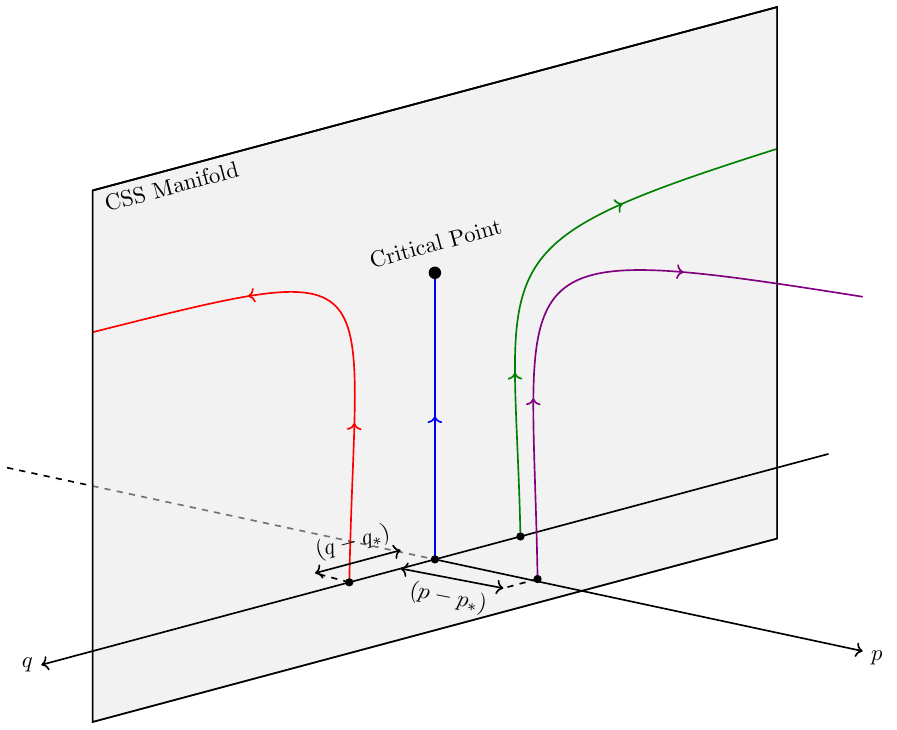}
	\caption{Cartoon of the RG flow for the CSS-preserving and CSS-breaking perturbations. The flows indicated in red, blue and green are identified as in Figure \ref{fig:phaseD} and the purple flow is the CSS-breaking perturbation. Note that the CSS manifold does \textit{not} have codimension 1 in the full phase space.}
	\label{fig:RGflow}
\end{figure}

As a point of reference, we present a cartoon of the phase space associated with this problem in Figure \ref{fig:RGflow}.

In this section we will first establish the connection between the Lyapunov exponent, associated with linear perturbations of the self-similar solution, and these critical exponents through a Renormalisation Group flow. We will then briefly discuss the critical exponent associated with preserving CSS, before turning our attention to the case where it is broken. 

Due to the complexity of the equations when we break the continuous self-similarity, we will begin by reviewing the exact analytics found in $D=4$ by Frolov \cite{Frolov:1997uu}, to see that the early and late time asymptotics are sufficient to completely reproduce these results. We will then use this insight to compute the Lyapunov exponent in an arbitrary number of dimensions for breaking CSS. 

Finally, we will discuss the spectrum of Lyapunov exponents that arises from this calculation and its implications for the structure of the full phase space, including the relative positions of the CSS and DSS manifolds.

\subsection{Renormalisation Group Flow} \label{sec:4.1}

As with many critical exponents in both quantum and statistical field theory, the mass scaling exponent can be understood in terms of a Renormalisation Group flow. To set up the theory, we follow the discussion of this as in \cite{Koike:1995jm}, specialising it to our system.

Let us consider the general scenario where CSS has not yet been imposed, and collect all metric and matter fields under one umbrella for convenience as $F(x,s)$, e.g $F(x,s)=\left\{\rho(x,s),\Sigma(x,s),\phi(x,s)\right\}$, with $x$ the self-similar time and $s$ the scaling variable defined in \eqref{eqn:scalingcoords}\footnote{We could also find the same results by using the coordinate $\omega$ defined in \eqref{eqn:scalingcoordw}, but we choose to use $s$ for two reasons. First, the equations are easier to solve using $s$. Second, we can connect with the existing literature more easily.}. The EFEs and EoM \eqref{eqn:PDEs} for the scalar field define a set of coupled, partial differential equations, that when given boundary conditions, determine a specific flow in phase space.

These differential equations are autonomous --- they are invariant under translations in both of the coordinates. This means that there are two renormalisation group transformations that we could construct here relating to translations in each coordinate. The transformation corresponding to the coordinate $x$ will generate flows, when using our CSS critical solution, that maintain the symmetry. We will make a few short comments about this at the end of the subsection to find this critical exponent $\gamma_{q}$. However, since we are primarily interested in modes that break the self-similarity, we focus on a flow in the scaling variable. 

Let us define an operator $O(s_{0})$ by its action on functions of $x$ at a specific scale $s$ as
\begin{equation}
	O(s_{0})\cdot F(x,s) = F(x,s+s_{0})\,.
	\label{eqn:RGT}
\end{equation}
This translation operation, that leaves the PDEs invariant, is called the Renormalisation Group Transformation (RGT) since it shifts the scale at which the function is evaluated. It forms a commutative, one-parameter semi-group with binary operation defined as 
\begin{equation}
	O(s_{0})\cdot O(s_{1}) = O(s_{0}+s_{1})\,.
\end{equation}

This one parameter $s$ is continuous and so we can define the infinitesimal generator of the RGT as usual for a semi-group
\begin{equation}
	\begin{aligned}
		\hat{L}_{O}F(x,s) &=\lim_{s_{0}\rightarrow0}\frac{O(s_{0})\cdot F(x,s)-F(x,s)}{s_{0}}\\
		&= \pdr{F(x,s)}{s}\,,
	\end{aligned}
	\label{eqn:RGTinf}
\end{equation}
with the full group action recovered as
\begin{equation}
	O(s_{0})\cdot F(x,s) = e^{s_{0}\hat{L}_{O}}F(x,s)\,.
\end{equation}

With this definition, we can interpret what a CSS \footnote{Note that we are also implicitly assuming the \textit{scale invariant} scalar field solution when we say CSS throughout this section.} solution means from the perspective of the Renormalisation Group. In Section 2, we saw that imposing CSS on the system forced all of the fields $F(x,s)$ to be functions of only the one variable, $x$. Consequently, a self-similar solution,  $F_{\text{ss}}(x)$, must be invariant under the action of the RGT
\begin{equation}
	O(s_{0})\cdot F_{\text{ss}}(x) = F_{\text{ss}}(x)\,.
	\label{eqn:SSRGT}
\end{equation}
Self-similar solutions are \textit{fixed points} of the Renormalisation Group Flow.

Working with this will allow us to determine the RG flow globally, but will involve solving a very difficult problem since it amounts to finding solutions to the full set of PDEs. However, since we care only about the nature of the flow away from the RG fixed point corresponding to CSS, we can simply consider the linearised problem instead as it will give us all of the information that we need according to the Hartman-Grobman Theorem, assuming that the fixed point is hyperbolic.

Let us take a solution that slightly breaks the CSS symmetry by an amount $\delta_{p}\equiv p-p_{*}$,
\begin{equation}
	F(x,s)=F_{\text{ss}}(x)+\delta_{p} G(x,s)\,.
	\label{eqn:breakingss}
\end{equation}

If we now act on this with the RGT, and treat $\delta_{p}$ as a small parameter, we can Taylor expand this operator to linear order

\begin{equation}
	\begin{aligned}
		O(s_{0})\cdot F(x,s) &=O(s_{0})\cdot\left(F_{\text{ss}}(x)+\delta_{p} G(x,s)\right)\\
		&\approx F_{\text{ss}}(x)+ \delta_{p} T(s_{0})G(x,s)+\order{\delta_{p}^{2}},
	\end{aligned}
\end{equation}
where we have defined the tangent map, the linearised RGT, near the CSS fixed point as
\begin{equation}
	T(s_{0})\cdot G(x,s)=\lim_{\delta_{p}\rightarrow0}\frac{O(s_{0})\cdot F(x,s)-F_{\text{ss}}(x)}{\delta_{p}}\,.
	\label{eqn:linRGT}
\end{equation}
We can once again define an infinitesimal action of the linearised operator through the generator
\begin{equation}
	\begin{aligned}
		\hat{L}_{T}G(x,s) &=\lim_{s_{0}\rightarrow0}\frac{\hat{T}(s_{0})\cdot G(x,s)-G(x,s)}{s_{0}}\\
		&= \pdr{G(x,s)}{s}\,,
	\end{aligned}
	\label{eqn:inflinRGT}
\end{equation}
recovering the full map as
\begin{equation}
	T(s_{0})\cdot G(x,s) = e^{s_{0}\hat{L}_{T}}G(x,s)\,.
\end{equation}

It is then convenient to decompose the linearised perturbation into a sum over the eigenmodes $g_{i}(x)$ of the infinitesimal tangent map with eigenvalues $k_{i}\in \mathbb{C}$ 
\begin{equation}
	G(x,s)=\sum_{i}e^{k_{i}s}g_{i}(x)\,,
	\label{eqn:Gdecomp}
\end{equation}
such that
\begin{equation}
	\begin{aligned}
		T(s_{0})\cdot G(x,s)&= \sum_{i}e^{s_{0}\hat{L}_{T}}e^{k_{i}s}g_{i}(x)\\
		&=\sum_{i}e^{k_{i}(s+s_{0})}g_{i}(x)\,.
	\end{aligned}
\end{equation}
Finally this gives us that the flow of the linearly perturbed solution is given by
\begin{equation}
	\begin{aligned}
		O(s_{0})\cdot F(x,s) &\approx F_{\text{ss}}(x)+ \delta_{p} T(s_{0})\cdot G(x,s)+\order{\delta_{p}^{2}}\\
		&\approx F_{\text{ss}}(x)+ \delta_{p}\sum_{i} e^{(s_{0}+s)k_{i}}g_{i}(x)+\order{\delta_{p}^{2}}\,.
	\end{aligned}
	\label{eqn:linactRGT}
\end{equation}
The spectrum of allowed eigenvalues are clearly the Lyapunov exponents and hence they must determine the nature of the fixed point of the RG flow. As $s_{0}\rightarrow +\infty$ we have the following qualitatively different descriptions of the eigenmodes based on the sign of the eigenvalue:
\begin{itemize}
	\item $Re(k_{i})<0$ --- These modes are called irrelevant. Although the initial data breaks CSS, these modes will result in the solutions converging back towards the CSS manifold.
	\item $Re(k_{i})=0$ --- These modes are called marginal. They correspond to perturbations that remain within the CSS manifold.
	\item $Re(k_{i})>0$ --- These modes are called relevant and will dominate the evolution of the system, driving it away from CSS manifold.
\end{itemize}
Due to this, it is clear that only the relevant modes (if they exist) will survive on the RHS of \eqref{eqn:linactRGT} and so the sum over $i$ can be approximated as a sum over only the relevant modes.

The universality observed by Choptuik, and in other systems such as the radiation fluid \cite{Choptuik:1992jv,Evans:1994pj}, can be understood as there being only \textit{one} relevant mode and hence a unique, positive, Lyapunov exponent. In the CSS breaking system however, as seen by Frolov \cite{Frolov:1999fv}, there is a spectrum of allowed Lyapunov exponents and so there is no universality.

Explicitly, we write the linear perturbations of each of the fields away from the critical self-similar solutions as
\begin{equation}
	\begin{aligned}
		\Sigma(s,x) &= \delta_{p}\sigma(x)e^{ks}\\
		\rho(s,x) &= \sqrt{y_{*}(x)}\left(1+\delta_{p} r(x)e^{ks}\right)\\
		\phi(s,x) &= \phi_{*}(x)+\delta_{p}\varphi(x)e^{ks}\,,
	\end{aligned}
	\label{eqn:pertans}
\end{equation}
where the subscript `$*$' indicates the critical solution. Note that all of the functions share the same Lyapunov exponent because there is only one true degree of freedom in the problem coming from the scalar field. We have also left the sum over relevant modes suppressed since we will be treating finding the Lyapunov exponents as an eigenvalue problem --- we will solve the perturbed differential equations and find the restrictions that this places on the allowed values of $k$.

In order to solve these perturbed differential equations however, we need to also impose boundary conditions. These boundary conditions are chosen as follows
\begin{equation}
	\begin{aligned}
		&r(-\infty) =0\,, \qquad &&\sigma(-\infty)=0\,, \qquad &&&\varphi(-\infty)=0\,,\\
		&r(\infty) <\tilde{A}\,, \qquad &&\sigma(\infty)<\tilde{B}\,, \qquad &&&\varphi(\infty)<\tilde{C} \frac{x}{\sqrt{\ve}}\,.
	\end{aligned}
\end{equation}
The initial time conditions are chosen to ensure that the perturbative solutions also satisfy the boundary conditions of the original problem \eqref{eqn:boundcond1}. The remaining boundary conditions are bounds on the growth rate of the perturbation functions at late times in $x$ such that they are always smaller than the background critical functions. This is to ensure that the Lyapunov exponent associated with growth in the scaling coordinate $s$ dominates --- the perturbation theory in $\delta_{p}$ can \textit{only} be broken by a growth in $e^{ks}$ and not in an exponential of $x$. This choice of boundary conditions ensures that $k=0$ is excluded from the spectrum so that the fixed point is hyperbolic and the Hartman-Grobman Theorem applies. Note that by definition this also excludes the possibility of seeing the Lyapunov exponent computed by Soda and Hirata \cite{Soda:1996nq} for the perturbations within CSS from this calculation and so a separate analysis must be done for them. In practice, we will see that these late-time conditions place a lower bound on this Lyapunov exponent and are hence irrelevant from the perspective of the critical exponent anyway. 

We can now make the connection between the Lyapunov exponent $k$ and the critical exponent $\gamma$ explicit. To do this, we look to find when an apparent horizon forms as a consequence of the perturbations. Taking the right hand side of \eqref{eqn:massfunc}, labelling it as $H$ for convenience, and substituting the perturbed fields we find
\begin{equation}
	\begin{aligned}
		H(x,s)& \equiv g^{\mu\nu}\partial_{\mu}\rho\partial_{\nu}\rho \\
		&\approx H_{0}(x)-\delta_{p} e^{ks}H_{1}(x)+\order{\delta_{p}^{2}}.
	\end{aligned}
	\label{eqn:horizonfunc}
\end{equation}
The $H_{0}(x)$ term is always strictly positive meaning that an apparent horizon does not form from the critical solution according to the condition \eqref{eqn:AHcond}. However, with the correct sign for $\delta_{p}$, the first order perturbation can grow large enough as $s$ increases that an apparent horizon does form.

At the apparent horizon we find that
\begin{equation}
	\begin{aligned}
		H_{0}(x_{\ah}) &\approx \delta_{p} e^{ks_{\ahe}}H_{1}(x_{\ah})\\
		&\approx \frac{\delta_{p}}{\left(\rho_{\ah}\right)^{k}}\tilde{H}_{1}(x_{\ah})\,,
	\end{aligned}
\end{equation}
where we have used that from our perturbation ansatzae we can write $\rho_{\ah}=e^{-s_{\ahe}}\left(e^{-x_{\ahe}}y_{*}(x_{\ah})\right)^{\frac{1}{2}}+\order{\delta_{p}}$. 

Since we have argued that the perturbation functions in $x$ must remain bounded, the only way in which these two terms can become comparable in size is when
\begin{equation}
	\delta_{p} \sim \left(\rho_{\ah}\right)^{k}\,.
\end{equation}
Then by using the scaling for the mass of the black hole with the geometric radius in $D$ dimensions, we can finally relate the Lyapunov exponent to the mass of the black hole
\begin{equation}
	\begin{aligned}
		M &\sim \left(\rho_{\ah}\right)^{D-3} \sim \delta_{p}^{\frac{D-3}{k}}\,.
	\end{aligned}
\end{equation}
The last thing to note is that although there is a spectrum of allowed Lyapunov exponents, the one with the largest real component will dominate the RG flow. For this reason the Lyapunov exponent that should be used in this formula is the one with the largest real part, $\kappa\equiv\sup(Re(k))$. Thus, by comparison with \eqref{eqn:massscalp}, we can write the critical exponent as
\begin{equation}
	\gamma_{p} = \frac{1-\ve }{\kappa \ve }\,.
	\label{eqn:critexp}
\end{equation}

\vspace{10pt}
Finally we make a couple of short remarks about the critical exponent associated with the mass of the black hole when the CSS is \textit{not} broken. 

In this case, a very similar procedure to above can be carried out with perturbations exclusively in the $x$ coordinate, where the fixed point of the flow is the saddle point\footnote{From this perspective, we can see just how fine-tuned the CSS critical solution is --- the critical end-state sits at the fixed point of both RG flows in $x$ and $s$.}. In fact, this allows us to directly read off the Lyapunov exponent for the perturbations from the linearisation about the saddle \eqref{eqn:linearisation}
\begin{equation}
	k_{q} = \frac{1}{\sqrt{2\ve}}\,.
\end{equation}
However, the equation that links this to the critical exponent is slightly different when compared to breaking CSS since it is the growth in $x$ that will now lead to apparent horizon formation. As we will see, this coordinate scales with the radius in a slightly different way. 

Repeating the same procedure for a perturbed fields in $x$, we find
\begin{equation}
	\begin{aligned}
		H(x,s) \approx H_{0}(x)- \delta_{q} e^{k x}\tilde{H}_{1}(x)+\order{\delta_{q}^{2}}\,.
	\end{aligned}
\end{equation}
Using the same logic as before, at the apparent horizon
\begin{equation}
	\begin{aligned}
		H_{0}(x_{\ah}) &\approx \delta_{q} e^{kx_{\ahe}}H_{1}(x_{\ah})\\
		&\approx \frac{\delta_{q}}{\left(\rho_{\ah}\right)^{2k}} e^{-2ks}y_{*}^{k}\tilde{H}_{1}(x_{\ah})\,,
	\end{aligned}
\end{equation}
and hence the critical exponent has an extra factor of two in relation to the Lyapunov exponent
\begin{equation}
	\begin{aligned}
		\gamma_{q} &=\frac{1-\ve}{2\ve k_{q}}\\
		&=\frac{D-3}{\sqrt{2(D-2)}}\,,
	\end{aligned}
\end{equation}\label{page:PresExp}
in complete agreement with the analysis of Soda and Hirata \cite{Soda:1996nq}, despite the use of difference coordinates.

\subsection{Review of Four Dimensions} \label{sec:4.2}
We now briefly review what is known about these CSS breaking perturbations in four spacetime dimensions by Frolov \cite{Frolov:1999fv} for two reasons. First, it will allow us to use these solutions to check that our general dimensional result reduces to the correct expression in $D=4$. Second, it will provide some insight into how to perform the calculation in a higher number of dimensions. The important take away is that:
\begin{center}
	\textit{The asymptotic nature of the functions at the past and future boundaries is sufficient to recover the full set of restrictions on the Lyapunov exponent.}
\end{center}

Taking the perturbation ansatzae in $D=4$ and substituting them into the field equations \eqref{eqn:PDEs} gives us at first order in $\delta_{p}$ 
\begin{subequations}\label{eqn:pert1}
	\begin{align}
		2\eta(1+\eta)\varphi''+2\left(k+\eta(1+k)\right)\varphi'+k\varphi+2\sqrt{2}\eta r'+\sqrt{2}k r&=0\,,\label{eqn:pert1a} \\
		4k\eta(1+\eta)r'+2k(k\eta+k-1)r+4\eta(1+\eta)\sigma'+2k(\eta+2)\sigma+\sqrt{2}k\eta\varphi&=0\,,\label{eqn:pert1b}\\
		2(1+\eta)r''+2r'-2\sigma'+\sqrt{2}\varphi'&=0\,, \label{eqn:pert1c}\\
		\eta(1+\eta)r''+\left(k-1+\eta(1+k)\right)r'-(1-k)r+\sigma&=0\,, \label{eqn:pert1d}
	\end{align}
\end{subequations}
where $\eta=2e^{x}$ and we have made the constraint \eqref{eqn:pert1b} explicitly first order.

The first thing to note is that \eqref{eqn:pert1c} is a total derivative, allowing us to solve for $\sigma$
\begin{equation}
	\sigma = C+(\eta +1) r'+\frac{1}{\sqrt{2}}\varphi\,.
\end{equation}
Substituting this into our constraint equation and evaluating it at the past boundary, $\eta=0$ gives
\begin{equation}
	C(k-1)=0\,.
\end{equation}
This means that the analysis should split into two different cases depending on whether $C=0$ or if $k=1$.

Substituting this solution for $\sigma$ into the remaining differential equations, and making the change of variables $\varphi(\eta)=\sqrt{2}(r(\eta)-z(\eta))$ gives a decoupled, hypergeometric differential equation for $z$
\begin{equation}
	\eta(1+\eta)z''+2(C+k+\eta(1+k))z'-2(2-k)z=0\,.
\end{equation}
From here, it is easy to see that taking $k=1$ will lead to a solution that can never satisfy the boundary conditions, and so we focus on the $C=0$ option, with solution given by
\begin{equation}
	z(\eta)= A_{z} \eta ^{1-k}\,_2 F_1 {\small\left.\left[ \begin{matrix} \ a_{4}+1, & b_{4}+1 \\ \multicolumn{2}{c}{2-k} \end{matrix}\ \right| -\eta \right]}+B_{z}\,_2 F_1 {\small\left.\left[ \begin{matrix} \ -b_{4}, & -a_{4} \\ \multicolumn{2}{c}{k} \end{matrix}\ \right| -\eta \right]}\,,
\end{equation}
where
\begin{equation}
	a_{4}=-\frac{k}{2}+\frac{1}{2}\sqrt{4-2k+k^{2}}\,,\qquad b_{4}=-\frac{k}{2}-\frac{1}{2}\sqrt{4-2k+k^{2}}\,.
	\label{eqn:hyparg}
\end{equation}
We now impose the early time boundary conditions on this solution, $z(0)=0$. Expanding the solution about $\eta=0$ 
\begin{equation}
	z(\eta)\sim A_{z} \eta^{1-k}+B_{z}+\order{\eta}\,,
\end{equation}
gives $B_{z}=0$ and that $\Re(k)<1$. 
\begin{center}
	\textit{The upper bound on the real part of the Lyapunov exponent is set by the early time boundary condition.}
\end{center}

Now expanding the solution about $\eta=+\infty$ to apply the late time boundary condition we find
\begin{equation}
		z(\eta)\sim A_{z}\Gamma(2-k)\left(\frac{\Gamma (b_{4}-a_{4}) \eta^{-a_{4}-k}}{\Gamma (b_{4}+1) \Gamma (-a_{4}-k+1)}+\frac{\Gamma (a_{4}-b_{4}) \eta^{-b_{4}-k}}{\Gamma (a_{4}+1) \Gamma (-b_{4}-k+1)}\right)\,,
\end{equation}
meaning that for this function to be bounded at late times, we require that both
\begin{equation}
	\Re(a_{4}+k)>0\,,\qquad \Re(b_{4}+k)>0\,.
\end{equation}
This requirement can be seen to translate to a restriction on the possible values that $k$ can take in the complex plane. Combining this with the previous restriction from the past boundary, one finds
\begin{equation}
	\begin{aligned}
		 \frac{1}{2}<\Re(k)<1\,,\qquad \abs{\Im(k)}>\sqrt{\frac{2(2-\Re(k))\Re(k)^{2}}{2\Re(k)-1}}\,.
	\end{aligned}
\end{equation}
\begin{center}
	\textit{The lower bound on the real part of the Lyapunov exponent (and the restriction of the imaginary value) is given by the late time boundary condition.}
\end{center}

The other functions can be computed trivially from here and imposing the boundary conditions on them leads to no further restrictions on the Lyapunov exponent --- as can be expected given that there is only one true degree of freedom in this problem. 

Consequently there is a full spectrum of allowed Lyapunov exponents, with all allowed values requiring a non-zero imaginary part. These two properties can be interpreted in the following ways:
\begin{itemize}
	\item The continuous spectrum of Lyapunov exponents indicates that the CSS manifold does not have codimension 1, and hence there is no universality.
	\item The complex nature of the Lyapunov exponent with the largest real part indicates that, in $D=4$, the perturbed CSS solution sits in the basin of attraction of the intermediate DSS attractor. The imaginary part naturally leads to oscillations of the functions at a frequency that is close to the one observed by Choptuik \cite{Frolov:1999fv}. 
\end{itemize} 

Taking the Lyapunov exponent with the largest real part, we compute the critical exponent in four dimensions to be
\begin{equation}
	\gamma_{p} = 1\,.
\end{equation}

Finally, in order to check against the general dimension results in the next subsection, the full solutions for the perturbation functions are
\begin{equation}
	\begin{aligned}
		r(\eta)&= \frac{1}{(2-k)(1-k)}\frac{1}{1+\eta}\left(2(1-k)\left(\,_2 F_1 {\small\left.\left[ \begin{matrix} \ a_{4}, & b_{4} \\ \multicolumn{2}{c}{1-k} \end{matrix}\ \right| -\eta \right]}-1\right)\right.\\
		&\qquad\qquad\qquad\qquad\qquad\qquad\qquad\left.-\eta\,_2 F_1 {\small\left.\left[ \begin{matrix} a_{4}+1, & b_{4}+1 \\ \multicolumn{2}{c}{3-k} \end{matrix}\ \right| -\eta \right]} \right)\\[0.5em]
		\sigma(\eta) &= \frac{2(1-k)}{2-k}\eta^{-k}\left((1+a_{4})\,_2 F_1 {\small\left.\left[ \begin{matrix} \ a_{4}, & b_{4} \\ \multicolumn{2}{c}{1-k} \end{matrix}\ \right| -\eta \right]}-a_{4}\, \,_2 F_1 {\small\left.\left[ \begin{matrix} 1+a_{4}, & b_{4} \\ \multicolumn{2}{c}{1-k} \end{matrix}\ \right| -\eta \right]}-1\right) \\[0.5em]
		\varphi(\eta)&= \frac{\sqrt{2}}{(2-k)(1-k)(1+\eta)}\eta^{1-k}\left((1-k)\left(2 \,_2 F_1 {\small\left.\left[ \begin{matrix} \ a_{4}, & b_{4} \\ \multicolumn{2}{c}{1-k} \end{matrix}\ \right| -\eta \right]}\right.\right.\\
		&\qquad\qquad\qquad\qquad\qquad\qquad\qquad\left.\left.-(2-k)(1+\eta)\,_2 F_1 {\small\left.\left[ \begin{matrix} 1+a_{4}, & 1+b_{4} \\ \multicolumn{2}{c}{2-k} \end{matrix}\ \right| -\eta \right]}-2\right)\right.\\
		&\qquad\qquad\qquad\qquad\qquad\qquad\qquad\left.-\eta \,_2 F_1 {\small\left.\left[ \begin{matrix} 1+a_{4}, & 1+b_{4} \\ \multicolumn{2}{c}{3-k} \end{matrix}\ \right| -\eta \right]}\right),
	\end{aligned}
	\label{eqn:pert4sol}
\end{equation}
where the overall amplitude has been absorbed into $\delta_{p}$. 

\subsection{General Dimensions}

Repeating the same as before, we find the perturbation equations in general dimensions with $\mu=\frac{y_{*}'}{y_{*}}$ as
\begin{subequations}\label{eqn:pertgen}
	\begin{align}
		4\ve\varphi''+2(2\ve k+\mu)\varphi'+k(1+\mu)\varphi+4\phi_{*}'r'+2k\phi_{*}'r &=0\,, \label{eqn:pertgena} \\[5pt]
		2k r'+k(k+\mu)r+2\sigma'+k(1-\mu)\sigma+2\ve k\phi_{*}'\varphi&=0\,, \label{eqn:pertgenb}\\[5pt]
		r''+\mu r'-(1+\mu)\sigma'+2\ve\phi_{*}'\varphi'&=0\,, \label{eqn:pertgenc}\\[0.5pt]
		2\ve r''+2(k\ve+\mu)r'+\left(k(1+\mu)-\frac{2(1-\ve)}{y_{*}}\right)r+\frac{2(1-\ve)}{y_{*}}\sigma&=0\,, \label{eqn:pertgend}
	\end{align}
\end{subequations}

Taking what we have learnt from the calculation in $D=4$, we split the computation based on the two asymptotic limits $x\rightarrow-\infty$ and $x\rightarrow+\infty$ since it is the boundary conditions at these limits that give us the constraints on the eigenvalue $k$. This also allows us to make an argument about the critical exponent that is in principle \textit{exact}\footnote{Assuming that the matching provides no additional restrictions. Note that we could try to check this using \largeD techniques with some form of `Method of Multiple Scales' analysis, but the large freedom in the possible scaling of the frequency of the oscillations makes this difficult in practice.} in the number of dimensions as the boundary conditions themselves are exact in $\ve$.

We start with the past asymptotic boundary. The critical solutions at this boundary are
\begin{equation}
	\begin{aligned}
		y_{*}(x) &\sim 1+\frac{1}{2}e^{-x} ~~\text{as}~~x\rightarrow-\infty\\
		\phi_{*}(x) &\sim2^{\frac{1}{\ve} }s^{\frac{1}{2 \ve }} \sqrt{1-\ve } \, e^{\frac{x}{2 \ve }}~~\text{as}~~x\rightarrow-\infty\,.
	\end{aligned}
\end{equation}
It is convenient to eliminate $\phi_{*}$ entirely from the differential equations \eqref{eqn:pertgen} by defining $\psi(x)=\phi_{*}'(x)\varphi(x)$ and through the use of the background equations. After doing this, we use the Frobenius method to determine the leading order solutions at the boundary as a series solution. Carefully ensuring that all equations are satisfied at the leading orders gives the following asymptotic solutions as $x\rightarrow-\infty$
\begin{equation}
	\begin{aligned}
		r(x) &\sim r_{0} e^{x \left(\frac{1}{\ve }-k\right)}\,,\qquad \sigma(x)\sim \frac{\ve k (k-1)r_{0}}{2(1-2k \ve)}e^{x\left(\frac{1}{\ve}-k\right)}\,,\\
		\varphi(x) &\sim -\frac{2(1-k\ve)(1-\ve(k+1))r_{0}}{\ve\sqrt{1-\ve}(1-2k\ve)(4s)^{\frac{1}{2\ve}}}e^{x\left(\frac{1}{2\ve}-k\right)}\,.
	\end{aligned}
\end{equation}
These can be seen to agree with the asymptotic solution of the full equations in $D=4$ \eqref{eqn:pert4sol} with the identification that
\begin{equation}
	r_{0} = \frac{1}{2-k}\,.
\end{equation}
In order to satisfy the boundary conditions, we can see that the Lyapunov exponent has an upper bound of
\begin{equation}
	k<\frac{1}{2\ve}\,.
\end{equation}
To find the lower bound and the restriction in the complex plane, we turn our attention now to the future asymptotic boundary. The calculation here proceeds in a very similar fashion to the way that it does for the full solution in $D=4$. The critical solutions at this boundary are
\begin{equation}
	\begin{aligned}
		y_{*}(x) &\sim 2(1-\ve) ~~\text{as}~~x\rightarrow+\infty\\
		\phi_{*}(x) &\sim \frac{x}{2\sqrt{\ve}}~~\text{as}~~x\rightarrow+\infty\,.
	\end{aligned}
\end{equation}
Upon substituting these into the differential equations \eqref{eqn:pertgen}, we find that once again \eqref{eqn:pertgenc} is a total derivative
\begin{equation}
	\sigma(x) =C+\sqrt{\ve}\varphi+r'\,,
\end{equation}
allowing us to eliminate $\sigma$ from the remaining equations as before. After defining $\varphi(x)=\frac{1}{\sqrt{\ve}}\left(r(x)-z(x)\right)$, we find a decoupled equation for $z$
\begin{equation}
	2 \ve  z''+2 k \ve  z'+\frac{1}{2} (k-2) z + C=0\,.
\end{equation}
The solution to this is given by 
\begin{equation}
	z(x) \sim A_z e^{-x (a_{\ve}+k)}+B_z e^{-x (b_{\ve}+k)}-\frac{2 C}{k-2}\,,
\end{equation}
where in analogy with \eqref{eqn:hyparg} we have defined
\begin{equation}
	a_{\ve}=-\frac{k}{2}-\frac{\sqrt{k^2 \ve -k+2}}{2 \sqrt{\ve }}\,,\qquad b_{\ve}=-\frac{k}{2}+\frac{\sqrt{k^2 \ve -k+2}}{2 \sqrt{\ve }}\,.
	\label{eqn:expargs}
\end{equation}
We can now go through the process of evaluating the original fields to find
\begin{equation}
	\begin{aligned}
		r(x) &\sim \frac{C}{2-k}+A_{z}e^{-x(a_{\ve}+k)}+B_{z}e^{-x(b_{\ve}+k)}+e^{-kx}B_{r}\,,\\
		\sigma(x) &\sim \frac{(1-k)C}{2-k}-\frac{k}{2}\left(A_{z}e^{-x(a_{\ve}+k)}+B_{z}e^{-x(b_{\ve}+k)}\right)+(1-k)B_{r}e^{-kx}\,,\\
		\varphi(x) &\sim -\frac{C}{\sqrt{\ve}(2-k)}+\frac{\sqrt{2-k(1-k\ve)}}{2\ve}\left(B_{z}e^{-x(b_{\ve}+k)}-A_{z}e^{-x(a_{\ve}+k)}\right)+\frac{1}{\sqrt{\ve}}B_{r}e^{-kx}\,,
	\end{aligned}
\end{equation}
in agreement once again with the full solutions in $D=4$ \eqref{eqn:pert4sol} with the identifications 
\begin{equation}
	\begin{aligned}
		A_{z} &=\frac{2^{1+b_{4}}\Gamma(2-k)\Gamma(b_{4}-a_{4})}{(2-k)\Gamma(b_{4})\Gamma(2+b_{4})}\,,\\
		B_{z} &=\frac{2^{1+a_{4}}\Gamma(2-k)\Gamma(a_{4}-b_{4})}{(2-k)\Gamma(a_{4})\Gamma(2+a_{4})}\,,\\
		B_{r}&=-\frac{2^{1-k}}{2-k}\,,\\
		C&=0\,.
	\end{aligned}
\end{equation}

\begin{figure}[H]
	\centering
	\begin{subfigure}[b]{0.49\textwidth}
		\includegraphics[width=\textwidth]{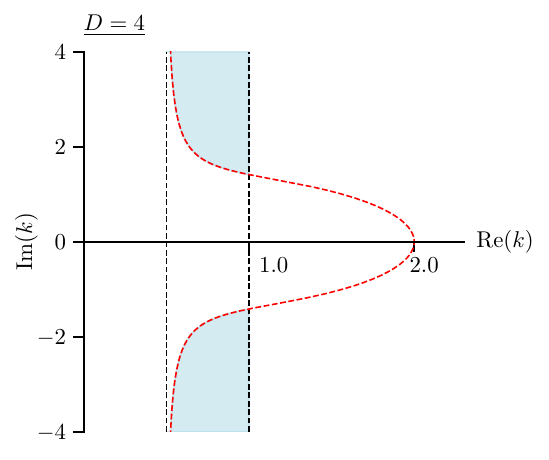}
	\end{subfigure}
	\begin{subfigure}[b]{0.49\textwidth}
		\includegraphics[width=\textwidth]{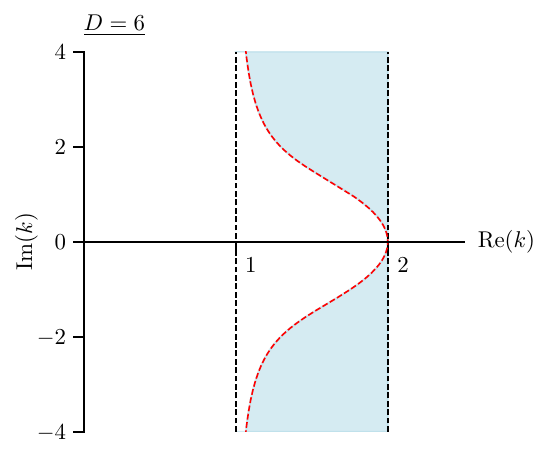}
	\end{subfigure}
	\\
	\begin{subfigure}[b]{0.49\textwidth}
		\includegraphics[width=\textwidth]{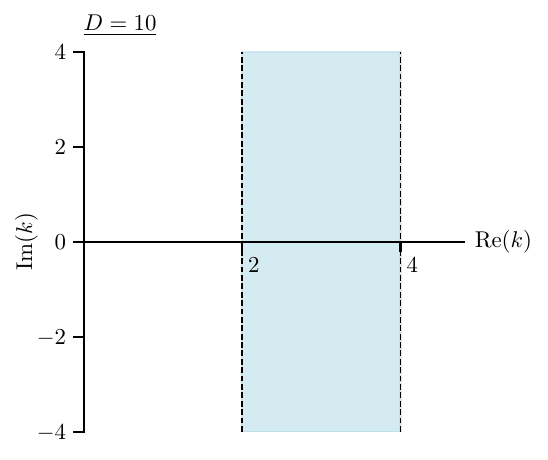}
	\end{subfigure}
	\begin{subfigure}[b]{0.49\textwidth}
		\includegraphics[width=\textwidth]{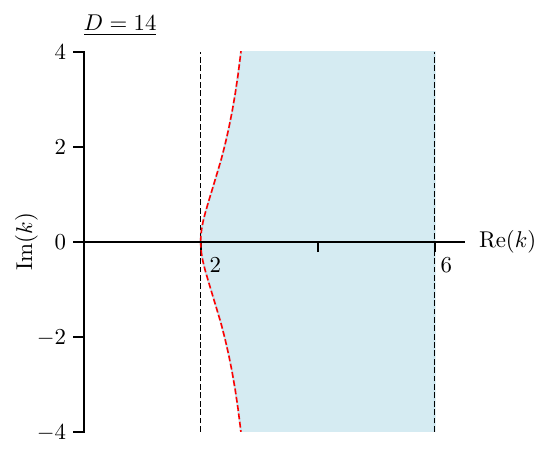}
	\end{subfigure}
	\caption{Plots of the allowed spectrum of Lyapunov exponents (highlighted in blue) for different spacetime dimensions.}
	\label{fig:lyapunovD}
\end{figure}

In order for these function to be bounded at late times, we require that both
\begin{equation}
	\Re(a_{\ve}+k)>0\,,\qquad \Re(b_{\ve}+k)>0\,.
\end{equation}
Combining these restrictions with the upper bounds found from the left boundary, we have that the Lyapunov exponent is constrained to the following values depending on the number of spacetime dimensions
\begin{equation}
	\begin{aligned}
		\ve>\frac{1}{8}&: \begin{cases}
			\frac{1}{4\ve}<\Re(k)<\frac{1}{2\ve}&~~ \abs{\Im(k)}> 2 \Re(k) \sqrt{\frac{\ve(2-\Re(k))}{4\Re(k)\ve-1}}  \\
			2<\Re(k)<\frac{1}{2\ve} & ~~ \Im(k)\in \mathbb{R} \,,
		\end{cases}\\
		\ve=\frac{1}{8}&: \,\,\,\,\, 2<\Re(k)<4 ~~~~~~~\, \Im(k)\in \mathbb{R}\,, \\
		\ve<\frac{1}{8}&: \begin{cases}
			2<\Re(k)<\frac{1}{4\ve}&~~ \abs{\Im(k)}< 2 \Re(k) \sqrt{\frac{\ve(\Re(k)-2)}{1-4\Re(k)\ve}} \\
			\frac{1}{4\ve}<\Re(k)<\frac{1}{2\ve}& ~~ \Im(k)\in \mathbb{R} \,.
		\end{cases}
	\end{aligned}
\end{equation}\label{page:SpectrumLyap}
As we can see, the $D=4$ result is exactly recovered by these inequalities. Furthermore, these results appear to suggest that there are two critical dimensions \footnote{It should be noted that our method is not entirely complete since we only ever compute asymptotic solutions. In principle, a full solution could impose additional constraints on the Lyapunov exponent spectrum through matching, but this computation is outside the scope of this paper.}:
\begin{itemize}
	\item $D=6$ --- this is the first time that the largest Lyapunov exponent \textit{can} be entirely real;
	\item $D=10$ --- this is the number of dimensions for which there is no restriction on the imaginary value of the Lyapunov exponent. For dimensions smaller than this, the restriction in the complex plane is to be above the curve and for dimensions larger it is to be  inside the curve as we show in Figure \ref{fig:lyapunovD}. It is important to note however, that this critical dimension only affects the structure of \textit{subdominant} perturbations. 
\end{itemize}\label{page:CritDim}

This result has an interesting consequence for the full phase space and the relative positions of the CSS and DSS critical manifolds.

In \cite{Frolov:1999fv}, Frolov analyses these perturbations as wave packets on the CSS background and concludes that they are drawn toward the DSS intermediate attractor in $D=4$. This conclusion is based on identifying the dominant Lyapunov mode --- the one with the largest real part and smallest allowed imaginary component --- which gives rise to a characteristic periodicity closely matching the frequency observed by Choptuik.

Assuming that this logic also holds in higher dimensions, we observe that for $D \geq 6$, the corresponding frequency becomes zero. Following this reasoning, we can conclude:
\begin{center}
	\textit{The perturbed CSS solution does not sit in the basin of attraction for the DSS intermediate attractor in $D\geq6$.}
\end{center}
This result is consistent with the observed impact of increasing the number of dimensions on the CSS phase space \ref{fig:phaseD}. As $D$ increases, phase space trajectories diverge from the critical manifold more rapidly than in lower dimensions. Consequently, resolving the influence of the saddle region, Region $\cB$, requires fine-tuning the initial scalar field amplitude to lie even closer to criticality. This behaviour suggests that a similar mechanism may be at play here as well: in higher dimensions, the perturbed CSS solution may lie too far from the DSS critical manifold for its influence to be observable.

A physical interpretation for $D=10$ as a critical dimension is more challenging, as it influences only the structure of subdominant perturbation modes. Nevertheless, its emergence may be related to the fact that $D=10$ also marks a critical threshold in the CSS merger solution describing the transition between black holes and black strings \cite{Kol:2005vy}, as discussed in Section \ref{sec:Introduction}. Perturbations of that merger solution are both unique and acquire imaginary components for $D<10$, exhibiting features reminiscent of both CSS-preserving and CSS-breaking modes \cite{Kol:2002xz}. This mixing of features may arise from the Wick rotations used to relate the two systems (changing their stability structures), or from the differing boundary conditions imposed in each case. A more detailed understanding of the connection between these scenarios may help clarify the origin of the $D=10$ critical dimension observed in CSS gravitational collapse.

Finally, taking \eqref{eqn:critexp}, we find the critical exponent to be
\begin{equation}
	\begin{aligned}
		\gamma_{p} &= 2(1-\ve)\\
		&=\frac{2(D-3)}{D-2}\,,
	\end{aligned}
\end{equation}\label{page:BreakExp}
which asymptotes to a finite value of $2$ in infinite dimensions --- a clear contrast to the diverging critical exponent coming from perturbing inside CSS. This gives further evidence that the critical exponent associated with DSS really could asymptote to a finite value in infinite dimensions as the numerical results suggest \cite{Bland2007}.
\newpage
\section{Outlook}\label{sec:5}
\label{sec:Outlook}

In this paper, we demonstrated that the \largeD expansion provides a powerful analytic framework for understanding gravitational collapse. The strength of this approach stems from the fact that it leads to a separation of scales, allowing for an `effective' description of the dynamics in distinct regions where different physical effects dominate. By analysing the structure of the governing differential equations, we identified the characteristic timescale associated with each of these regions and found that the initial conditions play an important role in defining their characteristics. In particular, we observed that when the boundary conditions are finely tuned to near-criticality, the system spends an uncharacteristically long time in one of these regions.

We then performed asymptotic expansions in $1/D$ within each of these regions to construct solutions to the one-parameter family of spacetimes in our model. This allowed us to construct analytically matched solutions for both the critical and very supercritical regimes. Additionally, we constructed solutions for the marginally subcritical and supercritical cases as numerically matched perturbations of the critical solution.

Finally, we computed the spectrum of allowed Lyapunov exponents corresponding to supercritical perturbations that break the continuous self-similarity. We identified two critical dimensions, $D=6$ and $D=10$, which change the structure of the spectrum such that for $D\geq6$, the perturbed CSS critical solution no longer resides within the basin of attraction of the DSS intermediate attractor. Using the Lyapunov exponent with the largest real part, we computed the mass scaling exponent and found that it asymptotes to a finite, rational value in the limit of infinite dimensions.

There are two challenges left that would complete the study of this system:
\begin{itemize}
	\item Finding a full analytic solution to the differential equation governing Region $\cB$
	\begin{equation}
		f''+\frac{1}{4}\left(f'\right)^{2}-f=0,
	\end{equation}
	would allow us to construct uniform solutions for \textit{every} member of the one-parameter family that we have discussed. Although this is not essential, such a solution would enhance the completeness of the analysis.
	\item The perturbation spectrum computed in Section \ref{sec:4} could be explored further. First, our conclusions could be confirmed by performing a more comprehensive calculation of the Lyapunov spectrum that we have found using asymptotics. This could involve performing numerics, or carrying out the analysis using wave packets as in \cite{Frolov:1999fv}. Second, the physical significance of the $D=10$ critical dimension remains unclear. The relation of this system to the critical merger solution between black holes and black strings suggests that maybe this could be understood with a deeper understanding of the connection between the two systems.
\end{itemize}

The main criticism of our analysis so far, is that we have only demonstrated the  power of \largeD techniques in a gravitational collapse model where $D=4$ is simpler. Nonetheless, we hope that this work can serve as a foundation for extending these techniques to more complex systems that are analytically intractable in four dimensions. We outline some of these potential applications below:

\begin{itemize}
	\item \textbf{Scale-varying scalar field:} The phase space is much richer if we allow the CSS scalar field to vary with scale \cite{Christodoulou:1994hg}. Describing the critical solutions in the \largeD limit for this case would be particularly interesting because the attractor solution corresponds to a naked singularity. Analytic solutions to this system do not exist, and perturbations away from the CSS manifold have not yet been studied.
	\item \textbf{Perfect fluid collapse:} Numerical evidence suggests that for a system modelled as a perfect fluid, the continuously self-similar critical manifold has codimension 1 in the full phase space for the full range of values of $0\leq w\leq1$ \cite{Evans:1994pj,Neilsen:1998qc,Brady:1998ir,Koike:1999eg}\footnote{Although this may appear to contradict the results obtained by Choptuik --- given that a massless scalar field is formally equivalent to a stiff fluid with equation of state $w=1$ --- there is, in fact, no inconsistency \cite{Brady:2002iz}. The DSS critical solution of the scalar field features regions where the effective energy density and pressure become negative, a behaviour that cannot be realised in perfect fluid models with a timelike fluid velocity.}, making it the universal attractor. It would be interesting to explore how the spectrum of allowed Lyapunov exponents for the perturbations changes as $w$ is varied. In particular, the solution for a radiation fluid is of interest for primordial black hole formation during the radiation dominated era of the universe \cite{Musco:2008hv,Carr:2020gox,LISACosmologyWorkingGroup:2023njw}.
	\item \textbf{Discrete self-similarity:} Extending these methods to systems with discrete self-similarity would require the development of additional techniques. This would likely involve introducing a Boltzmann-like hierarchy of equations, one for each partial wave of the scalar field. While this hierarchy is expected to be strongly coupled in four dimensions, it would be interesting to see whether or not it becomes weakly coupled in the \largeD limit, something similar to what was found for quasinormal modes \cite{Emparan:2014aba}. If so, the fractal behaviour of the Choptuik attractor should be calculable in a $1/D$ expansion.
	\item \textbf{The NFW Profile:} Another challenging application involves the structure of dark matter halos, in which a gas of dark matter particles virialises into a bound structure with very simple power-law behaviour. So far, these structures have been primarily understood through large $N$-body simulations \cite{Navarro:1995iw,Navarro:1996gj}. It would be intriguing to see whether the power-law behaviour, which varies with scale, could be computed analytically in the \largeD limit.
\end{itemize}

The results presented here mark an initial step in applying the \largeD expansion to gravitational collapse. We expect that this framework will lead to many interesting results in the near future, providing a tool to uncover universal features of the gravitational interaction in its nonlinear regime.\\

\newpage
\noindent{\bf Acknowledgments}

We thank Diego Blas, Bruno Bucciotti, Serban Cicortas, Roberto Emparan, Barak Kol, Frans Pretorius, Salvatore Raucci, Marija Toma\v{s}evi\'{c}, Chiara Toldo, Evita Verheijden, Tom Westerdijk and Umberto Zannier for insightful discussions and helpful comments. We also thank Diego, Roberto, Barak and Tom for their comments on a draft of this paper.

CC and GLP are supported by Scuola Normale, by INFN (IS GSS-Pi), and by the ERC (NOTIMEFORCOSMO, 101126304). 
Views and opinions expressed are, however, those of the author(s) only and do not necessarily reflect those of the European Union or the European Research Council Executive Agency. 
Neither the European Union nor the granting authority can be held responsible for them. GLP is further supported by a Rita-Levi Montalcini Fellowship from the Italian Ministry of Universities and Research (MUR), as well as under contract 20223ANFHR (PRIN2022). This research was supported in part by grant NSF PHY-2309135 to the Kavli Institute for Theoretical Physics (KITP). 

\appendix
\newpage
\section{Method of Regions}\label{app:MoR}

In this appendix, we discuss a `Method of Regions' approach to this problem. This method relies on the existence of the integral representation of the differential equation
\begin{equation}
	x= -\int f(y)\diff{y},
\end{equation}
where 
\begin{equation}
	f(y) = \frac{1}{\sqrt{y^{2}-2y+y^{2}\left(\frac{2q}{y}\right)^{\frac{1}{\ve}}}}.
\end{equation}
While this reliance means that this approach is not always applicable to problems in GR, we demonstrate that it can yield the same results for the Very Supercritical regime in this model.

First, we consider Region $\cA$ where $y$ is very large. The boundary condition for this region is given by
\begin{equation}
	\lim_{y\to\infty}x\to-\log(2y).
\end{equation}
This allows us to expand $f(y)$ under the assumption that the third term under the square root is small compared to the other two
\begin{equation}
	f(y) = \sum_{n=0}^{\infty}\binom{-\frac{1}{2}}{n}y^{2n}(y(y-2))^{-\frac{1}{2}-n}\left(\frac{2q}{y}\right)^{\frac{n}{\ve}}.
\end{equation}
This can be integrated term-by-term to find
\begin{equation}
	\xa+C = \sum_{n=0}^{\infty}\frac{2^{\frac{1}{2}+n}q^{\frac{n}{\ve}}}{2n-1}\binom{-\frac{1}{2}}{n}(y-2)^{\frac{1}{2}-n} \,_2 F_1 {\small\left.\left[ \begin{matrix} \ \frac{1}{2}-n, & \frac{1}{2}-n+\frac{n}{\ve} \\ \multicolumn{2}{c}{\frac{3}{2}-n} \end{matrix}\ \right| 1-\frac{y}{2} \right]},
\end{equation}
where $\,_2 F_1$ is the hypergeometric function. The first term of this sum, the leading-order part, gives the solution to Region $\cA$ that we found from the differential equation
\begin{equation}
	\xa \sim -2\operatorname{arcsinh}\left(\sqrt{\frac{y-2}{2}}\right),
\end{equation}
where the integration constant has been fixed by the boundary conditions. Consequently, the remaining terms must correspond to the exponentially suppressed corrections that we could not compute from the differential equation alone. Note that if they are included, however, there is no way to invert the solution to find $y(x)$. 

The boundary condition fixes an additive constant to each of these terms such that overall the solution in Region $\cA$ where 
\begin{equation}
	y\gtrsim\max(2,2q),
\end{equation}
can be expressed as a sum of incomplete beta functions $B_{z}(a,b)$
\begin{equation}
	\begin{aligned}
		\xa =& -2\operatorname{arcsinh}\left(\sqrt{\frac{y-2}{2}}\right)\\
		&~~+\sum_{n=1}^{\infty}\binom{-\frac{1}{2}}{n}\left(i (-1)^{n}B_{1-\frac{y}{2}}\left(\frac{1}{2}-n,-\frac{n}{\ve }+n+\frac{1}{2}\right)+\frac{\Gamma\left(\frac{1}{2}-n\right)\Gamma\left(\frac{n}{\ve}\right)}{\Gamma\left(\frac{1}{2}-n+\frac{n}{\ve}\right)}\right).
	\end{aligned}
\end{equation}
The solution written in this form is highly numerically unstable due to the functions and matching constants both giving large values that approximately cancel, ultimately leading to a small result.

We now move on to Region $\cC$ and apply similar procedure as for Region $\cA$, except now assuming that the third term under the square root is large in comparison to the other two. After integrating, we find solutions with a similar structure
\begin{equation}
	\xcbar = \sum_{n=0}^{\infty}\left((-1)^{n}\binom{-\frac{1}{2}}{n}q^{-\frac{1+2n}{2\ve}}B_{1-\frac{y}{2}}\left(1+n,\frac{1+2n-2n\ve}{2\ve}\right)-\bar{c}_{n}\right),
\end{equation}
where $\bar{c}_{n}$ are matching coefficients and $B_{z}(a,b)$ are incomplete beta functions once again.

The first term in this sum corresponds precisely to the near-singularity approximation \eqref{eqn:nearsing}, with the higher-order terms improving it. This solution is valid for
\begin{equation}
	0<y\lesssim 2q,
\end{equation}
meaning that it is valid up to where Region $\cA$ approximately ends.

The integral representation gives us a faster way to find solutions for these regions, to an arbitrary degree of accuracy, by expanding the integrand when certain terms are small or large. However, matching becomes complicated because the domains of validity of the solutions do not overlap. We could follow a procedure similar to Appendix \ref{app:Matching}, but even with this, there is no possibility of constructing a uniform solution due to this lack of overlap.

For this sake of completeness, we briefly show that Region $\cC$ can be constructed such that this overlap exists by using the near-horizon resummation from before \eqref{eqn:ycreparameter} 
\begin{equation}
	y(x)=(2^{1-2\ve}q)^{\frac{1}{1-\ve}}Y^{\ve}(x),
\end{equation}
and that this will give us the same solution as from the differential equation. Substituting this change of variables into the integral, we find
\begin{equation}
	x = -\ve\int\frac{ \diff{Y}}{\sqrt{Y \left(2^{\frac{\ve }{1-\ve }} q^{\frac{1}{\ve -1}} \left(2-Y^{1-\ve }\right)+Y\right)}},
\end{equation}
This can be expanded in $\ve$ and integrated at leading order, giving
\begin{equation}
	\xcbar+\bar{c} = 2\ve \sqrt{\frac{q}{q-1}} \log \left(\sqrt{(q-1) Y+2}-\sqrt{(q-1) Y}\right),
\end{equation}
which is precisely the same solution that we found from the differential equation \eqref{eqn:Y0sol}. We can continue to compute higher order corrections, but we will be unable to invert the solution if they are included.

This can now be matched to Region $\cA$ since the solutions overlap. Taking the leading order solution for Region $\cA$ and expanding it in $\ve$ when written with the $Y$-variable, we find
\begin{equation}
	\begin{aligned}
		\xa &\sim -2\operatorname{arcsinh}\left(\sqrt{\frac{(2^{1-2\ve}q)^{\frac{1}{1-\ve}}Y^{\ve}-2}{2}}\right)\\
		&\sim \log \left(2 q-2 \sqrt{q(q-1)} -1\right)-\sqrt{\frac{s}{s-1}}\log\left(\frac{Ys}{2}\right)\ve +\order{\ve^{2}}.
	\end{aligned}
\end{equation}
Taking the solution for Region $\cC$ and expanding in terms of the outer variable, or equivalently as $Y\to\infty$, we find
\begin{equation}
	\lim_{Y\to\infty}(\xcbar +\bar{c}) \xrightarrow{\sim}  -\sqrt{\frac{q}{q-1}} \ve  \left(\log \left(\frac{q Y}{2}\right)+\log \left(\frac{4 (q-1)}{q}\right)\right)+\order{\ve^{2}}
\end{equation}
Comparing order-by-order, we determine that the matching coefficient is given by
\begin{equation}
	\bar{c} = \log\left(-1+2q+\sqrt{q(q-1)}\right)+ \ve \sqrt{\frac{q}{q-1}}\log\left(\frac{q}{4(q-1)}\right),
\end{equation}
which in agreement with the result found from the differential equation \eqref{eqn:cmatching}.

\newpage
\section{Second Order Differential Equation for Region $\pmb{\cC}$} \label{app:SecondOrd}

In this appendix, we show that the same leading-order result can be obtained for Region $\cC$ using the second-order differential equation, without relying on any prior knowledge of the first-order differential equation. The key subtlety in performing the computation this way is that the unfixed boundary condition must be incorporated within the asymptotic ansatz in order for it to yield the correct results.

Starting with the second-order differential equation
\begin{equation}
	2 \ve yy''+(1-2\ve)(y')^{2} + 2(1-\ve) y - y^{2}=0,
\end{equation}
and keeping only the divergent terms as $y\rightarrow0$, we recover the same form of the solution as the near-singularity expression from before \eqref{eqn:nearsing}. Thus we can still conclude that the time-scale near the singularity is $x\sim\ve$ and that a convenient resummation in $\ve$ is given by 
\begin{equation}
	y= \tilde{q}Y^{\ve},
	\label{eqn:resumqtilde}
\end{equation}
where the significance of $\tilde{q}$ is not yet determined by this computation.

Substituting this ansatz into the second-order differential equation and changing to the short time variable $z=\frac{x}{\ve}$, we find
\begin{equation}
	2\tilde{q} Y^{1+\ve}Y''-\tilde{q}Y^{\ve}\left(Y'\right)^{2}+Y^{2}(2(1-\ve)-\tilde{q}Y^{\ve})=0,
\end{equation}
where the constant $\tilde{q}$ has not dropped out due to the fact that the differential equation is not invariant under rescalings of $y$.

We make the standard asymptotic ansatz for $Y$ and expanding the constant $\tilde{q}$ as 
\begin{equation}
	\tilde{q}\sim \tilde{q}_{0}+\ve \tilde{q}_{1}+\order{\ve^{2}},
\end{equation}
resulting in the following differential equation for $Y_{0}$
\begin{equation}
	2\tilde{q}_{0}Y_{0}Y_{0}''-\tilde{q}_{0} (Y_{0}')^{2}+(2-\tilde{q}_{0})Y_{0}^{2}=0.
\end{equation}
The solution to this differential equation is
\begin{equation}
	Y_{0}(x)=C\left(\frac{1}{2}e^{\sqrt{\frac{\tilde{q}_{0}-2}{\tilde{q}_{0}}}\left(\frac{x+\tilde{c}}{\ve}\right)}+\frac{1}{2}e^{-\sqrt{\frac{\tilde{q}_{0}-2}{\tilde{q}_{0}}}\left(\frac{x+\tilde{c}}{\ve}\right)}-1\right),
\end{equation}
which, upon comparison with our original solution \eqref{eqn:Y0sol}, matches precisely when $\tilde{q}_{0}=2q$. The remaining constant $C$ that multiplies the whole solution can simply be absorbed into $\tilde{q}_{1}$ as
\begin{equation}
	C^{\ve} \sim 1 +\ve \log(C) +\order{\ve^{2}}.
\end{equation}

Hence, we can see that the inclusion of this constant, $\tilde{q}$ in the resummation equation \eqref{eqn:resumqtilde}, is crucial for obtaining the correct result using the second-order differential equation. Intuitively, this makes sense: the constant $\tilde{q}$ is related to the undetermined boundary condition and, therefore, the size of the apparent horizon. Due to the nature of the resummation, any overall constant that multiplies the solution, such as $C$ here, can only ever contribute at $\order{\ve}$, which ties the position of the apparent horizon to always be $\order{1}$. Since we know that this is not the case, we must account for the unfixed boundary condition within the asymptotic ansatz from the beginning.

\newpage
\section{The Second-Order Correction To Region $\pmb{\cC}$}\label{app:HigherOrd}

In this appendix, we solve for the next-order correction to Region $\cC$ and match it to Region $\cA$. We will then use this correction to show that it improves the leading-order approximation and estimate the size of Region $\cC$.

Let us take the leading-order solution in Region $\cC$ from \eqref{eqn:Y0sol}
\begin{equation}
	Y_{0}(\xc) = \frac{1}{q-1}\left(\frac{1}{2}e^{\xce}+\frac{1}{2}e^{-\xce}-1\right),
\end{equation}
where $\xc=\sqrt{\frac{q-1}{q}}\frac{x+c}{\ve}$, and substitute it into the next-order differential equation that governs $Y_{1}(\xc)$ from \eqref{eqn:Yexp}, giving
\begin{equation}
	\begin{aligned}
		\sinh(\xc)Y_{1}'-\cosh(\xc)Y_{1}=\frac{1}{(q-1)^{2}}&\left((q-1)(\cosh(\xc)-1)\log\left(\frac{q}{2}\right)\right.\\
		&~~\left.+2\log\left(\frac{q}{2(q-1)}(\cosh(\xc)-1)\right)\sinh^{4}\left(\frac{\xc}{2}\right)\right).
	\end{aligned}
\end{equation}
This has a solution given by
\begin{equation}
	\begin{aligned}
		Y_{1}(\xc)= &\frac{\sinh(\xc)}{4(q-1)^{2}}\left[\xc\left(4-\xc-4\log(1-e^{\xc})+2\log\left(\frac{q}{2(q-1)}(\cosh(\xc)-1)\right)\right)\right.\\
		&~~\left.+4\operatorname{Li}_{2}(e^{-\xc})-4\left(q\log\left(\frac{q}{2}\right)+\log\left(\frac{1}{q-1}(\cosh(\xc)-1)\right)\right)\tanh\left(\frac{\xc}{2}\right)\right]\\
		&~~+c_{2} \sinh(\xc),
	\end{aligned}
\end{equation}
where $\operatorname{Li}_{2}(x)$ is the dilogarithm function, and $c_{2}$ is a matching coefficient.

\begin{figure}[H]
	\centering
	\includegraphics[width=0.9\textwidth]{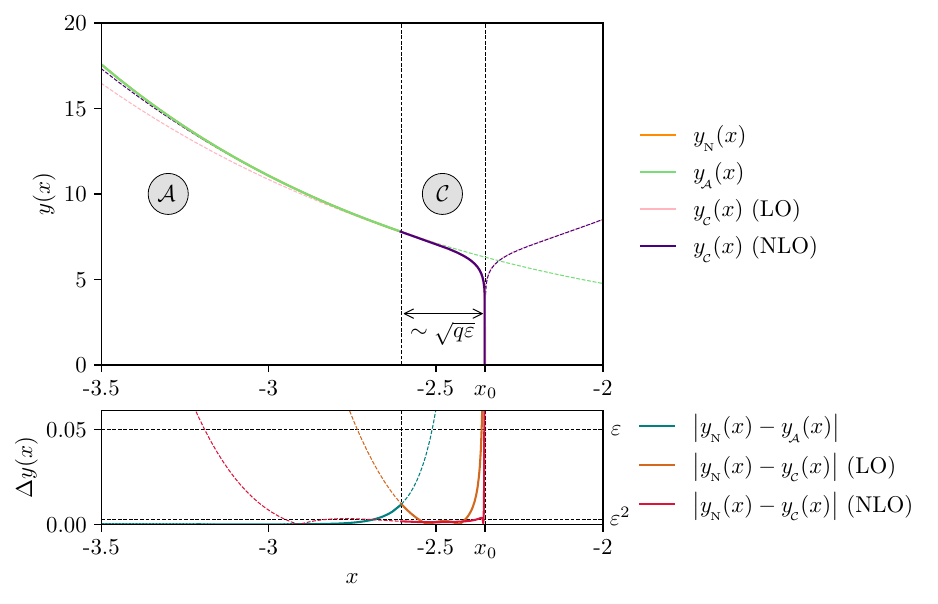}
	\caption{Comparison of the LO and NLO results for the very supercritical solution approximations with the numerical result for $\ve=0.05$. The NLO approximation is clearly an improvement over the LO result.}
	\label{fig:nloyc}
\end{figure}

Next, we match this solution to Region $\cA$. Expanding the solution in terms of the outer variable $x$, or equivalently as $\xc\to -\infty$, we find
\begin{equation}
	\begin{aligned}
		\lim_{\xc\to-\infty}Y_{1}(\xc)\to \frac{e^{-\xc}}{8(q-1)^{2}}&\left(\xc^{2}-2\xc\log\left(\frac{q}{4(q-1)}\right)-4(q-1)^{2}c_{2}\right.\\
		&~~~~~~~~~~~~~~~\left.-\frac{4\pi}{3}+4\log(2)-4q\log\left(\frac{q}{2}\right)\right)
	\end{aligned}
\end{equation}
Substituting this back into \eqref{eqn:ycreparameter}, we find that
\begin{equation}
	\begin{aligned}
		\lim_{\xc\to\infty}\yc \xrightarrow{\sim} &2q+2q\ve\left(\hspace{-0.25em}\log\left(\frac{q}{4(q-1)}\right)-\xc\hspace{-0.3em}\right)+\frac{q\ve^{2}}{6(q-1)}\left[3(2q-1)\left(\hspace{-0.25em}\log\left(\frac{q}{4(q-1)}\right)-\xc\hspace{-0.3em}\right)^{2}\right.\\
		&\left.-12\log\left(\frac{q}{4(q-1)}\right)-3\log^{2}\left(\frac{q}{4(q-1)}\right)-4(\pi^{2}+3(q-1)^{2}c_{2})\right]+\order{\ve^{3}}.
	\end{aligned}
\end{equation}
Now, we take the solution in Region $\cA$ and expand it in terms of the inner variable $\xc$
\begin{equation}
	\begin{aligned}
		\ya \sim & \frac{1}{2}e^{-c_{0}}(1+e^{c_{0}})^{2}+\frac{1}{2}e^{-c_{0}}(e^{2c_{0}}-1)\left(c_{1}-\sqrt{\frac{q}{q-1}}\xc\right)\ve\\
		&+\frac{1}{4}e^{-c_{0}}(1+e^{2c_{0}})\left(c_{1}-\sqrt{\frac{q}{q-1}}\xc\right)^{2}\ve^{2}+\order{\ve^{3}}.
	\end{aligned}
\end{equation}
By comparing these expansions order-by-order in $\ve$, we can determine the matching coefficients to be
\begin{equation}
	\begin{aligned}
		c_{0} &= \log\left(-1+2q+\sqrt{q(q-1)}\right),\\
		c_{1} &= \sqrt{\frac{q}{q-1}}\log\left(\frac{q}{4(q-1)}\right),\\
		c_{2} &= -\frac{1}{12(q-1)^{2}}\left[4\pi^{2}+3\log\left(\frac{q}{4(q-1)}\right)\left(4+\log\left(\frac{q}{4(q-1)}\right)\right)\right].
	\end{aligned}
\end{equation} 
We plot this next-to-leading (NLO) solution in Figure \ref{fig:nloyc} and compare it to the leading-order (LO) solution. 
Although this is clearly an improvement over the LO approximation, it does not improve the approximation of the singularity time $x_{0}$ since
\begin{equation}
	\lim_{x\to c_{0}+\ve c_{1}}Y_{1}(x)\to 0.
\end{equation}
Consequently, we can conclude that corrections to the approximation of the singularity time \eqref{eqn:singapprox} must come from the $Y_{2}$ corrections or higher orders.

Finally, we can use $Y_{1}$, to approximate the size of Region $\cC$. To do this, we look for the point at which perturbation theory breaks down as the solutions approach Region $\cA$. This breakdown occurs when
\begin{equation}
	Y_{0}(\xc)\sim \ve Y_{1}(\xc) ~~\text{as}~~ \ve\to0^{+}, ~~ \xc\to-\infty.
\end{equation}
Taking only the leading-order terms as $\xc\to-\infty$, we find
\begin{equation}
	\frac{1}{2(q-1)}e^{-\xc}\sim \ve\frac{e^{-\xc}\xc^{2}}{8(q-1)^{2}}.
\end{equation}
Rearranging, we find that the breakdown occurs at 
\begin{equation}
	x \sim -c -2\sqrt{q\ve},
\end{equation}
giving the characteristic size of Region $\cC$ that we labelled in Figure \ref{fig:vsupcritapprox} for Region $\cC$.
\newpage
\section{Matching The Critical Solutions} \label{app:Matching}

Due to our inability to solve the governing equation in Region $\cB$, we must use an alternative method to match the approximate solution to Region $\cA$. In this appendix we use the integral representation to find the matching constant for both the metric function and scalar field at criticality.

\subsection{The Metric Function} \label{app:MatchingMet}

We take the integral as defined in \eqref{eqn:defint}
\begin{equation}
	\begin{aligned}
		\xbcap-\xacap &=-\int_{y(\xacape)}^{y(\xbcape)}f(z)\diff{z}\,,
	\end{aligned}
	\label{eqn:defint2}
\end{equation}
where `$\xacap$' is a point in Region $\cA$, `$\xbcap$' is a point in Region $\cB$ and we have defined for convenience
\begin{equation}
	f(z)=\frac{1}{\sqrt{z^{2}-2z+z^{2}\ve\left(\frac{2(1-\ve)^{1-\ve}}{z}\right)^{\frac{1}{\ve}}}}.
\end{equation}

We now choose the points $\xacap$ and $\xbcap$ to be at the past and future asymptotic boundaries respectively since we know the values of $y(\xacap)$ and $y(\xbcap)$ there exactly. They are given by the boundary conditions \eqref{eqn:bdcondy} and \eqref{eqn:boundcrit}
\begin{equation}
	\lim_{\xacape\to-\infty}y(\xacap)\to\frac{1}{2}e^{-\xacape},~~\lim_{\xbcape\to\infty}y\to2(1-\ve).
	\label{eqn:ablimits}
\end{equation}
Hence, let us define the definite integral over the entire domain up to these boundaries as
\begin{equation}
	I=-\int^{\infty}_{2(1-\ve)}f(z)\diff{z}.
	\label{eqn:Iint}
\end{equation}
However, this integral is logarithmically divergent at each limit
\begin{equation}
	\lim_{z\to+\infty}f(z)\to\frac{1}{z},~~ \lim_{z\to2(1-\ve)}f(z)\to -\frac{\sqrt{2\ve}}{z-2(1-\ve)}.
\end{equation}
In order to match, we need to extract a finite number from this integral, Consequently, we regularise our integral, by adding these divergences back in with the opposite sign as `counterterms' so that they cancel
\begin{equation}
	I_{\text{reg}}=I+\int_{2(1-\ve)}^{\infty} \frac{1-\sqrt{2\ve}}{z}\diff{z}+\int_{2(1-\ve)}^{\infty}\frac{\sqrt{2\ve}}{z-2(1-\ve)}\diff{z}.
\end{equation}
Note that the coefficient multiplying the $\frac{1}{z}$ term has been chosen to cancel the contribution from the second term as $z\to\infty$.

To evaluate the regularised integral, we now split the integration domain into two distinct regions, where different terms dominate. This is precisely the splitting that we discovered in Section \ref{sec:3} and so unsurprisingly these regions correspond to Region $\cB$ and Region $\cA$
\begin{equation}
	\int_{2(1-\ve)}^{\infty} = \underbrace{\int_{2(1-\ve)}^{\delta}}_{\Ib}+\underbrace{\int_{\delta}^{\infty}}_{\Ia}.
\end{equation}
The result should be independent of the precise value chosen for $\delta$. This typically serves as a consistency check to confirm that the expansions have been performed correctly. However, due to the complexity of $f(z)$ and its expansions, we set
\begin{equation}
	\delta = 2(1+\ve),
\end{equation}
and verify numerically that the result remains consistent.

We first focus on the domain of integration corresponding to Region $\cB$. The term without the pole in this domain can be directly integrated and expanded in $\ve$ to find
\begin{equation}
	\int^{2(1+\ve)}_{2(1-\ve)} \frac{1-\sqrt{2\ve}}{z}\diff{z}\sim 2\ve +\order{\ve^{\frac{3}{2}}}.
\end{equation}

The next two terms require additional care since they both contain divergences over this domain. Expanding $f(z)$ about $z=2(1-\ve)$ would immediately cancel these divergences and hence seems to be the correct approach. However, this leaves a complicated finite piece that cannot be expressed in a general closed form. 

To circumvent this issue, we instead choose to expand both terms about $z=2$, even though this generates an asymptotic series that must be optimally truncated. The complication that we find with these terms reflects once again the underlying difficulty of Region $\cB$.

The functional part for each of these expansions is given by 
\begin{equation}
	\int_{2(1-\ve)}^{2(1+\ve)} (z-2)^{n} = \frac{2^{n+1} \left(1+(-1)^{n}\right)}{n+1}\ve ^{n+1} ~~\forall n\geq0.
\end{equation}
This allows us to evaluate the pole integral as
\begin{equation}
	\int^{2(1+\ve)}_{2(1-\ve)}\frac{\sqrt{2\ve}}{z-2(1-\ve)}\diff{z} = \sum_{n=0}^{\infty}(-1)^{n}\frac{1+(-1)^{n}}{1+n}\sqrt{2\ve},
\end{equation}
which is a divergent sum, as expected. 

We can then Taylor expand $f(z)$ about $z=2$ to find
\begin{equation}
	\int_{2(1-\ve)}^{2(1+\ve)} f(z)\diff{z} \sim \sqrt{e}\left(\sum_{n=0}^{\infty}\frac{1+(-1)^{n}}{2^{n}(1+n)n!}\gamma_{n}\right)\sqrt{\ve}+\order{\ve^{\frac{3}{2}}},
\end{equation}
where $\gamma_{n}$ diverges as $n$ grows and is given by
\begin{equation}
	\gamma_{n}=-1+\sum_{i=1}^{n}(-1)^{i+1}\left(\frac{(2i+1)^{n-i-1}(n+1-i)\binom{n}{i-1}(2i+1)!!}{i}\right)e^{i}.
	\label{eqn:gamma}
\end{equation}
The divergence from this sum and the divergence from the previous term cancel in a highly non-trivial way due to the choice of expanding about $z=2$.

Combining all terms, we obtain for this part of the domain
\begin{equation}
	\Ib \sim \left(\sum_{n=0}^{\infty}\frac{1+(-1)^{n}}{1+n}\left(\sqrt{2}(-1)^{n}+\sqrt{e}\frac{\gamma_{n}}{2^{n}n!}\right)\right)\sqrt{\ve}+2\ve+\order{\ve^{\frac{3}{2}}}.
\end{equation}

We now turn our attention to the domain of integration corresponding to Region $\cA$. Here, we expand $f(z)$ by treating the third term under the square root as small,
\begin{equation}
	f(z) = \sum_{n=0}^{\infty} \ve^{n}\left(2(1-\ve)^{1-\ve}\right)^{\frac{n}{\ve}}\binom{-\frac{1}{2}}{n}z^{n\left(\frac{2\ve-1}{\ve}\right)}(z(z-2))^{-\frac{1}{2}-n}.
\end{equation}
The zeroth-order term contains the divergence over this domain of integration. We isolate it and combine it with the regularising terms to find upon integration that
\begin{equation}
	\begin{aligned}
		\int^{\infty}_{2(1+\ve)}\left(\frac{1-\sqrt{2\ve}}{z}+\frac{\sqrt{2\ve}}{z-2(1-\ve)}-\frac{1}{\sqrt{z^{2}-2z}}\right)\diff{z} \sim & -2\log(2)+\sqrt{2\ve}(\sqrt{2}-\log(2\ve))\\
		&-\ve+\order{\ve^{\frac{3}{2}}}.
	\end{aligned}
\end{equation}

The remaining terms can also be integrated easily. First, we remove the $\ve$ dependence from the integration limits using the change of variables $z=2(1+\ve)u^{\ve}$. This allows us to expand in  $\ve$ before integrating. The final result is
\begin{equation}
	-\int^{\infty}_{2(1+\ve)}\left(f(z)-\frac{1}{\sqrt{z^{2}-2z}}\right)\diff{z}\sim -\sqrt{\ve}\sum^{\infty}_{n=1}\binom{-\frac{1}{2}}{n}e^{-n}E_{n+\frac{1}{2}}(n)+\order{\ve^{\frac{3}{2}}},
\end{equation}
where $E_{n}(x)$ is the generalised exponential integral. This sum converges quickly to a finite value and is not divergent.

Combining all of the contributions from Region $\cA$, we find
\begin{equation}
	\Ia \sim -2\log(2)+\sqrt{\ve}\left(2-\sum^{\infty}_{n=1}\binom{-\frac{1}{2}}{n}e^{-n}E_{n+\frac{1}{2}}(n)-\sqrt{2}\log(2\ve)\right)-\ve+\order{\ve^{\frac{3}{2}}}.
\end{equation}

We now have the result for the full regularised integral
\begin{equation}
	I_{\text{reg}} \sim -2\log(2)+\sqrt{2\ve}(\tilde{b}_{0}+\log(2)-\log(\ve))+\ve +\order{\ve^{\frac{3}{2}}},
\end{equation}
where $\tilde{b}_{0}$ is given by 
\begin{equation}
	\begin{aligned}
		\tilde{b}_{0} =\frac{1}{\sqrt{2}}\bigg(2 -\sum_{n=1}^{\infty}\binom{-\frac{1}{2}}{n}e^{-n}  E_{n+\frac{1}{2}}(n)&+\sqrt{e}\sum_{n=0}^{\infty}\frac{1+(-1)^{n}}{2^{n}(1+n)n!}\gamma_{n}\\
		&+\sqrt{2}\sum_{n=0}^{\infty} (-1)^{n}\frac{1+(-1)^{n}}{1+n} \bigg)-2\log(2).
	\end{aligned}
\end{equation}

Now we use this to find how $\tilde{b}_{0}$ relates to our matching coefficient $b_{0}$. We begin by noting that
\begin{equation}
	{\raisebox{1ex}{$\lim$}}_{\hspace{-5ex}\substack{\xacape\rightarrow -\infty\\\xbcape\rightarrow-\infty}}\int^{y(\xacape)}_{y(\xbcape)}\left(-f(z)+\frac{1-\sqrt{2\ve}}{z}+\frac{\sqrt{2\ve}}{z-2(1-\ve)}\right)\diff{z}\to I_{\text{reg}},
\end{equation}
and so upon evaluating the left-hand side algebraically (before taking the limits), we find
\begin{equation}
	\xacap-\xbcap+\log\left(\frac{y(\xacape)}{y(\xbcape)}\right)+\sqrt{2\ve}\left(\log(y(\xbcap))-\log(y(\xacap))+\log\left(\frac{y(\xacape)-2(1-\ve)}{y(\xbcape)-2(1-\ve)}\right)\right).
\end{equation}
Now taking the limits, and using the boundary conditions \eqref{eqn:ablimits} where we do not encounter a divergence, we obtain
\begin{equation}
	\lim_{\xbcap\to\infty}\left(-\xbcap -2\log(2)+\ve+\sqrt{2\ve}\Big(\log(2)-\log\big(y(\xbcap)-2(1-\ve)\big)\Big)+\order{\ve^{\frac{3}{2}}}\right) \xrightarrow{\sim} I_{\text{reg}}.
\end{equation}
Using the expression for the regularised integral and rearranging terms for $y(\xbcap)$, we find an expression for the solution in Region $\cB$ near the asymptotic future boundary
\begin{equation}
	\lim_{\xbcap\to\infty}y(\xbcap) \xrightarrow{\sim} 2(1-\ve)+\ve e^{-\frac{1}{\sqrt{2\ve}}(\xbcape+\sqrt{2\ve}\tilde{b}_{\scriptscriptstyle{0}})}+\order{\ve^{2}}.
\end{equation}
By comparing this with the ansatz for Region $\cB$ \eqref{eqn:critsol}, we identify the matching constant as
\begin{equation}
	b_{0}=\tilde{b}_{0}.
\end{equation}
Optimal truncation of the sum gives the numerical value
\begin{equation}
	b_{0} \approx -0.281221.
\end{equation}

This method can easily be extended to also find the matching coefficient at higher orders in $\ve$. In fact, doing so shows that at leading order in $x$, the $\order{\ve^{2}}$ correction will be another decaying exponential of the same form.

\subsection{The Scalar Field} \label{app:MatchingSca}

Given our discussion at the beginning of Section \ref{sec:3.3}, it may seem strange that the same method works for the scalar field --- the shift in $x$ is already fixed by the previous computation. However, for the scalar field, the matching constant \textit{still} corresponds to a shift symmetry. Consequently we can use the same method if we write the differential equation in the right way.

To do this, we take the first integrated form of the equation of motion of the scalar field \eqref{eqn:phiampexpl} and perform the chain rule to find an autonomous differential equation for $y$ in terms of $\phi$
\begin{equation}
	y'(\phi) = -\frac{2\sqrt{\ve}}{(2(1-\ve))^{\frac{1}{2\ve}}}\sqrt{y^{\frac{1+2\ve}{\ve}}-2y^{\frac{1+\ve}{\ve}}+2^{\frac{1}{\ve}}\ve(1-\ve)^{\frac{1-\ve}{\ve}}y^{2}}.
	\label{eqn:phiampexpl2}
\end{equation}
From here, the computation proceeds almost identically to the metric function.
After integrating once, we find 
\begin{equation}
	\phiacap-\phibcap = \int^{y(\phiacape)}_{y(\phibcape)}g(z)\diff{z},
\end{equation}
where $\phiacap=\phi(\xacap)$ and $\phibcap=\phi(\xbcap)$ are the values of the scalar field evaluated at a point in Region $\cA$ and at a point in Region $\cB$ respectively. We have also defined for convenience
\begin{equation}
	g(z)= \frac{(2(1-\ve))^{\frac{1}{2\ve}}}{2\sqrt{\ve}}\frac{1}{\sqrt{z^{\frac{1+2\ve}{\ve}}-2z^{\frac{1+\ve}{\ve}}+2^{\frac{1}{\ve}}\ve(1-\ve)^{\frac{1-\ve}{\ve}}z^{2}}}.
\end{equation}
We now fix these two points to be at the asymptotic boundaries once again, which are given by
\begin{equation}
	\lim_{\phiacap\to0}y(\phiacap)\to \infty, ~~ \lim_{\phibcap\to\infty}y(\phibcap)\to 2(1-\ve),
	\label{eqn:boundcondscal}
\end{equation}
Allowing us to define the definite integral for this problem as
\begin{equation}
	J = -\int^{\infty}_{2(1-\ve)}g(z)\diff{z}.
	\label{eqn:Jint}
\end{equation}
This time, the integral is only logarithmically divergent at one boundary
\begin{equation}
	\lim_{z\to 2(1-\ve)}g(z) \to \frac{1}{\sqrt{2}}\frac{1}{(z-2(1-\ve))}.
\end{equation}
Adding this as a counterterm to the integral on its own will create another logarithmic divergence as $z\to \infty$, forcing us to introduce a second counterterm to regularise it. The regularised integral can hence be written
\begin{equation}
	J_{\text{reg}}=J + \int^{\infty}_{2(1-\ve)}\frac{1}{\sqrt{2} (z-2(1- \ve ))}\diff{z}-\int^{\infty}_{2(1-\ve)}\frac{1}{\sqrt{2} z}\diff{z}
\end{equation}

To evaluate the regularised integral, we split the integration domain into two distinct regions once again
\begin{equation}
	\int_{2(1-\ve)}^{\infty} = \underbrace{\int_{2(1-\ve)}^{2(1+\ve)}}_{\Jb}+\underbrace{\int_{2(1+\ve)}^{\infty}}_{\Ja},
\end{equation}
where we have chosen the position of the splitting to be the same as with the metric function computation.

Let us first focus on the domain of integration corresponding to Region $\cB$. The term without the pole in this domain can be directly integrated and expanded in $\ve$ to find 
\begin{equation}
	\int_{2(1-\ve)}^{2(1+\ve)}\frac{1}{\sqrt{2}z}\diff{z}\sim \sqrt{2}\ve +\order{\ve^{3}}.
\end{equation}
The remaining two terms once again require care and for the same reasons we expand $g(z)$ about $z=2$. This allows us to evaluate the term that regularises the integral in this domain as
\begin{equation}
	\int_{2(1-\ve)}^{2(1+\ve)}\frac{1}{\sqrt{2} (z-2(1- \ve ))}\diff{z} = \frac{1}{\sqrt{2}}\sum_{n=0}^{\infty} \frac{(-1)^{n}(1+(-1)^{n})}{(1+n)},
\end{equation}
and the $g(z)$ term as
\begin{equation}
	\int_{2(1-\ve)}^{2(1+\ve)}g(z)\diff{z} \sim 1+\sum_{n=2}^{\infty}\frac{1+(-1)^{n}}{(1+n)}\frac{\bar{\gamma}_{n}}{4n!},
\end{equation}
where
\begin{equation}
	\bar{\gamma}_{n}=2\sum_{i=0}^{n-1}\left(\frac{-e^{i}}{2}\right)^{i+1}\binom{n}{i+1}(2i+1)!!(i+1)^{n-i-1}.
	\label{eqn:gammabar}
\end{equation}
The sum is once again only asymptotic.

Combining all terms, we obtain the result for this part of the domain
\begin{equation}
	\Jb \sim \sqrt{2}-1+\sum_{n=2}^{\infty}\frac{1+(-1)^{n}}{n+1}\left(\frac{(-1)^{n}}{\sqrt{2}}+\frac{\bar{\gamma}_{n}}{4n!}\right) +\order{\ve}.
\end{equation}

We now turn our attention to the domain of integration corresponding to Region $\cA$. Since there is no divergence associated with this domain from $g(z)$, we can immediately integrate the counterterms to find
\begin{equation}
	\int_{2(1+\ve)}^{\infty}\left(\frac{1}{\sqrt{2} (z-2(1- \ve ))}-\frac{1}{\sqrt{2} z}\right)\diff{z} \sim -\frac{1}{\sqrt{2}}\log(2\ve).
\end{equation}
Now, we expand $g(z)$ by treating the third term under the square root as small
\begin{equation}
	g(z)=\sum_{n=0}^{\infty}\binom{-\frac{1}{2}}{n}  2^{\frac{2 n-2 \ve +1}{2 \ve }} (1-\ve )^{\frac{2 n (1-\ve )+1}{2 \ve }} \ve ^{n-\frac{1}{2}} z^{-\frac{-(2 n-1) \ve +2 n+1}{2 \ve }}(z-2)^{-n-\frac{1}{2}}.
\end{equation}
These terms can be integrated easily by first removing the $\ve$ dependence from the integration limits using the change of variables $z=2(1+\ve)u^{\ve}$ and expanding in $\ve$ before integrating. The final result is
\begin{equation}
	\int_{2(1+\ve)}^{\infty}g(z)\diff{z} \sim \frac{1}{2}\sum_{n=0}^{\infty}e^{-n-\frac{1}{2}} \binom{-\frac{1}{2}}{n} E_{n+\frac{1}{2}}\left(n+\frac{1}{2}\right) + \order{\ve},
\end{equation}
where $E_{n}(x)$ is the generalised exponential integral.

Combining all of the contributions in Region $\cA$, we find
\begin{equation}
	\Ja \sim -\frac{1}{\sqrt{2}}\log(2\ve)-\frac{1}{2}\sum_{n=0}^{\infty}e^{-n-\frac{1}{2}} \binom{-\frac{1}{2}}{n} E_{n+\frac{1}{2}}\left(n+\frac{1}{2}\right) + \order{\ve}.
\end{equation}

We can now write the result for the full regularised integral as 
\begin{equation}
	J_{\text{reg}} \sim -\frac{1}{\sqrt{2}}\log(2\ve) + \tilde{a}_{0} +\order{\ve},
\end{equation}
where
\begin{equation}
	\tilde{a}_{0} = \sqrt{2}-1+\sum_{n=2}^{\infty}\frac{1+(-1)^{n}}{n+1}\left(\frac{(-1)^{n}}{\sqrt{2}}+\frac{\bar{\gamma}_{n}}{4n!}\right)-\frac{1}{2}\sum_{n=0}^{\infty}e^{-n-\frac{1}{2}} \binom{-\frac{1}{2}}{n} E_{n+\frac{1}{2}}\left(n+\frac{1}{2}\right).
\end{equation}

Now we use this to find how $\tilde{a}_{0}$ relates to our matching coefficient $a_{0}$. We begin by noting that
\begin{equation}
	{\raisebox{1ex}{$\lim$}}_{\hspace{-5ex}\substack{\phiacape\rightarrow -\infty\\\phibcape\rightarrow-\infty}}\int^{y(\phiacape)}_{y(\phibcape)}\left(-g(z)+\frac{1}{\sqrt{2} (z-2(1- \ve ))}-\frac{1}{\sqrt{2} z}\right)\diff{z}\to J_{\text{reg}},
\end{equation}
and so upon evaluating the left hand side algebraically (before taking the limits), we find
\begin{equation}
	\phiacap-\phibcap +\frac{1}{\sqrt{2}}\left(\log(y(\phibcap))-\log(y(\phibcap)-2(1-\ve))+\log\left(\frac{y(\phiacap)-2(1-\ve)}{y(\phiacap)}\right)\right).
\end{equation}
Now taking the limits using the boundary conditions \eqref{eqn:boundcondscal} and our solution to the metric function in Region $\cB$, we find
\begin{equation}
	\lim_{\xbcape\to\infty}\left(-\phi(\xbcap)+\frac{1}{\sqrt{2}}\left(\log(2)-\log(\ve)+\frac{x+\sqrt{2\ve}b_{0}}{\sqrt{2\ve}}\right)+\order{\ve}\right)\xrightarrow{\sim} J_{\text{reg}}
\end{equation}
After rearranging, we find an expression for the solution in Region $\cB$ for the scalar field near the asymptotic future boundary
\begin{equation}
	\lim_{\xbcape}\phi(\xbcap)\xrightarrow{\sim} \sqrt{2}\log(2)-\tilde{a}_{0}+\frac{x+\sqrt{2\ve}b_{0}}{\sqrt{2\ve}} +\order{\ve}.
\end{equation}
By comparing this with the ansatz \eqref{eqn:scalarcritans}, we identify the matching constant as
\begin{equation}
	a_{0} = \sqrt{2}\log(2)-\tilde{a}_{0}.
\end{equation}
Optimal truncation of the sum gives the numerical value
\begin{equation}
	a_{0} \approx 0.801973.
\end{equation}
\newpage
\section{Table of Notation}\label{app:Notation}
\begin{longtable}[htbp]{@{}clc@{}}
	\toprule
	\textbf{Symbol} & \textbf{Meaning} & \textbf{Reference} \\ 
	\bottomrule
	\addlinespace[3pt]
	$D$ & The number of dimensions & \textsection \ref{sec:Section 2} \\
	$\ve$  & The small parameter related to the inverse number of dimensions & \eqref{eqn:defeps}\\
	$\gamma$  & Mass Critical Exponent & \eqref{eqn:critexpdef}\\
	$q$ & The amplitude of the incoming scalar field & \eqref{eqn:1y2}\\
	$\delta_{q}$ & The distance from criticality (in CSS) & \eqref{eqn:distcritq}\\
	$p$ & The CSS-breaking parameter & \eqref{eqn:massscalp} \\
	$\delta_{p}$ & The distance from criticality (breaking CSS) & \eqref{eqn:breakingss}\\
	\midrule
	$R$  & The Ricci Scalar & \eqref{eqn:action}\\
	$\phi$  & The Scalar Field & \eqref{eqn:action}\\
	$\rho$  & The geometric radius metric function & \eqref{eqn:metricp}\\
	$\Sigma$ & Component of the metric & \eqref{eqn:metricp}\\
	$\cR$ & The conformal radius function & \eqref{eqn:redefinedmet}\\
	$y$ & Redefined conformal radius function & \eqref{eqn:ydef} \\
	$Y$ & Redefined conformal radius function for Region $\cC$ & \eqref{eqn:ycreparameter}\\
	\midrule
	$\theta_{i}$ & Angular coordinates & \textsection \ref{sec:2.1}\\
	$u$  & Retarded time, null coordinate & \textsection \ref{sec:2.1}\\
	$v$ & Advanced time, null coordinate & \textsection \ref{sec:2.1}\\
	$x$ & Self-similar time & \eqref{eqn:selfsimilartime}\\
	$s$ & Scaling coordinate & \eqref{eqn:scalingcoords}\\
	$\omega$ & Alternative scaling coordinate & \eqref{eqn:scalingcoordw}\\
	$\eta$ & Redefined self-similar time variable & \textsection \ref{sec:4.2}\\
	\midrule
	$\cA$ & Label used for the flat space region & \textsection \ref{sec:3}\\
	$\cB$ & Label used for the transition region near the saddle point &\textsection \ref{sec:3}\\
	$\cC$ & Label used for the near-horizon/near-singularity region &\textsection \ref{sec:3} \\
	\midrule
	$M,m$   & Mass/ Misner Sharp mass function & \eqref{eqn:massfunc}\\
	$\xi$ & Homothetic Vector Field (HVF) & \eqref{eqn:hvfmet}\\
	$U^{\mu}$ & The fluid velocity & \eqref{eqn:fluidvel}\\
	$f$ & Redefined metric function for Region $\cB$ & \eqref{eqn:feq}\\
	\midrule
	$\Delta$ & DSS Period& \textsection \ref{sec:2.1}\\
	$x_{0}$ & The \textit{exact} time at which a singularity forms & \eqref{eqn:nearsing}\\
	$\Lambda$ & A rescaled amplitude of the incoming scalar field for near-criticality & \eqref{eqn:sclosecrit}\\
	$c$ & Very supercritical metric matching constant & \eqref{eqn:Y0sol} \\
	$b_{0}$ & Critical metric matching constant  & \eqref{eqn:critsol} \\
	$d_{i}$ & Numerically determined metric matching constants & \eqref{eqn:margsub} \\
	$\alpha_{n}$ & Non-linear recursion coefficients for $y$ & \eqref{eqn:alphares} \\
	$\beta_{n}$ & Non-linear recursion coefficients for $\phi$ & \eqref{eqn:betacoef}\\
    \bottomrule
\end{longtable}
\begin{longtable}[htbp]{@{}clc@{}}
	\toprule
	\textbf{Symbol} & \textbf{Meaning} & \textbf{Reference} \\ 
	\bottomrule
	\addlinespace[3pt]    
	$\hat{A}$ & Scalar field amplitude & \eqref{eqn:scalarampd}\\
	$\hat{B}$ & Scalar field shift in Region $\cA$ & \eqref{eqn:scalarina}\\
	$a_{0}$ & Critical scalar field matching constant & \eqref{eqn:scalarcritans}\\
	\midrule
	$F$ & The set of all fields in a gravitational collapse problem & \textsection \ref{sec:4.1}\\
	$F_{\text{ss}}$ & The set of self-similar fields in a gravitational collapse problem & \eqref{eqn:SSRGT}\\
	$G$ & Set of linearly perturbed fields breaking CSS & \eqref{eqn:breakingss}\\
	$g_{i}$ & Linearly perturbed fields with only $x$-dependence & \eqref{eqn:Gdecomp}\\
	$O$ & Renormalisation Group Transformation (RGT) & \eqref{eqn:RGT}\\
	$\hat{L}_{O}$ & Infinitesimal generator of the RGT & \eqref{eqn:RGTinf}\\
	$T$ & The linearised RGT (tangent map) & \eqref{eqn:linRGT}\\
	$\hat{L}_{T}$ & Infinitesimal generator of the tangent map & \eqref{eqn:inflinRGT}\\
	$k$ & Lyapunov exponent for perturbations from criticality & \eqref{eqn:linactRGT} \\
	$\kappa$ & The largest real part of the spectrum of Lyapunov exponents & \eqref{eqn:critexp} \\
	$H$ & Horizon formation function & \eqref{eqn:horizonfunc}\\
	\midrule
	$r$ & The perturbation of the radial metric function & \eqref{eqn:pertans}\\
	$\varphi$ & The perturbation of the scalar field & \eqref{eqn:pertans} \\
	$\sigma$ & The perturbation of the other metric component & \eqref{eqn:pertans}\\
	$a_{4}$ & A constant controlling the spectrum of perturbations in $D=4$& \eqref{eqn:hyparg}\\
	$b_{4}$ & A constant controlling the spectrum of perturbations in $D=4$ & \eqref{eqn:hyparg} \\
	$a_{\ve}$ & A constant controlling the spectrum of perturbations in general $D$ & \eqref{eqn:expargs}\\
	$b_{\ve}$ & A constant controlling the spectrum of perturbations in general $D$ & \eqref{eqn:expargs}\\
	\midrule
	$I$ & Definite integral for matching the metric function at critical & \eqref{eqn:Iint}\\
	$J$ & Definite integral for matching the scalar field at critical & \eqref{eqn:Jint} \\
	$\gamma_{n}$ & Terms in a series for the metric function matching coefficient & \eqref{eqn:gamma} \\
	$\bar{\gamma}_{n}$ & Terms in a series for the scalar field matching coefficient & \eqref{eqn:gammabar}\\
	\midrule
	$W(x)$ & The Lambert function & \cite{mathworld}\\
	$E_{n}(x)$ & The generalised exponential integral & \cite{mathworld} \\
	$B_{z}(a,b)$ & The incomplete beta function & \cite{mathworld} \\
	$\,_2 F_1$ & Hypergeometric function & \cite{mathworld} \\
	\bottomrule
\end{longtable}

\clearpage
\phantomsection
\addcontentsline{toc}{section}{References}
\bibliographystyle{utphys}
{\linespread{1.075}
\bibliography{Refs}
}

\end{document}